\documentclass[prd,twocolumn,showpacs,floats,floatfix,nofootinbib]{revtex4}
\usepackage{dcolumn}
\usepackage{amssymb}
\usepackage{natbib}
\usepackage{bm}
\usepackage[pdftex]{graphicx}
\usepackage{epstopdf}
\def\spose#1{\hbox to 0pt{#1\hss}}
\def\lta{\mathrel{\spose{\lower 3pt\hbox{$\mathchar"218$}}
     \raise 2.0pt\hbox{$\mathchar"13C$}}}
\def\gta{\mathrel{\spose{\lower 3pt\hbox{$\mathchar"218$}}
     \raise 2.0pt\hbox{$\mathchar"13E$}}}
\newcommand{\be}{\begin{equation}}
\newcommand{\en}{\end{equation}}
\newcommand{\bea}{\begin{eqnarray}}
\newcommand{\ena}{\end{eqnarray}}
\newcommand{\ex}{\mbox{e}}
\newcommand{\dd}{\mbox{d}}

\newcommand{\cL}{c_{_\mathrm{L}}}
\newcommand{\cT}{c_{_\mathrm{T}}}
\newcommand{\mrm}[1]{\mathrm{#1}}
\newcommand{\mcl}[1]{\mathcal{#1}}
\newcommand{\ee}{\end{equation}}
\newcommand{\eea}{\end{eqnarray}}

\newcommand{\lb}{\left(}
\newcommand{\rb}{\right)}
\newcommand{\lsb}{\left[}
\newcommand{\rsb}{\right]}

\font\openface=msbm10 at10pt

\def\Reals{{\hbox{\openface R}}}

\def\setR{\mathbb{R}}
\def\setC{\mathbb{C}}

\newcommand{\ie}{\textsl{i.e.}}

\newcommand{\eg}{\textsl{e.g.}}
\newcommand{\apriori}{\textsl{a priori }}
\newcommand{\nablas}{\overline{\nabla}}
\newcommand{\tr}{\mbox{Tr}}
\begin{document}
\title{Coupled currents in cosmic strings.}
\author{Marc Lilley}
\email{lilley@iap.fr} \affiliation{Institut d'Astrophysique de Paris,
  ${\cal G}\setR\varepsilon\setC{\cal O}$, FRE 2435-CNRS, 98bis
  boulevard Arago, 75014 Paris, France}
\author{Xavier Martin}
\email{xavier.martin@univ-tours.fr} \affiliation{Laboratoire de Math\'ematiques et Physique Th\'eorique CNRS, UMR 6083, F\'ed\'eration de Recherche Denis Poisson (FR 2964), Universit\'e Fran\c{c}ois Rabelais, Parc de Grandmont, 37200 Tours, France}
\author{Patrick Peter} \email{peter@iap.fr} \affiliation{Institut
d'Astrophysique de Paris, ${\cal G}\setR\varepsilon\setC{\cal
O}$, FRE 2435-CNRS, 98bis boulevard Arago, 75014 Paris, France}
\begin{abstract}
We first examine the microstructure of a cosmic string endowed with two simple Abelian currents.  This microstructure depends on two state parameters.  We then provide the macroscopic description of such a string and show that it depends on an additional Lorentz-invariant state parameter that relates the two currents. We find that in most of the parameter space, the two-current string is essentially equivalent to the single current-carrying string, \ie, only one field condenses onto the defect.  In the regions where two currents are present, we find that as far as stability is concerned, one can approximate the dynamics with good accuracy using an analytic model based on either a logarithmic (on the electric side, \ie, for timelike currents) or a rational (on the magnetic side, \ie, for spacelike currents) worldsheet Lagrangian.
\end{abstract}
\date{\today}
\maketitle
\section{Introduction}
Among all possible topological defects~\cite{kibble1,kibble2,book} that may have appeared as consequences of early phase transitions, only cosmic strings may still be compatible with cosmology.  They generically arise~\cite{BPRS,PBRS02} in the hybrid inflationary models~\cite{hybrid1,hybrid2} that are implemented in Grand Unified Theories~\cite{rachel1,rachel2,john} and are an unavoidable consequence of Grand Unified Theories and supersymmetry breaking when there exists a low-energy unbroken $R-$parity. More recently, cosmic strings have been the subject of renewed interest because it has been realized that such configurations could also exist as cosmologically relevant solutions in the context of string theory where $D3/\bar{D}3$ and $D3/D7$ brane inflationary models lead to the formation of D-strings at the end of the expanding phase.  It has been suggested~\cite{DKVP04} that these objects correspond to D-term cosmic strings in supergravity theory.\\

In many instances, \eg, in a supersymmetric theories~\cite{susystring}, the Lorentz invariance that is typical of Nambu-Goto strings~\cite{GN1,GN2} used in cosmological settings~\cite{network1,network2,network3,network4,network5,network6,network7}, can be broken by the appearance of currents flowing along the worldsheet~\cite{witten1,witten2}. This effect is known to halt cosmic string loop decay caused by dissipative effects and yield new equilibrium configurations named vortons~\cite{vortons0,vortons1,vortons2,vortons3,vortons4,vortons5}.  The density of these vortons is tightly constrained by the value of the normalized density $\Omega\equiv \rho /\rho_\mathrm{crit}$ today and by primordial nucleosynthesis~\cite{vortons6,vortons7}.  A non current-carrying string network also has a characteristic energy scale $m_\mathrm{cs}$ which is constrained directly by the time stability measurements of binary pulsars~\cite{BinPuls2007} and indirectly by the observation that the Cosmic Microwave Background (CMB) \cite{Komatsu:2008hk} is dominated by the signal produced by the amplification of primordial density fluctuations during inflation~\cite{CMBstring2007}.\\

Vorton (equilibrium) states can be destabilized~\cite{1D1,1D2,1D3} through various mechanisms, \eg, with a coupling of the Higgs field to an electromagnetic current~\cite{gpb,correcem,boucleem}, with the existence of shocks or in the presence of high curvature regions~\cite{2D,3D}.  The vortons are described by means of various possible internal equations of state~\cite{formal1,formal2,formal3,formal4,models} which stem from a detailed study of the internal microscopic structure of the vortices that arise from the coupling of the string-forming Higgs field to a bosonic current-carrier~\cite{current,neutral,enon0}. More recently, vorton instabilities have (to some extent) been confirmed in large field theory simulations~\cite{yl}.\\

The macroscopic formalism developed by Carter~\cite{formal1,formal2,formal3,formal4} and that describes general $p-$branes embedded in a $n$-dimensional spacetime can be used to study the dynamics of current-carrying cosmic strings. As it stands, the formalism applies to the case of a single current for which a single state parameter provides a complete description of the string.  However, it cannot be used for a system described by two or more state parameters as is the case when the Higgs field couples to fermions~\cite{fermions0}. It gets worse when massive modes are present in the spectrum~\cite{fermionsM}, because the current induces qualitatively different macroscopic properties when the state parameters are changed.  There is no complete description of a general ``many-current'' worldsheet and only the case of a ``cold'' superconducting current coupled to a ``hot'' entropy current appropriate for the ``warm'' string model has been treated~\cite{2c} (see however the objections concerning the use of the word ``superconducting'' string in this reference).\\

The two-current case provides the one-dimensional analog of the ordinary 3D Landau superfluid model~\cite{landau}.  It has been shown that hydrodynamical systems with more than one current present new~\cite{KH} instabilities, as, \eg, the two-stream plasma instability~\cite{Prix,instab2007} (see Ref.~\cite{AmJ} for a pedagogical presentation of this issue).  In order to test the stability of a vorton state against perturbations in more than a single current, we provide in this paper one more step towards a complete formalism for the many-current carrying cosmic string.  For that purpose, we consider in Section~\ref{micro} a specific model with two scalar fields trapped in the vortex core, leading to two conserved currents, and thus three boost-invariant state parameters.  Being strictly local, the solutions of the field equations in the microscopic theory for the Nielsen-Olesen vortex ansatz~\cite{NO} depend on only two out of the three state parameters, respectively proportional to the amplitude of the two independent currents. These two state parameters therefore completely determine the string structure at any point along the worldsheet.  On the other hand, the integrated quantities introduced in Section~\ref{macro} and the worldsheet dynamics described in Sections~\ref{secdyn} and \ref{sectwocurr} exhibit an additional dependence on a third state parameter, a quantity proportional to the scalar product of the two currents.  In fact, Section~\ref{secdyn} deals with the worldsheet description of a string endowed with a set of $N$ condensates, in which case there exists, in addition to the usual $N$ state parameters, an extra set of $N(N-1)/2$ state parameters given by the scalar products of all pairs of distinct currents.  This number is equal to 1 when $N=2$.  In Sections~\ref{secdyn} and \ref{sectwocurr}, we also consider the internal stability of a string endowed with $N$ condensates ($N=2$ in Section \ref{sectwocurr}) and show that, contrary to what the three-dimensional case suggests, no new instability is predicted when several currents are involved.\\

Finally, it is known that single current-carrying strings can be described by a macroscopic Lagrangian depending on a single string state parameter~\cite{neutral,models}. In the case at hand, and since the field equations only depend on two state parameters, one could think of similarly describing a two-current string by means of a sum of Lagrangians, one for each current.  In Section~\ref{simplifiedmodel}, we compare the dynamics of the string as given by the fully interacting theory of previous sections to the dynamics of this sum of individual Lagrangians that describe non-interacting fields and that provide the possibility to perform a fully analytic treatment of the string physics.  In principle, the two approaches do not describe the exact same physics, even when the coupling term of the fields  in fully interacting theory is set to zero, but our results exhibit satisfactory agreement between the two for zero and even for weak coupling offering the possibility to study strings with several weakly coupled field condensates in a fully analytic way.
\section{Two complex scalar fields model}
\label{micro}
In order to introduce a current-current coupling in a vorton state, we couple a Higgs field $H$, charged under a broken local $U(1)$ symmetry with associated gauge field $C_{\mu}$ and charge $q$, to two uncharged complex scalar fields, $\Phi$ and $\Sigma$, with global $U(1)$ symmetry,
\begin{widetext}
\begin{equation}
{\cal L}= - \frac{1}{2} \left(D_\mu H\right)^\dagger \left( D_\mu H\right) - \frac{1}{4} C_{\mu\nu}C^{\mu\nu} -\frac{\lambda}{8} \lb|H|^{2} - \eta^{2}\rb^{2} - \frac{1}{2} \partial_\mu \Phi^\star \partial^\mu \Phi- \frac{1}{2}\partial_\mu \Sigma^\star\partial^\mu \Sigma - V\lb H, \Phi, \Sigma\rb.
\label{lag}
\end{equation}
The interaction potential for $\Phi$ and $\Sigma$ is given by
\begin{equation}
V\lb H, \Phi, \Sigma\rb=\frac{1}{2} m_\phi^2 |\Phi|^{2} + \frac{1}{2} m_\sigma^2 |\Sigma|^{2}+ \frac{1}{2} \left(|H|^{2} - \eta^{2}\right) \left( f_\phi |\Phi|^2 + f_\sigma |\Sigma|^{2} \right) + \frac{1}{4}  \lambda_\phi |\Phi|^{4} + \frac{1}{4} \lambda_\sigma |\Sigma|^{4} + \frac{g}{2} |\Phi|^{2} |\Sigma|^{2}.
\label{V}
\end{equation}
\end{widetext}
The kinetic term for $H$ reads $D_{\mu}H = \lb\partial_{\mu} + iq C_{\mu}\rb H$ and $C_{\mu\nu}\equiv \partial_{\mu} C_{\nu} - \partial_{\nu} C_{\mu}$.  The vacuum is defined as usual as the minimum of the potential, \ie, through $\delta V/\delta H =\delta V/\delta \Phi =\delta V/\delta \Sigma =0$.  This leads to a system of three cubic equations for the field amplitudes $|H|$, $|\Phi|$ and $|\Sigma|$.  Although these equations are soluble in general, we shall only be interested in the regions of parameter space for which the absolute minimum is at $|\langle H\rangle_0|^2 = \eta^2$ (breaking the $U(1)^{\rm local}$ ymmetry) and $\Phi=\Sigma = 0$. Such a choice is made implicitely with the potential written as in Eq.~(\ref{V}).\\

There exist two additional $U(1)^{\rm global}$ symmetries, carried by $\Phi$ and $\Sigma$, that independently leave the vacuum unchanged. Before symmetry breaking, one therefore has $U(1)^{\rm local}\times U(1)^{\rm global}\times U(1)^{\rm global}$. In the case $f_\phi=f_\sigma$ and $\lambda_\phi=\lambda_\sigma = g$, one has $U(1)^{\rm global}\times U(1)^{\rm global}\rightarrow SU(2)^{\rm  global}$.\\

At zero temperature, the theory stemming from Eq.~(\ref{lag}) admits vortex-like solutions.  In order to study locally cylindrically symmetric cosmic string configurations, the reference frame is chosen locally aligned with the vortex, and the $z$ axis is defined to lie along the string. We denote by $r$ and $\theta$ the usual cylindrical coordinates centered on the string location and express the solution using the Nielsen-Olesen~\cite{NO} {\sl ansatz}
\begin{equation}
H\lb x^\alpha\rb = h(r) \ex^{i n\theta} \hbox{ and } C_\mu (x^\alpha)= \displaystyle{Q(r)-n\over q} \delta_\mu^\theta,
\label{NO}
\end{equation}
for a string with winding number $n$, $h(0)=0$, and $Q(0)=n$.  Far from the vortex, one recovers the vacuum state defined by $h(\infty)=|\langle H\rangle_0| = \eta$ and $Q(\infty)=0$.\\

In this solitonic background, the breaking of $U(1)^{\rm global}\times U(1)^{\rm global}$, dynamically generated as the bosonic fields $\Phi$ and $\Sigma$ condense in the string, can lead to the appearance of currents along the vortex. To see this, we repeat the analysis, first presented in Refs.~\cite{witten1,witten2} and based on a perturbative expansion in the fields $\Phi$ and $\Sigma$ which are assumed small to begin with. The field equations derived from Eq.~(\ref{lag}) can be written in the form of two-dimensional time-independent Schr\"odinger equations as follows. Given $\Phi = \phi (r,\theta) \ex^{iE_\phi t}$ and $\Sigma = \sigma (r,\theta) \ex^{i E_\sigma t}$ and neglecting nonlinear terms in the fields, one has
\begin{eqnarray}
-\Delta \phi + {\cal V}_\phi \phi &=& E_\phi^2 \phi,\\
-\Delta \sigma + {\cal V}_\sigma \sigma &=& E_\sigma^2  \sigma,
\label{Schroe}
\end{eqnarray}
with
\begin{equation}
{\cal V}_\phi = f_\phi (h^2-\eta^2)+m_\phi^2 , \qquad {\cal V}_\sigma =f_\sigma (h^2-\eta^2)+m_\sigma^2.
\label{Potentials}
\end{equation}
If these potentials are both negative definite, there exist bound states for both fields, \ie, solutions with $E_\phi^2 <0$ and $E_\sigma^2 <0$, leading to instabilities in these fields, and therefore to condensates. If the vacuum masses of $\Phi$ and $\Sigma$ vanish, this is certainly the case; it holds true as well in a neighbourhood of $(m_\phi^2,m_\sigma^2)=(0,0)$ provided the constraints
\begin{equation}
f_\phi\eta^2>m_\phi^2 , \qquad f_\sigma\eta^2>m_\sigma^2
\label{constraints}
\end{equation}
are imposed.  Although these conditions have been derived from a dynamical analysis, they can be recovered by examining the vacuum condition $\delta V/\delta \Phi =\delta V/\delta \Sigma =0$ at the string location, \ie, setting $H=0$.  Assuming neither $\Phi$ nor $\Sigma$ vanish, this gives
\begin{eqnarray*}
\lambda_\phi |\Phi|^2 + g |\Sigma|^2 &=& f_\phi \eta^2 -m_\phi^2, \\
g |\Phi|^2 + \lambda_\sigma |\Sigma|^2 &=& f_\sigma \eta^2 -m_\sigma^2,
\end{eqnarray*}
which is only possible in the range of parameters defined by Eq.~(\ref{constraints}). As we now make clear, although these constraints are necessary they are by no means sufficient.  Indeed, when one field condenses, inclusion of its nonlinear potential term $(g/2)|\Phi|^2|\Sigma|^2$ in the action modifies the other field Schr\"odinger equation in a way that cannot be determined analytically. This correction is positive definite so that a strong coupling between current-carrying fields tends to drive one of the condensates to zero.  In this work we focus our attention on the effect of this coupling on the string's energy density, tension and stability and we further restrict attention to the weak to moderate coupling case and impose the condition
\begin{equation}
\lambda_\phi \lambda_\sigma > g^2
\end{equation}
on the quartic coupling constants.
\section{Two-current-carrying configurations and numerical solutions}
\label{numeric}
For the cylindrically symmetric forms 
\begin{eqnarray}
\Phi(x^\alpha) &=& \phi (r) \ex^{i \psi_\phi} = \phi (r) \ex^{i(\omega_\phi t -k_\phi z) },
\label{ansatzPhi}\\
\Sigma(x^\alpha) &=& \sigma (r) \ex^{i \psi_\sigma} = \sigma (r)\ex^{i
(\omega_\sigma t -k_\sigma z)},
\label{ansatzSigma}
\end{eqnarray}
together with (\ref{NO}), the field equations read
\begin{widetext}
\begin{eqnarray}
{\dd^2 h\over \dd r^2}+{1\over r}{\dd h\over \dd r} &=& \left[{Q^2\over r^2} + {1\over 2}\lambda (h^2 -\eta^2)+ f_\phi\phi^2 + f_\sigma \sigma^2 \right] h,
\label{eqh}\\
{\dd^2 Q\over \dd r^2}-{1\over r}{\dd Q\over \dd r} &=& q^2 h^2 Q,
\label{eqQ} \\
{\dd^2 \phi\over \dd r^2}+{1\over r}{\dd \phi\over \dd r} &=& \big[ w_\phi + f_\phi (h^2-\eta^2) +m_\phi^2+\lambda_\phi \phi^2+ g\sigma^2\big] \phi,
\label{eqPhi}\\
{\dd^2 \sigma\over \dd r^2}+{1\over r}{\dd \sigma\over \dd r} &=& \big[ w_\sigma + f_\sigma (h^2-\eta^2)+m_\sigma^2+\lambda_\sigma \sigma^2 +g\phi^2\big] \sigma,
\label{eqSigma}\nonumber
\end{eqnarray}
\end{widetext}
where the two real state parameters
\begin{equation}
w_\phi \equiv k_\phi^2 - \omega_\phi^2 \quad \hbox{ and } \qquad
w_\sigma \equiv k_\sigma^2 - \omega_\sigma^2.
\label{state12}
\end{equation}
are Lorentz scalars.  Knowledge of these two parameters is sufficient to fully determine the microscopic structure of the defect. Since the currents are vectors, an additional Lorentz scalar can be built out of the vortex fields.  It represents the relative position of the two currents and arises in the two integrated functions (energy per unit length and tension) that are necessary to describe the string dynamics completely.
\subsection{Dimensionless quantities}
The coupled equations (\ref{eqh}) to (\ref{eqSigma}) can be solved using a Successively Over Relaxed (SOR) method, a procedure appropriate to this category of boundary condition nonlinear systems~\cite{current,neutral,enon0}. This method is presented in full detail in Ref.~\cite{sor}.  Defining the dimensionless distance variable $$\rho \equiv \sqrt{\lambda}\eta r,$$ and the rescaled field functions $X$, $Y$ and $Z$ through
\begin{equation}
h \equiv X \eta, \quad m_\phi Y\equiv\sqrt{\lambda_\phi}\phi, \quad m_\sigma Z\equiv\sqrt{\lambda_\sigma}\sigma,
\end{equation}
we renormalize the coupling constants as
\begin{equation}
\tilde q^2 \equiv \frac{q^2}{\lambda}, \qquad \tilde g\equiv {gm_\phi^2m_\sigma^2 \over \lambda \lambda_\phi \lambda_\sigma \eta^4},
\end{equation}
and
 \begin{equation}
\alpha_i \equiv {m_i^2 \over \lambda_i \eta^2},\quad \beta_i\equiv {f_i m_i^2 \over \lambda \lambda_i \eta^2},\quad \gamma_i \equiv {m_i^4 \over \lambda \lambda_i \eta^4},
\end{equation}
the index $i$ standing for either $\phi$ or $\sigma$.  In terms of the dimensionless variables and couplings, the action reads
\begin{widetext}
\begin{eqnarray}
  {\cal S} &=& -\pi\eta^2 \int \bigg[ \left({\dd X\over \dd\rho}\right)^2 +
  {X^2Q^2\over \rho^2} +{1\over \tilde q^2 \rho^2}\left( {\dd Q\over \dd\rho}
  \right)^2+{1\over 4} (X^2-1)^2 + \alpha_\phi \left({\dd Y\over \dd\rho}\right)^2
  + \alpha_\sigma \left({\dd Z \over \dd\rho}\right)^2 \nonumber \\
  & &+ (\tilde w_\phi +\gamma_\phi) Y^2 + (\tilde w_\sigma
  +\gamma_\sigma) Z^2+ \left( \beta_\phi Y^2 + \beta_\sigma Z^2\right)
  \left( X^2-1\right)+{1\over 2} \gamma_\phi Y^4 + {1\over
    2}\gamma_\sigma Z^4 +\tilde g Y^2 Z^2 \bigg] \rho \, \dd \rho,
\label{action}
\end{eqnarray} 
\end{widetext}
in which we have used the dimensionless form of the state parameters,
\begin{equation}
\tilde w_i \equiv {m_i^2 \over \lambda \lambda_i \eta^4} w_i,
\end{equation}
with $i=\phi,\sigma$.  The constraints~(\ref{constraints}), when expressed in terms of the dimensionless parameters, simply read $\beta_i > \gamma_i$.\\

\begin{figure}
\includegraphics[width=8cm]{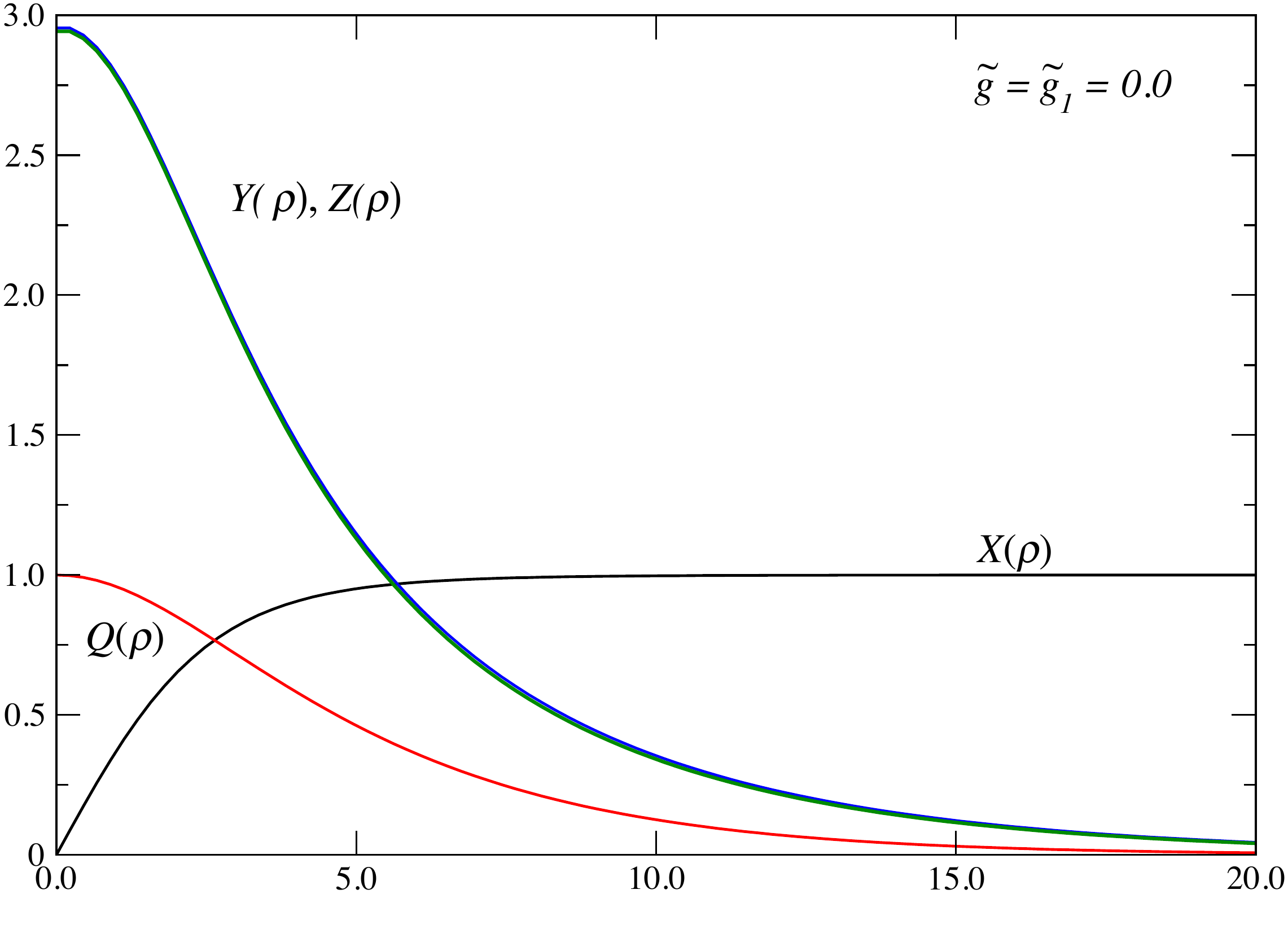}
\includegraphics[width=8cm]{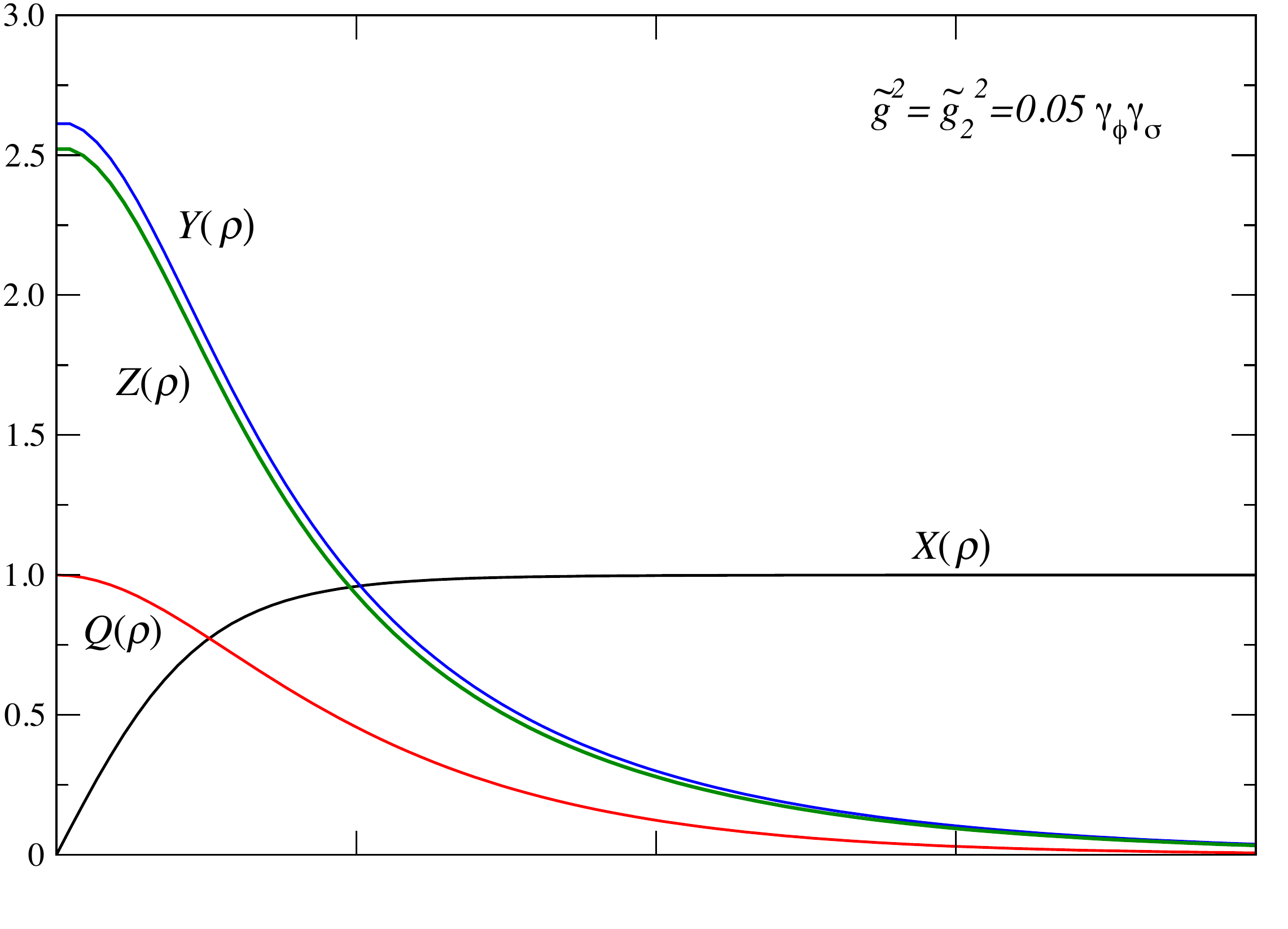}
\includegraphics[width=8cm]{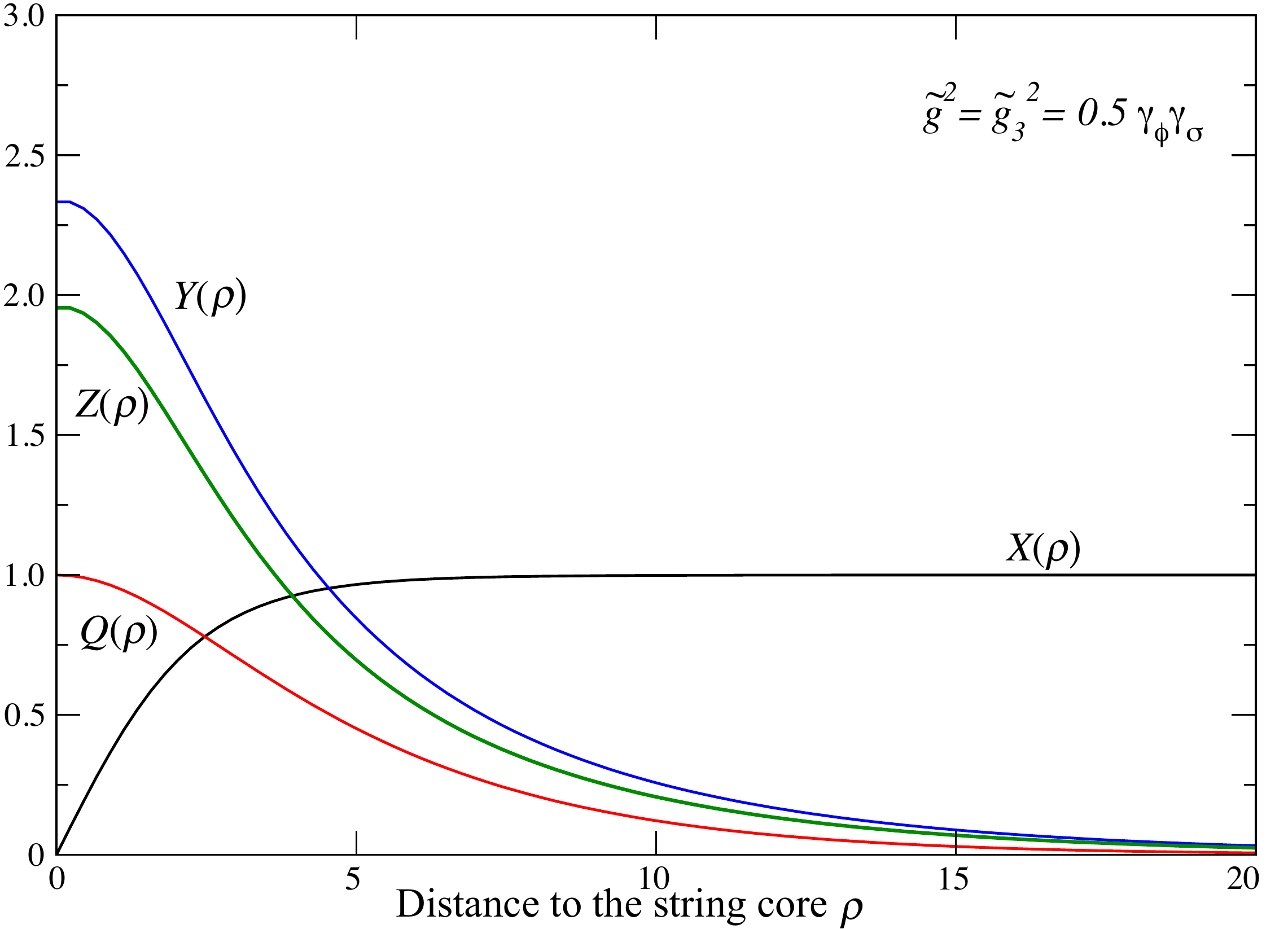}
\caption{Dimensionless field profiles obtained using $\alpha_{\phi}=1.8\times 10^{-2}$, $\alpha_{\sigma}=1.4\times 10^{-2}$, $\beta_{\phi}=1.0\times 10^{-2}$, $\beta_{\sigma}=0.8\times 10^{-2}$, $\gamma_{\phi}=5.5\times 10^{-4}$, $\gamma_{\sigma}=4.5\times 10^{-4}$, $\tilde{q}^2=1.0\times 10^{-1}$, $\tilde{w}_{\phi}=\tilde{w}_{\sigma}=\tilde{x}=0$ and finally from top to bottom $\tilde{g}^2=\tilde{g}_1^2=0.0$, $\tilde{g}^2=\tilde{g}_2^2= 0.05\, \gamma_{\phi}\gamma_{\sigma}$ and $\tilde{g}^2=\tilde{g}_3^2=0.5\, \gamma_{\phi}\gamma_{\sigma}$.}
\label{fig:fields}
\end{figure}

The coupling parameter space was explored by fixing $\alpha_i$, $\beta_i$ and $\tilde{q}$ and varying $\lambda_i$ and $\tilde{g}$ for light-like currents, \ie, for $\tilde w_i=0$.  Three typical solutions are presented in Fig.~\ref{fig:fields}.  These solutions were obtained with the following boundary conditions\footnote{For now on, a prime will represent a differentiation with respect to the rescaled distance variable $\rho$.}: $Y'(0)=Z'(0)=Y(\infty)=Z(\infty)=0$, while the vortex itself satisfies $X(0)=0$, $X(\infty)=1$, $Q(0)=n=1$ (we restrict attention to unit winding numbers strings in what follows) and $Q'(0)=0$.
\subsection{The state parameter space}
We now discuss the allowed range of values for $\tilde{w}_i$. We first consider the solution near the string core. Assuming two condensates to be present, we expand both functions as
\begin{equation}
Y = y_0 + y_2 \rho^2 + \cdots, \quad Z = z_0 + z_2\rho^2 + \cdots,
\label{expan}
\end{equation}
Using (\ref{expan}) in the field equations yields
\begin{equation}
\tilde w_i < \beta_i - \gamma_i.
\label{w_upper}
\end{equation}
From Eq.~(\ref{constraints}), we note that (making use of the one-current terminology~\cite{neutral})  this is constraint on  the ``magnetic'' side of the current, \ie, a constraint on the possible range of spacelike currents having $w\geq 0$.\\

For $\rho\rightarrow \infty$, the field equations read
\begin{eqnarray}
\alpha_\phi \left( Y''+{1\over \rho} Y'\right) &\sim &
\left(\tilde w_\phi +\gamma_\phi \right) Y, \nonumber \\
\alpha_\sigma \left( Z''+{1\over \rho} Z'\right) &\sim &
\left(\tilde w_\sigma +\gamma_\sigma \right) Z,
\end{eqnarray}
and their solutions are expressed in terms of Bessel functions.  As in the one-current case, there are phase frequency thresholds above which the currents radiate away from the vortex. This translates into a constraint on timelike currents,
\begin{equation} \tilde w_i > - \gamma_i.
\label{w_lower}
\end{equation}
Both constraints imply that the two state parameters from which the microscopic structure can be derived must satisfy $-m_i^2 < w_i < f_i \eta^2 - m_i^2$. These conditions are necessary to confine the current, but not sufficient to ensure there is indeed one. Indeed, because of the non-linear interactions between the fields, even when these conditions are met, there exists the possibility that the total energy be minimized by a vanishing current.\\

\section{Equation of state}
\label{macro}
In this section, we determine the macroscopic, \ie, integrated, quantities from which the string dynamics can be derived. These are the total currents associated with $\Phi(x^\alpha)$ and $\Sigma(x^\alpha)$ and the eigenvalues of the stress-energy tensor.
\subsection{Currents}
A conserved current is associated with each $U(1)^{\rm global}$ symmetry. Given (\ref{ansatzPhi}) and (\ref{ansatzSigma}), one has
\begin{equation}
{\cal J}_\mu^{(\phi)} = \phi^2 \partial_\mu \psi_\phi  , \qquad {\cal J}_\mu^{(\sigma)} = \sigma^2 \partial_\mu \psi_\sigma
\end{equation}
where
\begin{equation}
\partial_\mu \psi_i=\omega_i t_\mu - k_i z_\mu.
\label{defdmu}
\end{equation}
The unit vectors $t_\mu$ and $z_\mu$ are respectively timelike and spacelike and are defined by
\begin{equation}
t^\mu \equiv \pmatrix{1 \cr 0 \cr 0 \cr 0} \qquad \hbox{and}
\qquad z^\mu \equiv \pmatrix{0 \cr 1 \cr 0 \cr 0},
\label{tmuzmu}
\end{equation}
in the system of coordinates $\{ t, z, x, y\}$. Given this convention, the two state parameters Eq.~(\ref{state12}) are
\begin{equation}
w_i = g^{\mu\nu} \partial_\mu\psi_i \partial_\nu\psi_i 
\,\,\,\,\hbox{(no sum on $i=\sigma,\phi$)}.
\label{paramscalar}
\end{equation}
What matters in a macroscopic formalism such as that of Refs.~\cite{formal1,formal2,formal3,formal4} is the Lorentz square of the currents integrated in the transverse direction,
\begin{equation}
{\cal C} \equiv \sqrt{ \big| c_\mu  c^\mu \big| },
\label{Ctot}
\end{equation}
where
\begin{equation}
  c_\mu = c_\mu^{(\phi)} + c_\mu^{(\sigma)} \equiv \int \left[ {\cal
      J}_\mu^{(\phi)} + {\cal J}_\mu^{(\sigma)} \right]\dd^2 x^\perp .
\label{cimu}
\end{equation}
We denote the integrated currents by $c^i_\mu$, $i=\phi,\sigma$.  For a straight and static string configuration, we then have
\begin{widetext}
\begin{equation}
\left(\frac{{\cal C}}{2\pi}\right)^2=\Bigg| w_\phi \left(\int \phi^2
  r\, \dd r \right)^2 + w_\sigma \left( \int \sigma^2r\, \dd r
\right)^2+ 2 x \left( \int \phi^2 r\, \dd r \right) \left( \int
  \sigma^2r\, \dd r \right)\Bigg|, 
\label{Cexpand}
\end{equation}
\end{widetext}
which involves, in addition to $w_\phi$ and $w_\sigma$, a third state parameter, $x$, given by
\begin{equation}
x \equiv \partial \psi_\phi \cdot \partial \psi_\sigma = k_\phi k_\sigma -\omega_\phi \omega_\sigma .
\label{defx}
\end{equation}
We note that, as mentioned in the introduction, although it does not enter into the microscopic description, $x$ appears in the macroscopic description of the vortex dynamics.\\

In the macroscopic formalism, the integrals of the scalar fields over the transverse section are especially meaningful.  They read
 \begin{equation}\label{KphiKsigma}
{\cal K}_\phi \equiv -2\pi \int \phi^2(r)r\,\dd r 
\quad \hbox{and} \quad {\cal K}_\sigma \equiv -
2\pi \int \sigma^2 (r)r\,\dd r,
\end{equation}
and correspond to the partial derivatives of the macroscopic Lagrangian,
\begin{equation}
{\cal L} (w_i)\equiv \int {\cal L} \dd^2
  x^\perp,
\label{lag2}
\end{equation}
 with respect to the currents.  Given that the solution of the field equations are determined with the 4-dimensional Lagrangian, variations with respect to the parameters $w_i$ yield directly
\begin{equation}
{\cal K}_i = 2 \frac{\partial \cal L}{\partial w_i}.
\label{defK1}
\end{equation}
The integrated current $\mcl{C}$ can of course be re-expressed using dimensionless variables.  Defining
\begin{equation}
\tilde {\cal C} \equiv \frac{{\cal C}}{\eta},
\end{equation}
we then have
\begin{widetext}
\begin{equation}
\tilde {\cal C} =\Bigg| \tilde w_\phi \frac{\gamma_\phi}{\alpha_\phi}
\left( \int Y^2 \rho\, \dd \rho \right)^2 + \tilde w_\sigma 
\frac{\gamma_\sigma}{\alpha_\sigma} \left( \int Z^2 \rho\, \dd \rho \right)^2+ 
2 \tilde x \sqrt{\frac{\gamma_\phi\gamma_\sigma}{\alpha_\phi\alpha_\sigma}} 
\left(\int Y^2 \rho\, \dd \rho \right) \left( \int Z^2 \rho\, \dd \rho \right)\Bigg|^{1/2},
\label{Cnodim}
\end{equation}
\end{widetext}
where the third dimensionless parameter $\tilde{x}$ reads
 \begin{equation}
   \tilde x \equiv \frac{m_\phi m_\sigma}{\lambda\sqrt{\lambda_\phi\lambda_\sigma} \eta^4}x.
\label{xtilde}
\end{equation}
\subsection{Stress-energy tensor}
The stress-energy tensor can be obtained from the Lagrangian through
 \begin{equation}
T^{\mu\nu}\equiv - 2 g^{\mu\alpha}g^{\nu
 \beta} \frac{\delta {\cal L}}{\delta g^{\alpha
 \beta}} + g^{\mu\nu} {\cal L}.
\label{Tmunu}
\end{equation}
It has two eigenvalues corresponding to a spacelike direction and a timelike direction of the worldsheet. Denoting by $u^{\mu}$ and $v^{\mu}$ the normalized, respectively timelike and spacelike, eigenvectors of $T^{\mu\nu}$, and setting $\eta^{\mu \nu}=-u^{\mu}u^{\nu}+v^{\mu}v^{\nu}$, the first fundamental tensor of (whose mixed form is the projector on) the string worldsheet, the eigenvalues, namely the energy per unit length $U$ and tension $T$, are obtained through the expression for the integrated stress-energy tensor~\cite{formal1,formal2,formal3,formal4}
\begin{eqnarray}
\overline T^{\mu\nu} &\equiv &\int T^{\mu\nu} \, \dd^2 x^\perp = U u^\mu u^\nu - T v^\mu v^\nu \nonumber\\
&=& (U-T) u^\mu u^\nu - T \eta^{\mu\nu}. \label{Tcan}
\end{eqnarray}
Unlike the case of a string with a single current, $T^{\mu\nu}$ obtained in Eq.~(\ref{Tmunu}) is not automatically diagonal.  We therefore define (in the worldsheet coordinates $t$ and $z$) $\tilde \eta^{ab} =\hbox{Diag}\,\{-1,+1\}$, the string metric tensor, and express the two-dimensional part of the stress-energy tensor as the sum of a diagonal part $T^{ab}_{_{\rm D}} = -A\tilde\eta^{ab}$ with
\begin{widetext}
\begin{eqnarray}
A &=& 2\pi \int \left\{ \frac{1}{2} \left[ \left(\frac{\dd h}{\dd r}\right)^2 +
{h^2 Q^2\over r^2} + {1\over q^2 r^2 } \left( {\dd Q \over \dd r}\right)^2+
\left( {\dd \phi\over\dd r}\right)^2 + \left( {\dd \sigma \over\dd r}\right)^2
\right]+ V(h,\phi,\sigma)\right\} r\dd r,
\label{defA}
\end{eqnarray} 
\end{widetext}
and a mixed non-diagonal part
\begin{equation}
T^{ab}_{_{\rm ND}} = \pmatrix{ B & C \cr C & B},
\end{equation}
\noindent with
\begin{equation}
B=\pi \int \left[ \left( k_\phi^2 +\omega_\phi^2 \right)\phi^2 +\left( k_\sigma^2 +\omega_\sigma^2 \right)\sigma^2 \right] r \dd r,
\label{B}
\end{equation}
\noindent and
\begin{equation}
C=2 \pi \int \left( k_\phi \omega_\phi \phi^2 + k_\sigma \omega_\sigma \sigma^2 \right) r \dd r.
\label{C}
\end{equation}
The eigenvalues of the stress-energy tensor are obtained through
\begin{equation}
\det \left( T^{ab} + \lambda \tilde \eta^{ab}\right) =0,
\end{equation}
and we find that the energy per unit length and tension read
\begin{eqnarray}
U&=&A+\sqrt{B^2-C^2},
\label{U}\\
  T&=&A-\sqrt{B^2-C^2}.
\label{T}
\end{eqnarray}
The quantity 
\begin{widetext}
\begin{equation}
B^2-C^2=\pi^2 \left[ \left( w_\phi \int\phi^2 r\dd r - w_\sigma \int \sigma^2 r\dd r\right)^2+4 x^2\left( \int\phi^2 r\dd r\right) \left(\int\sigma^2 r\dd r \right)\right]
\label{B2C2a}
\end{equation}
\end{widetext}
is obviously positive definite, so that $U$ and $T$ are well-defined.  Using (\ref{KphiKsigma}), this expression can be re-written as
\begin{equation}\label{B2C2K}
B^2-C^2 = \left( \frac{1}{2} w_\phi  \mathcal{K}_\phi - \frac{1}{2} w_\sigma  \mathcal{K}_\sigma \right)^2 + x^2 \mathcal{K}_\phi \mathcal{K}_\sigma.
\label{B2C2b}
\end{equation}
Note that the Nambu-Goto equation of state is recovered when $B^2=C^2$.\\

\begin{figure*}[t]
\includegraphics[width=7.5cm]{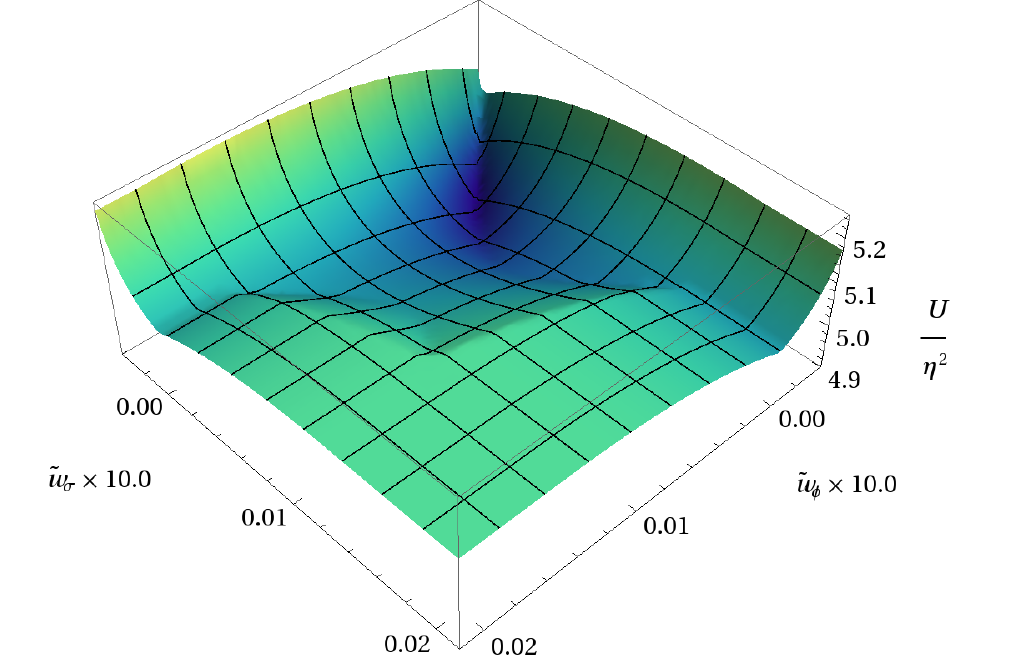}
\hspace{1cm}
\includegraphics[width=7.5cm]{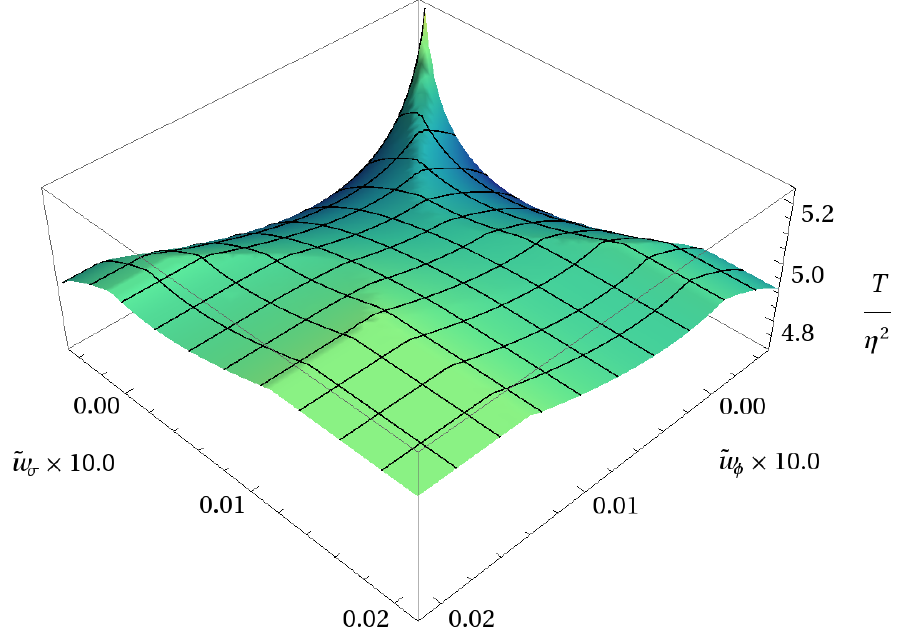}\\
\vspace{0.25cm}
\includegraphics[width=7.5cm]{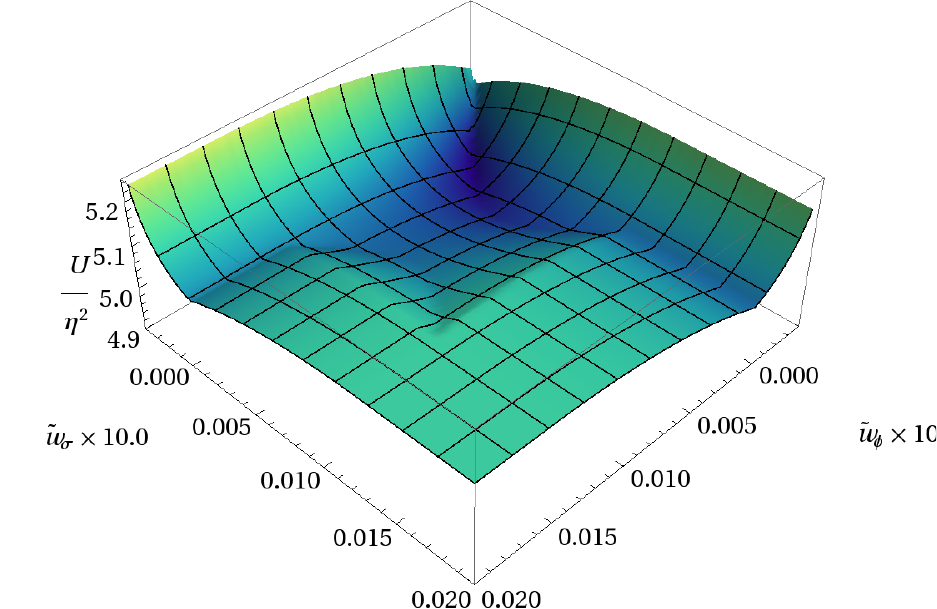}
\hspace{1cm}
\includegraphics[width=7.5cm]{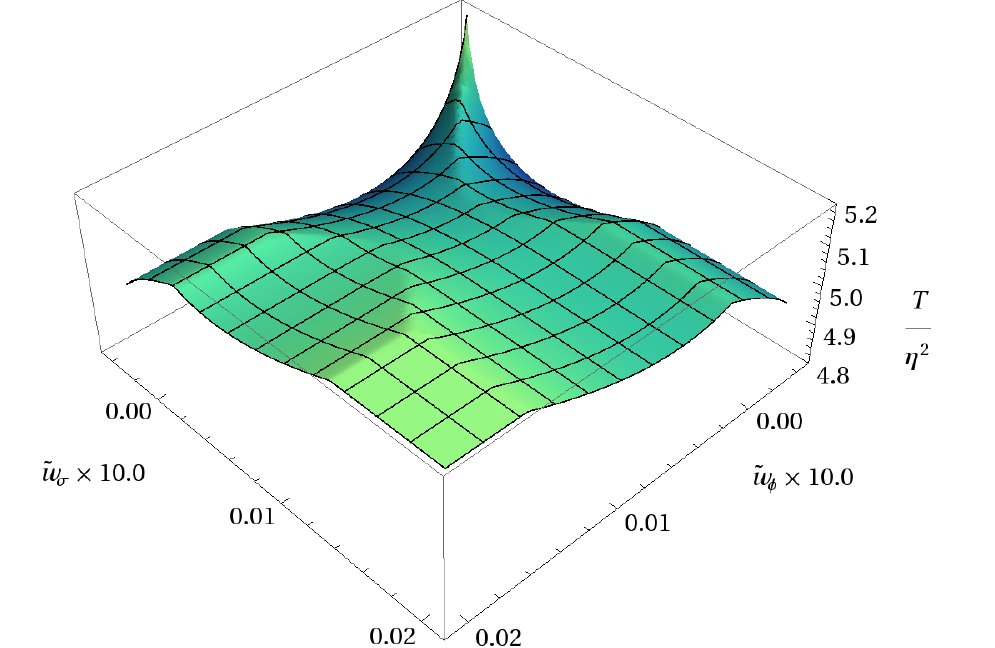}\\
\vspace{0.25cm}
\includegraphics[width=7.5cm]{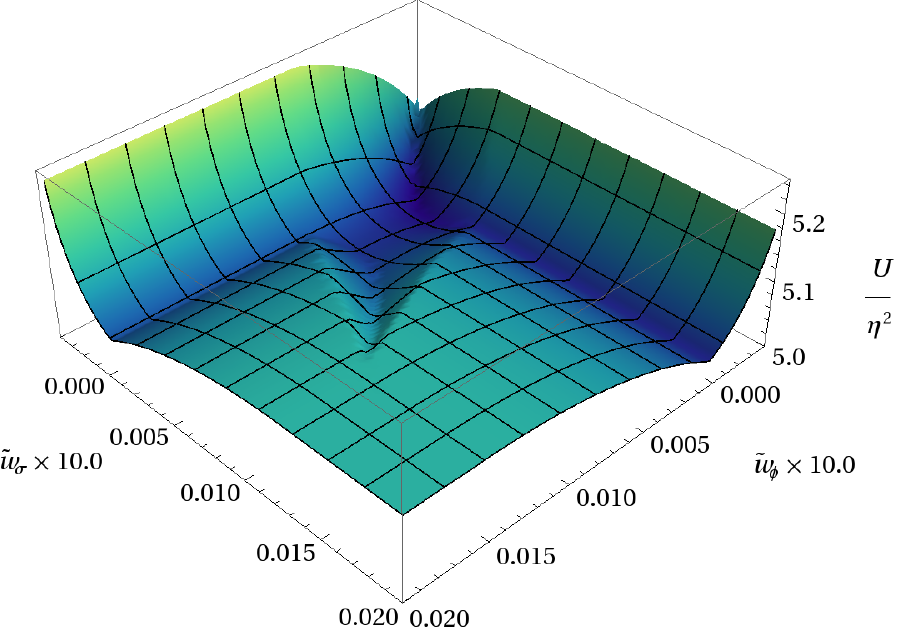}
\hspace{0.5cm}
\includegraphics[width=7.5cm]{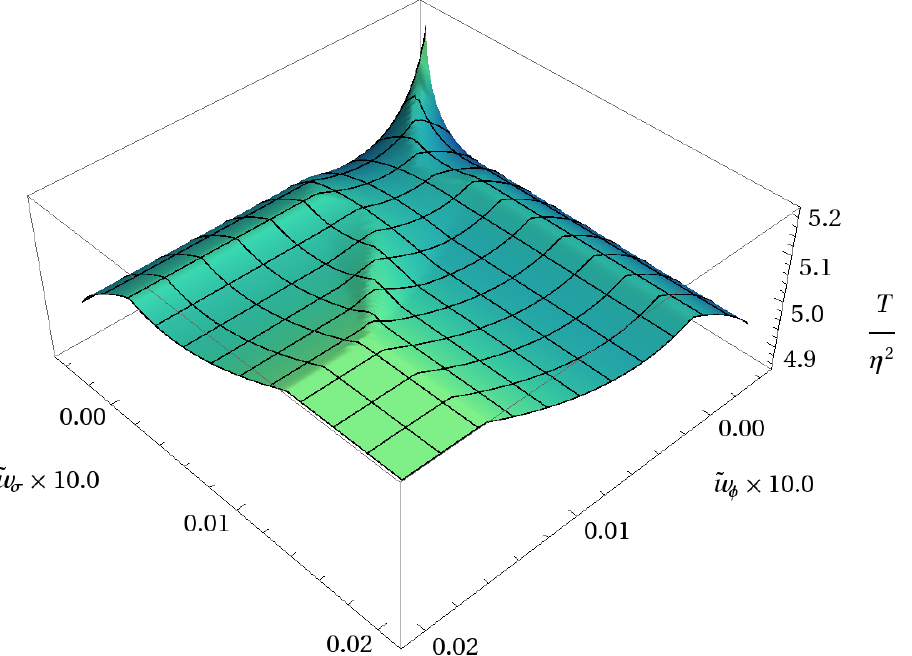}
\caption{Energy per unit length $U$ (left) and tension $T$ (right) as functions of $\tilde w_{\phi}$ and $\tilde{w}_{\sigma}$ for $\tilde{g}_1$, $\tilde{g}_2$ and $\tilde{g}_3$ from top to bottom respectively.  Both $U$ and $T$ diverge for $\tilde{w}_i\rightarrow -\gamma_i$.  Note that when both $\tilde{w}_{\phi}\rightarrow -\gamma_{\phi}$ and $\tilde{w}_{\sigma}\rightarrow -\gamma_{\sigma}$, $T$ and $U$ both diverge}
\label{fig:UT2D}
\end{figure*}

\begin{figure*}
\includegraphics[width=5.25cm]{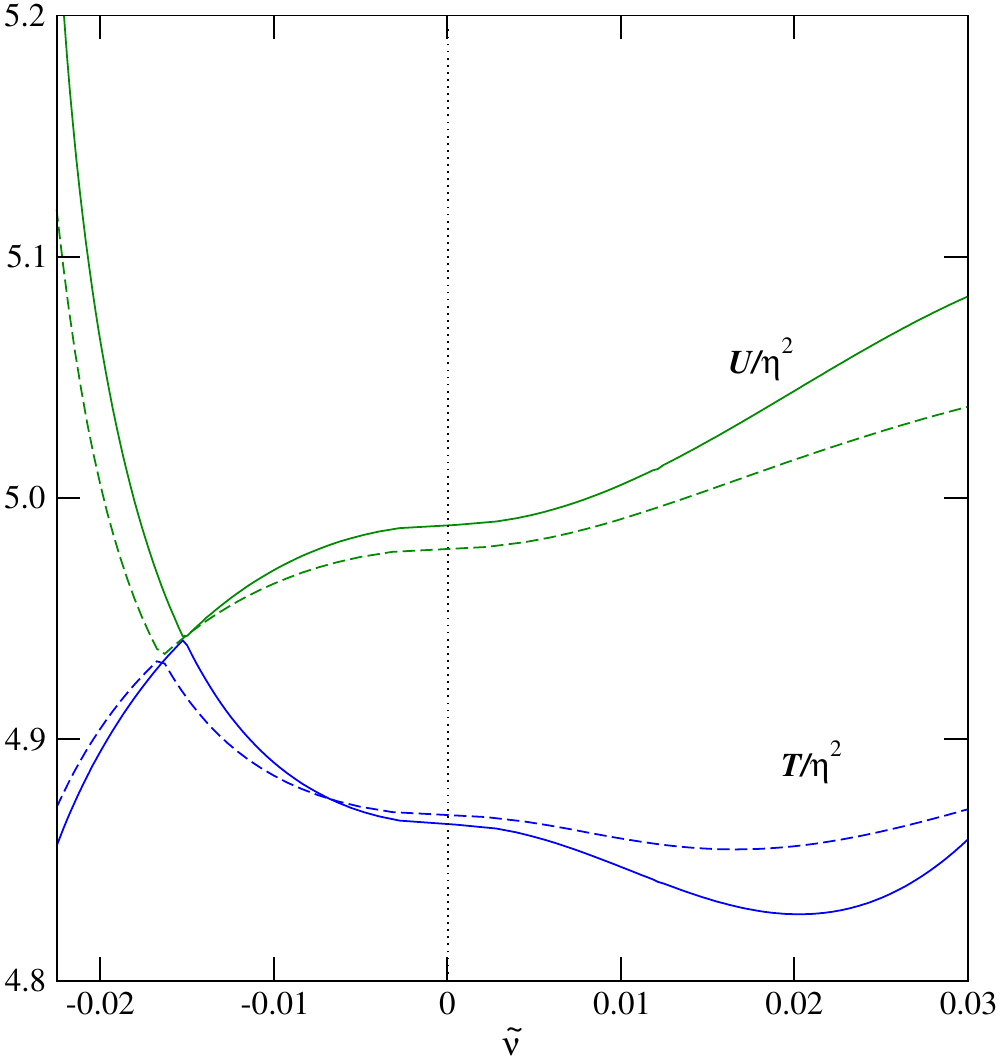}
\includegraphics[width=5.25cm]{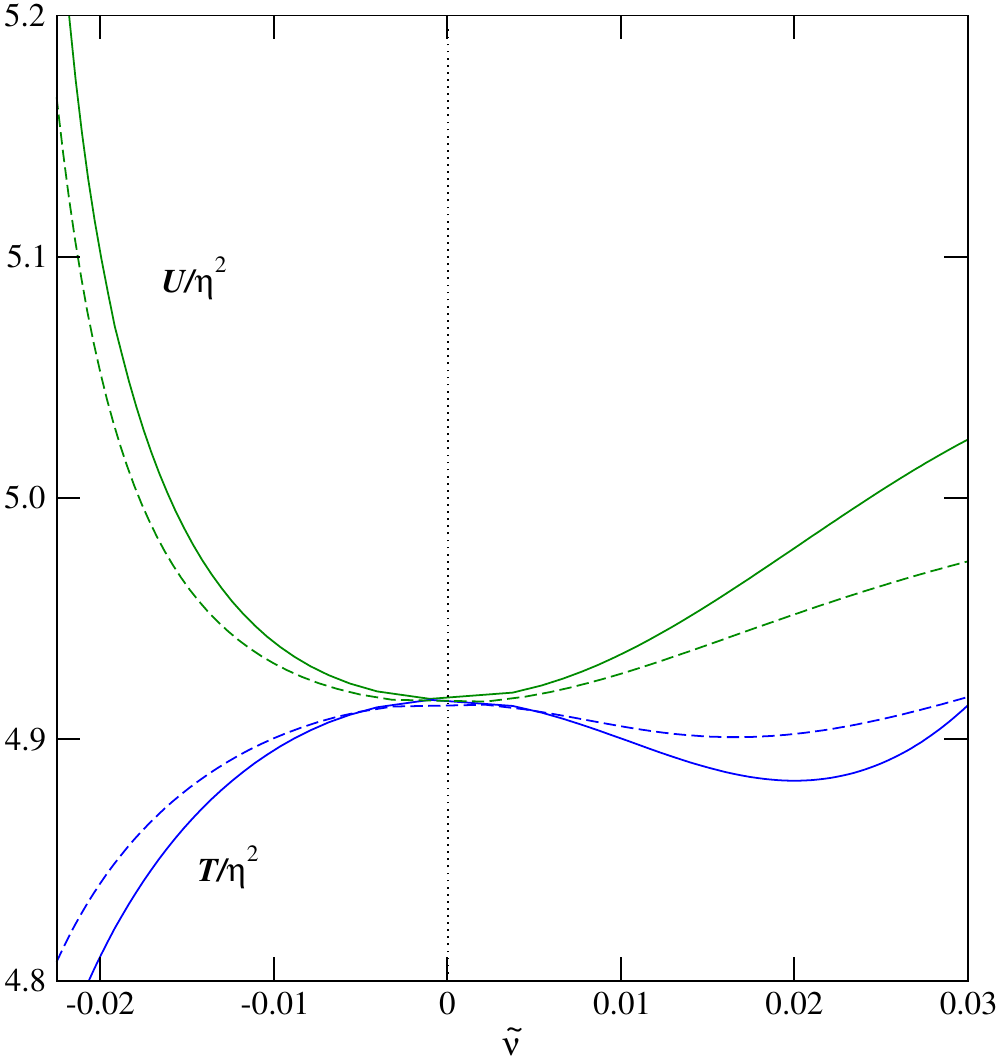}
\includegraphics[width=5.25cm]{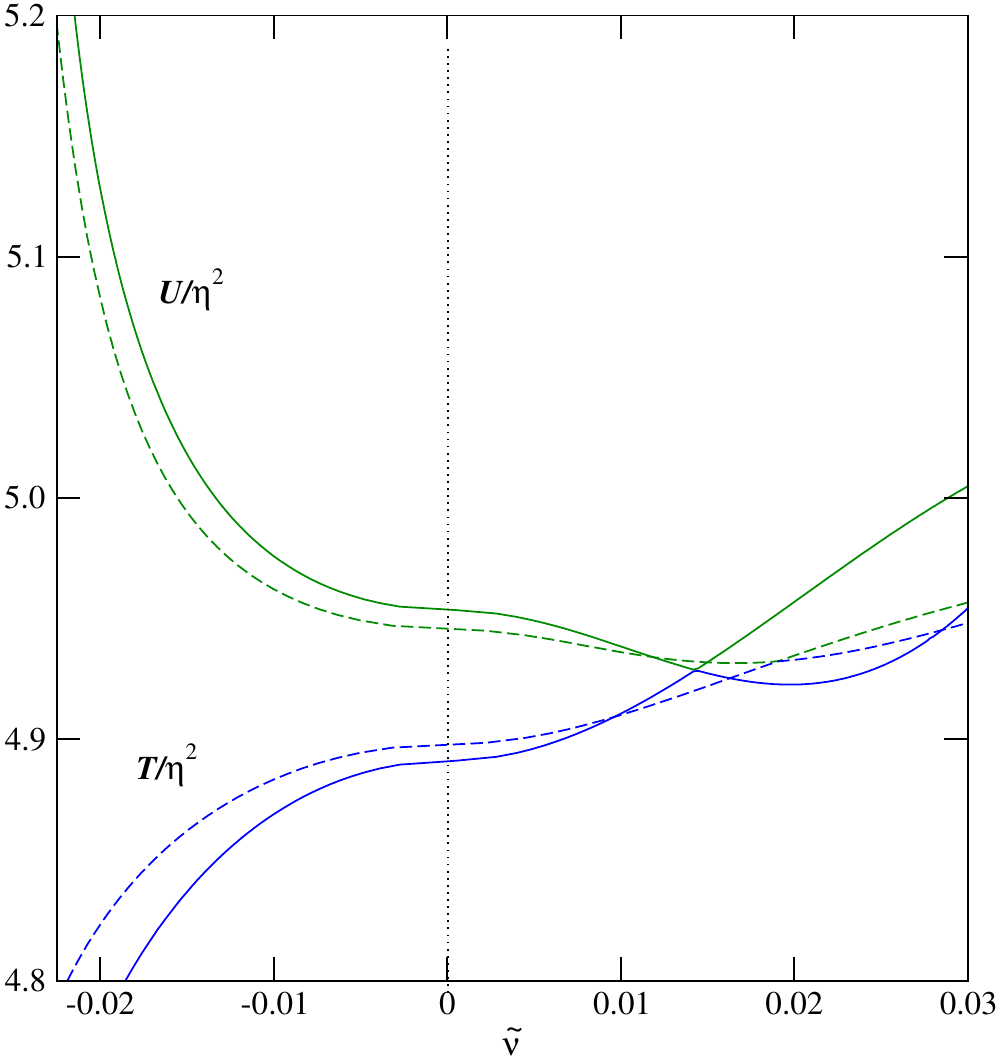}
\includegraphics[width=5.25cm]{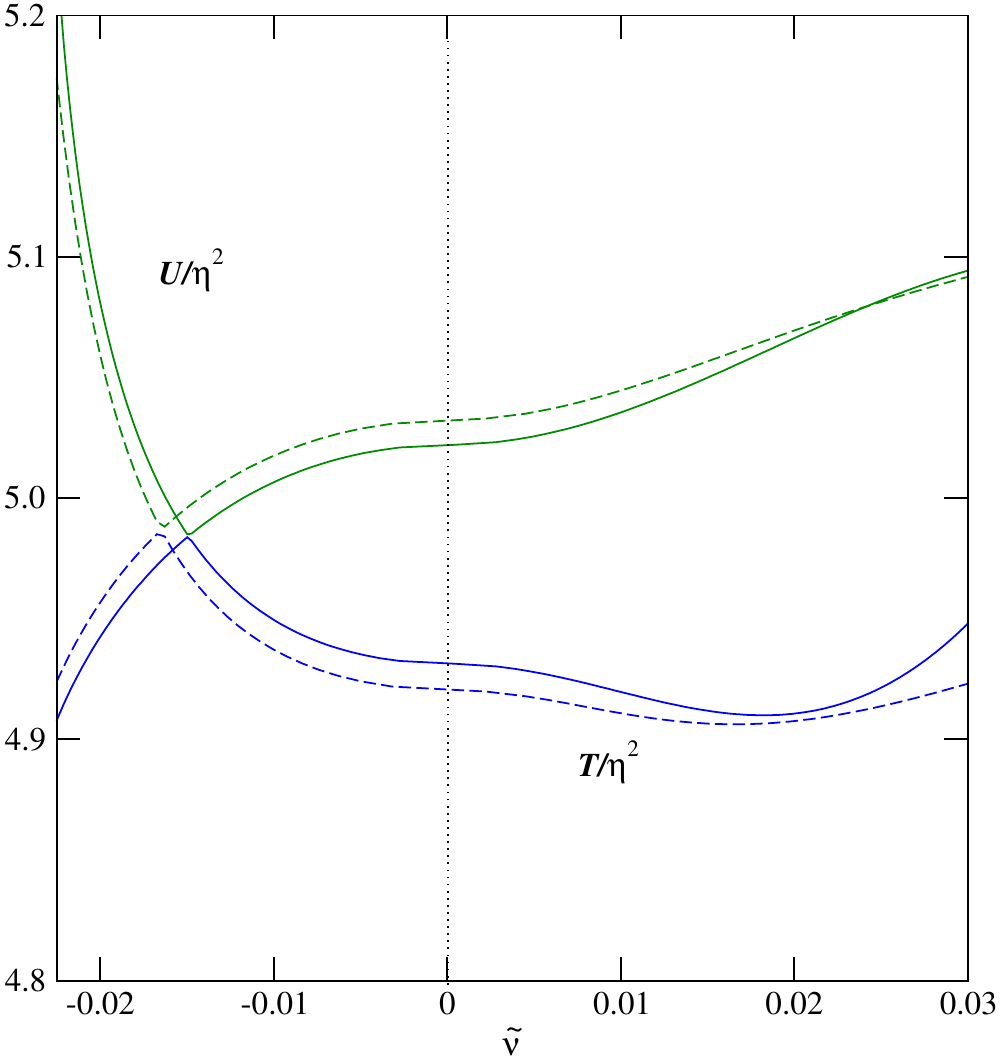}
\includegraphics[width=5.25cm]{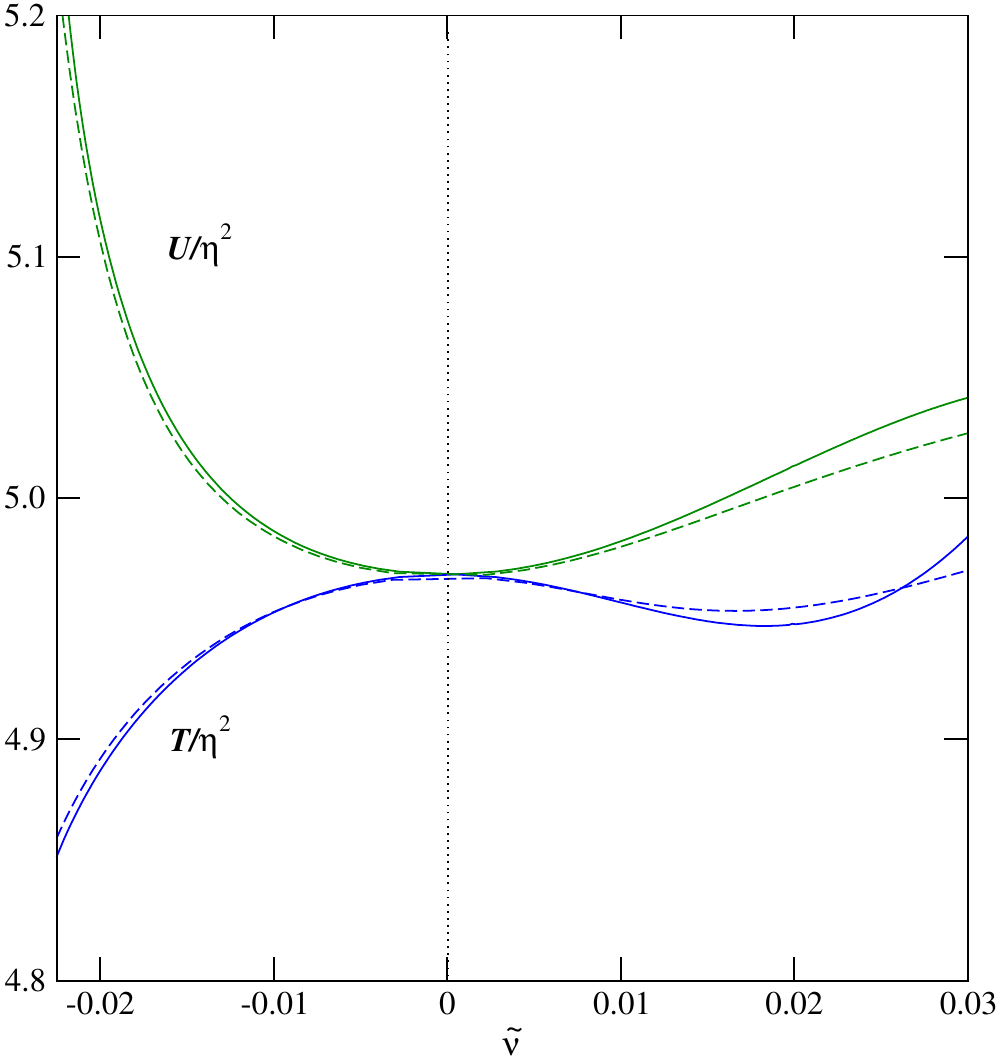}
\includegraphics[width=5.25cm]{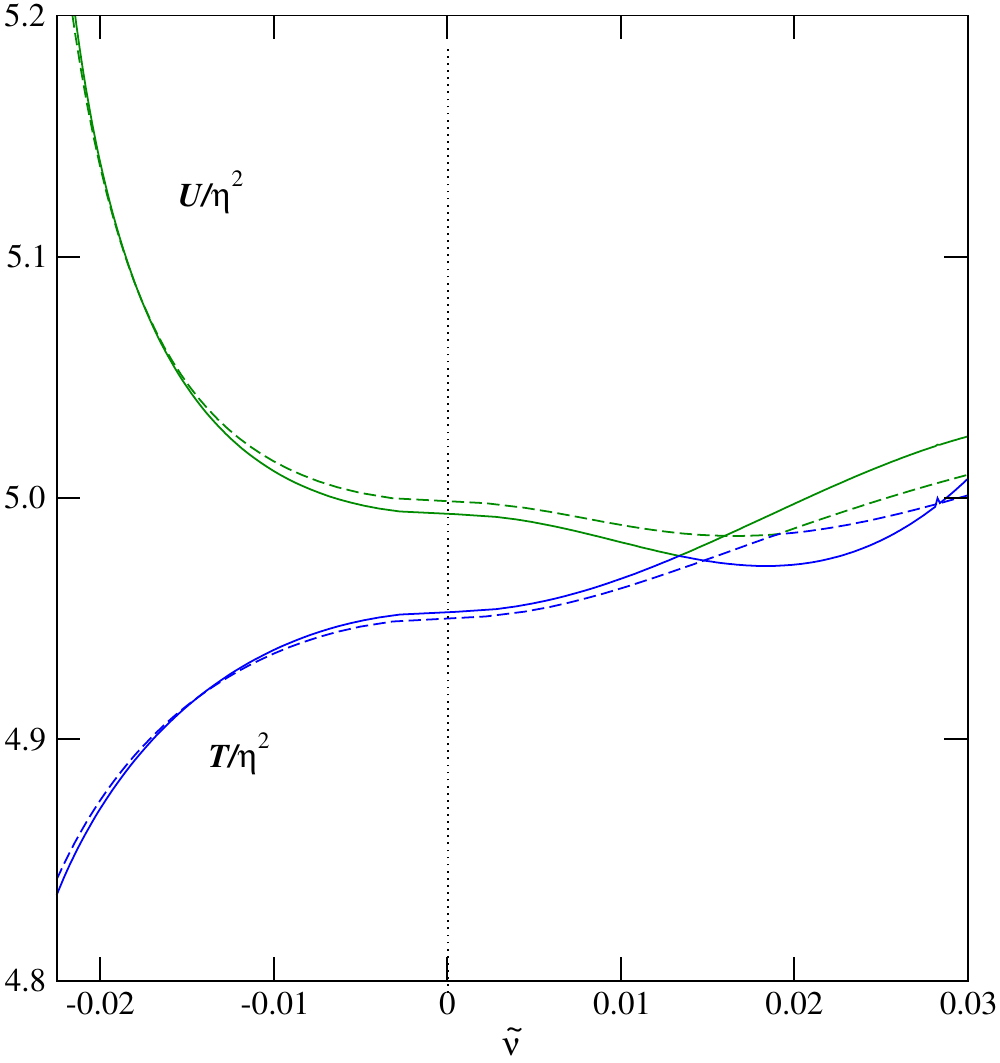}
\includegraphics[width=5.25cm]{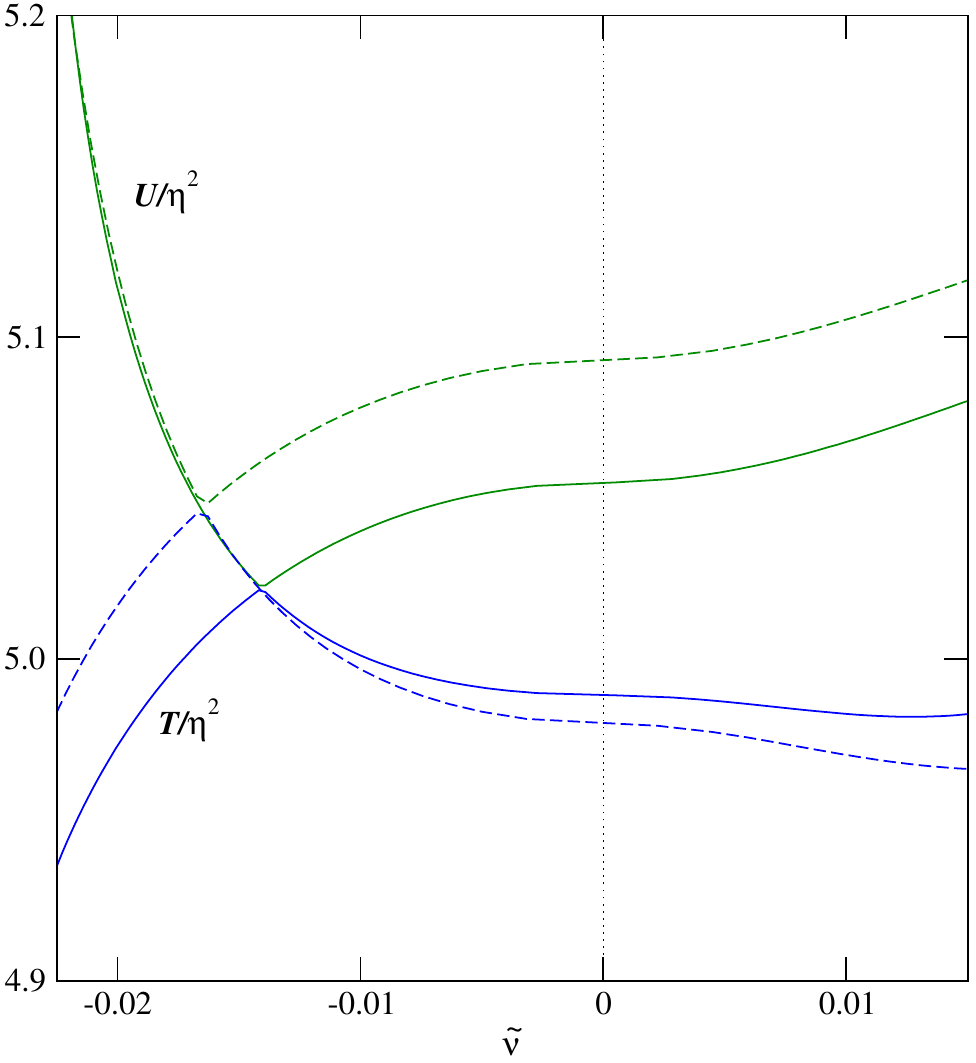}
\includegraphics[width=5.25cm]{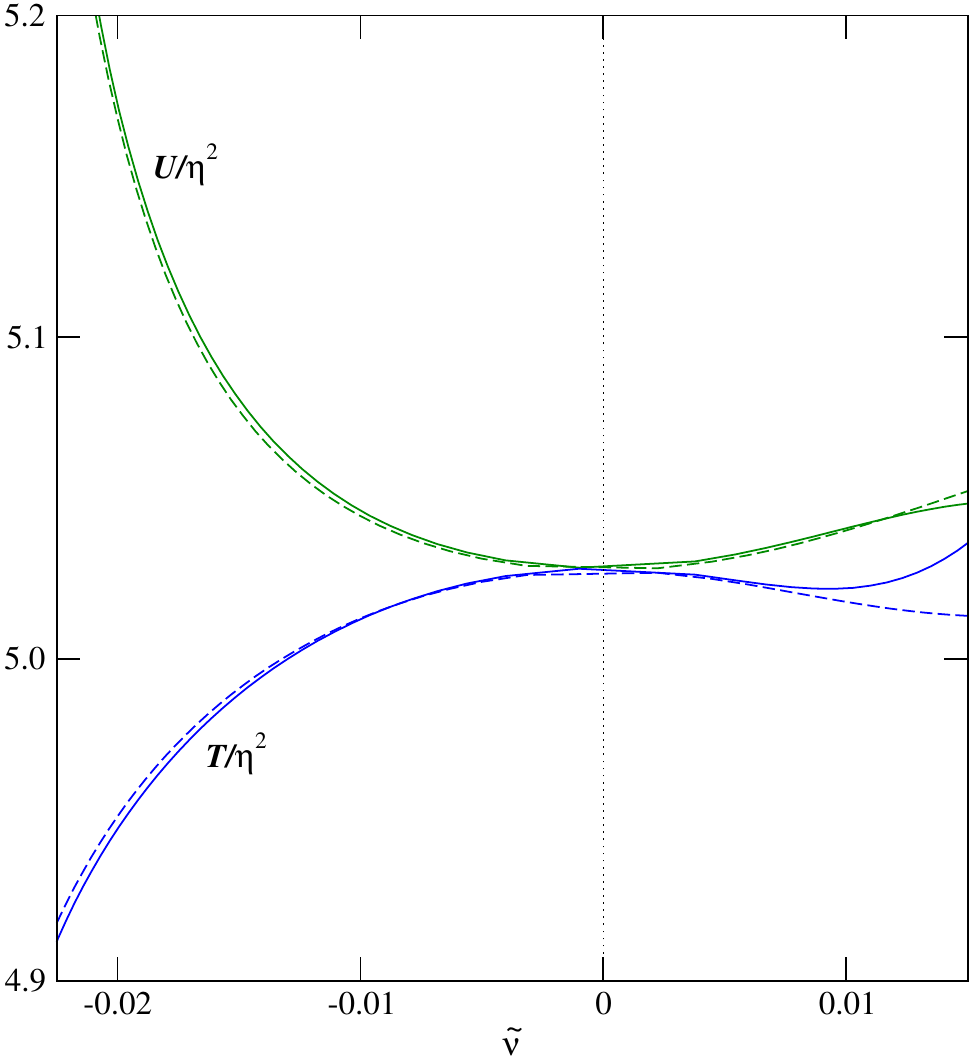}
\includegraphics[width=5.25cm]{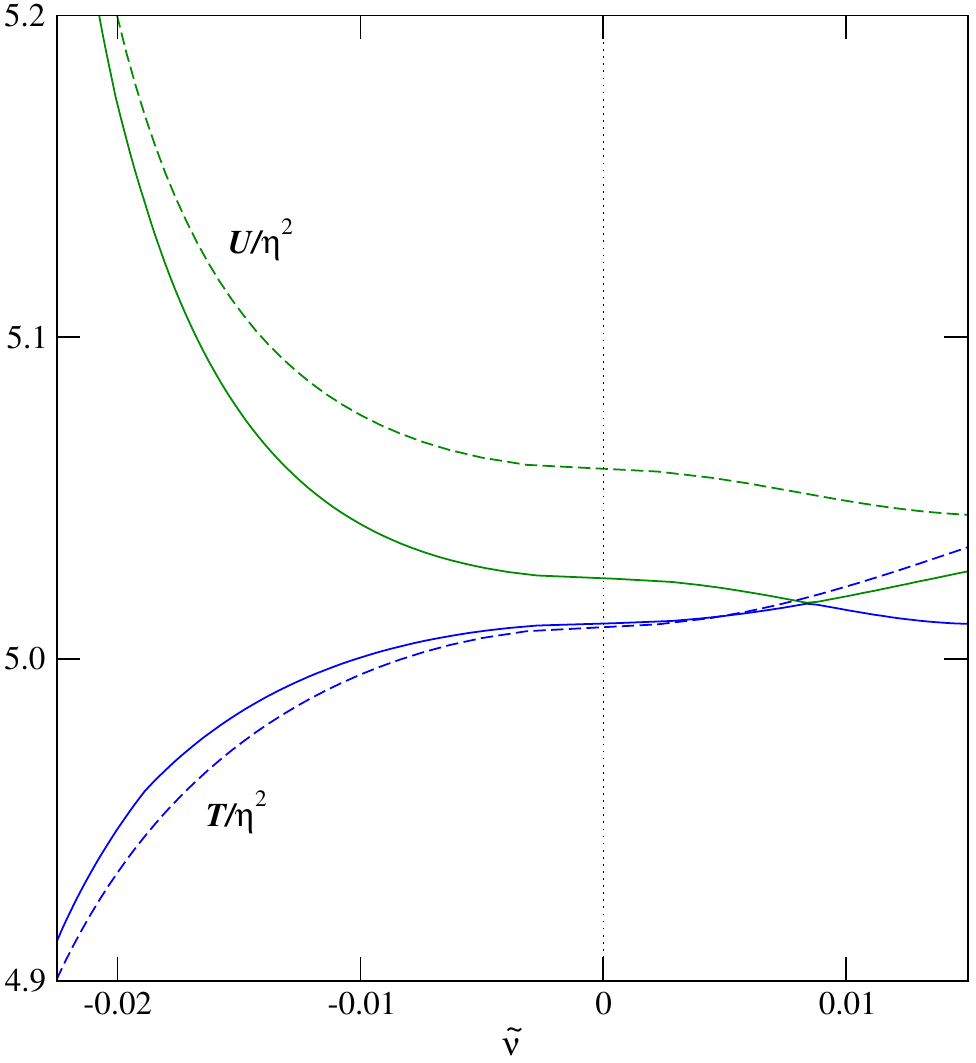}
\caption{U (upper curves) and T (lower curves) one-dimensional profiles along $w_{\phi}$ for $w_\sigma=-2.25\times 10^{-4}\sim -\gamma_i/2$, $0.0$  and $2.25\times 10^{-4}\sim \gamma_i/2$ (from left to right) and for $\tilde{g}_1$,  $\tilde{g}_2$, $\tilde{g}_3$ (from top to bottom).  The solid lines correspond to $U$ and $T$ obtained numerically while the dashed lines were obtained from the analytic model\cite{neutral,models} discussed in Section \ref{simplifiedmodel}.  Note the $x$-coordinate defined by $\tilde{\nu}=w_{\phi}/\sqrt{|w_{\phi}|}$ in order to emphasize the neighborhood of $\tilde{w}=0$.}
\label{fig:sectionsY}
\end{figure*}

\begin{figure*}
\includegraphics[width=5.25cm]{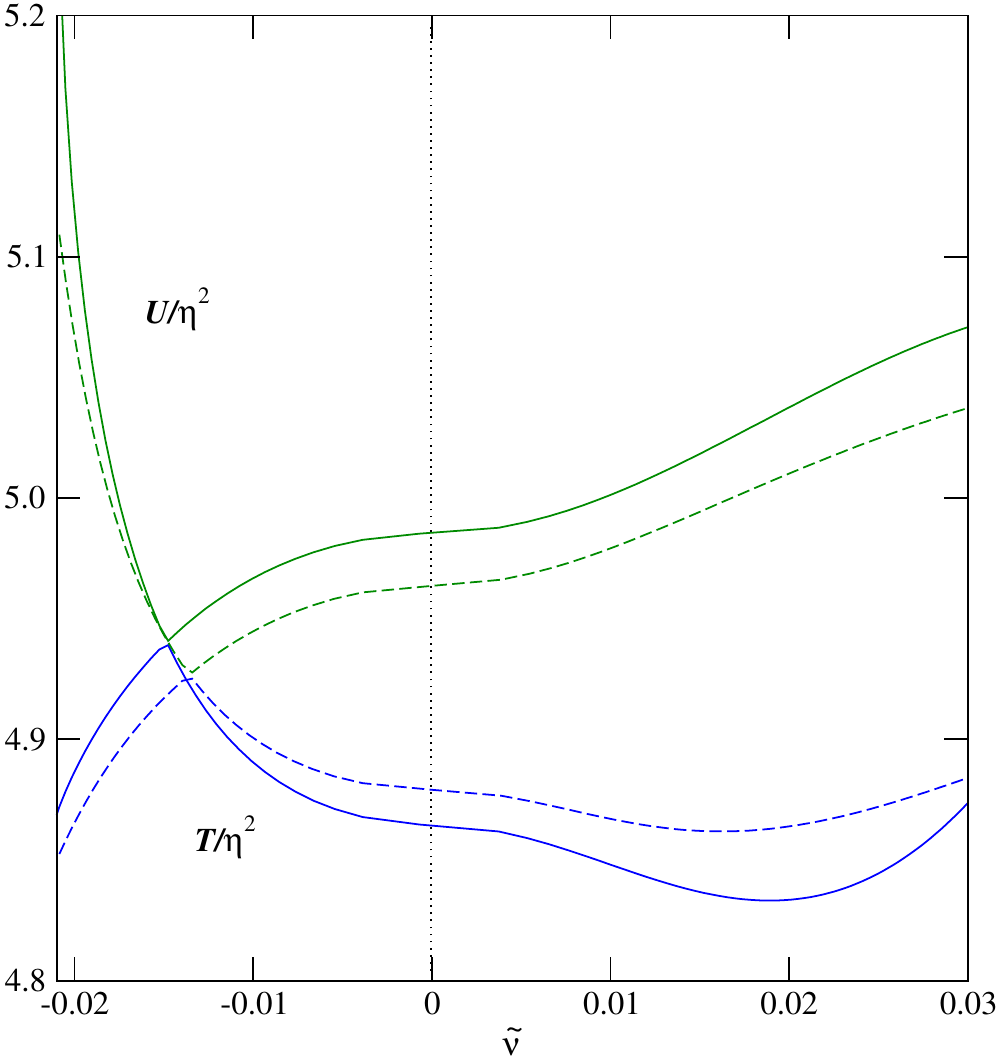}
\includegraphics[width=5.25cm]{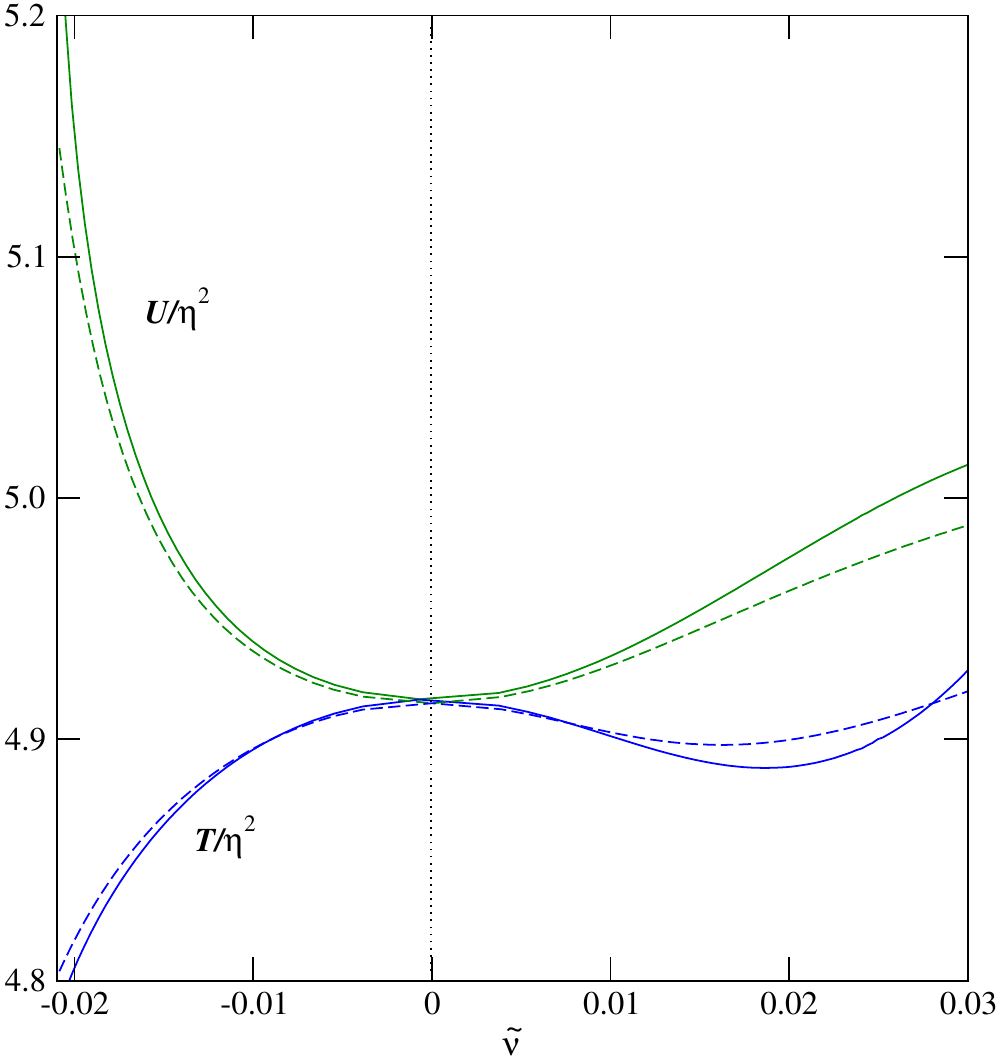}
\includegraphics[width=5.25cm]{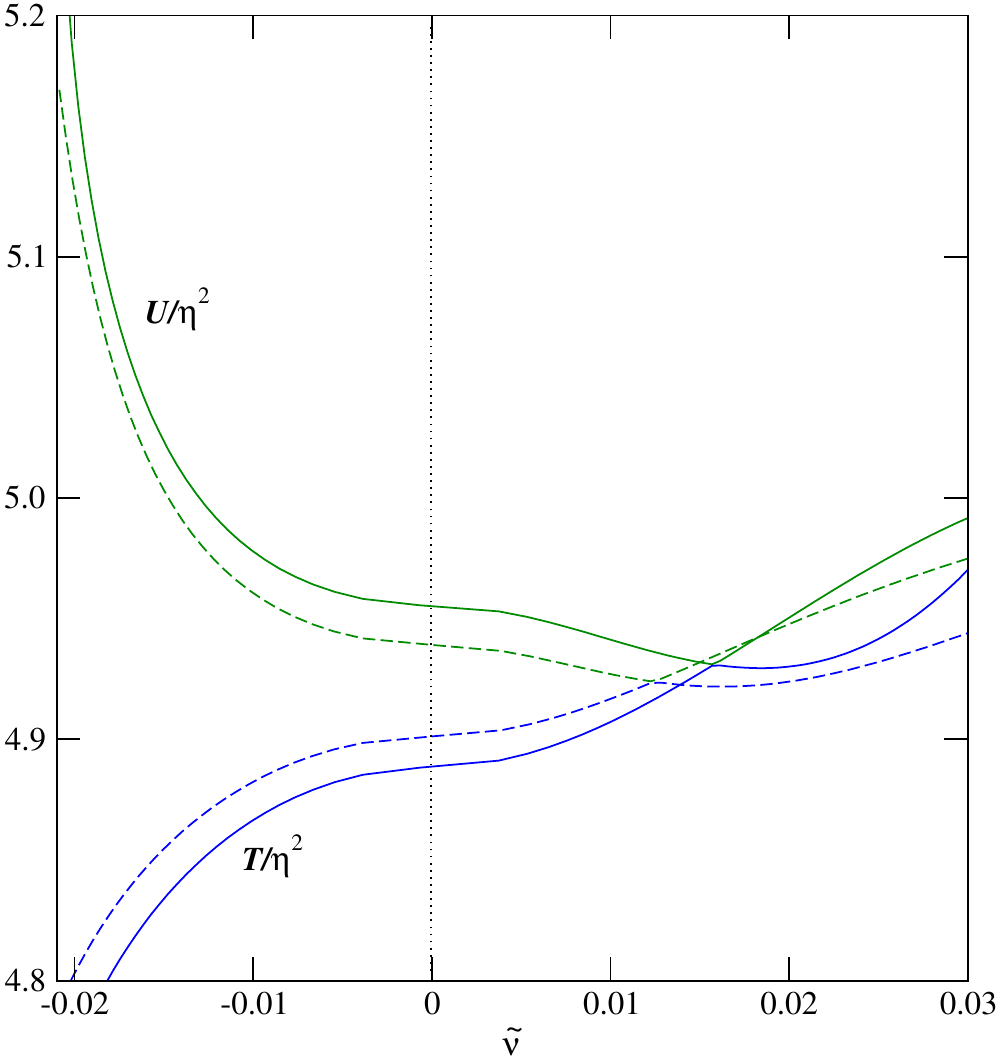}
\includegraphics[width=5.25cm]{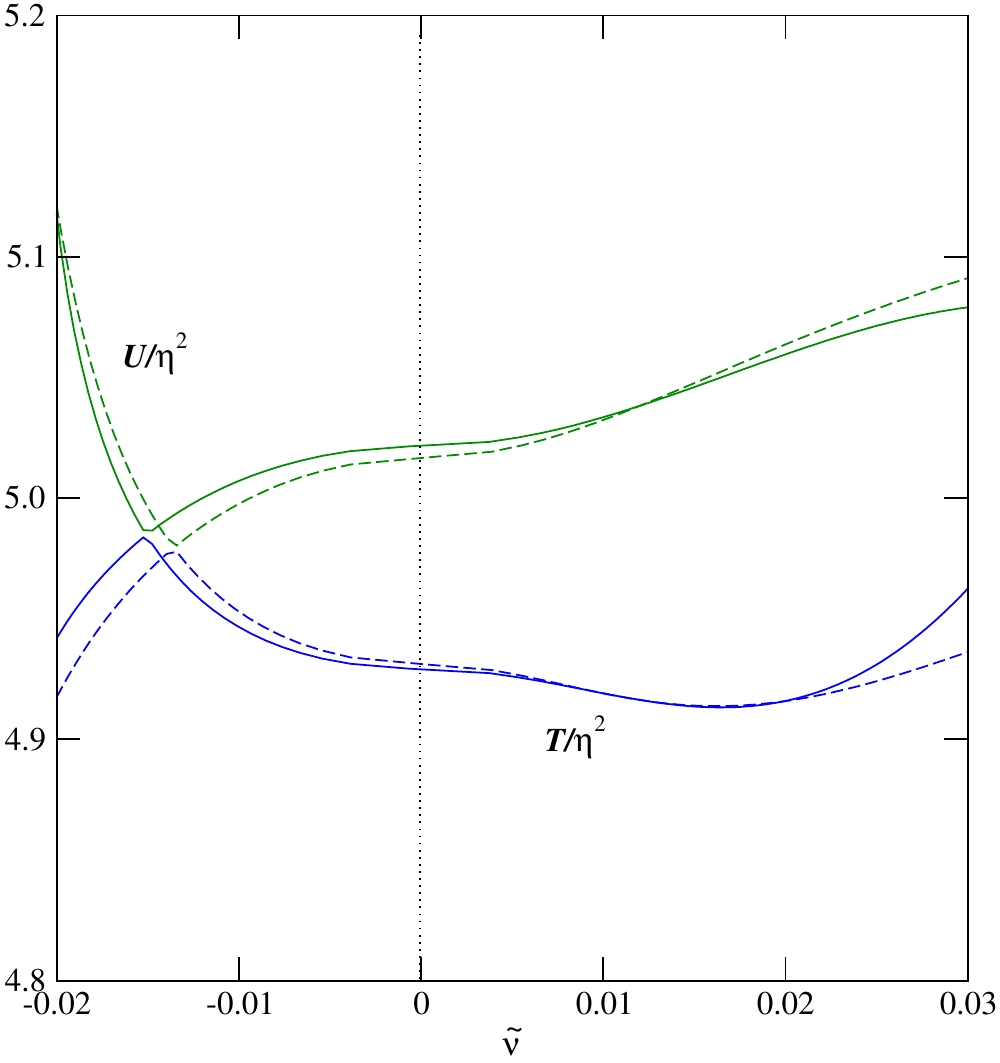}
\includegraphics[width=5.25cm]{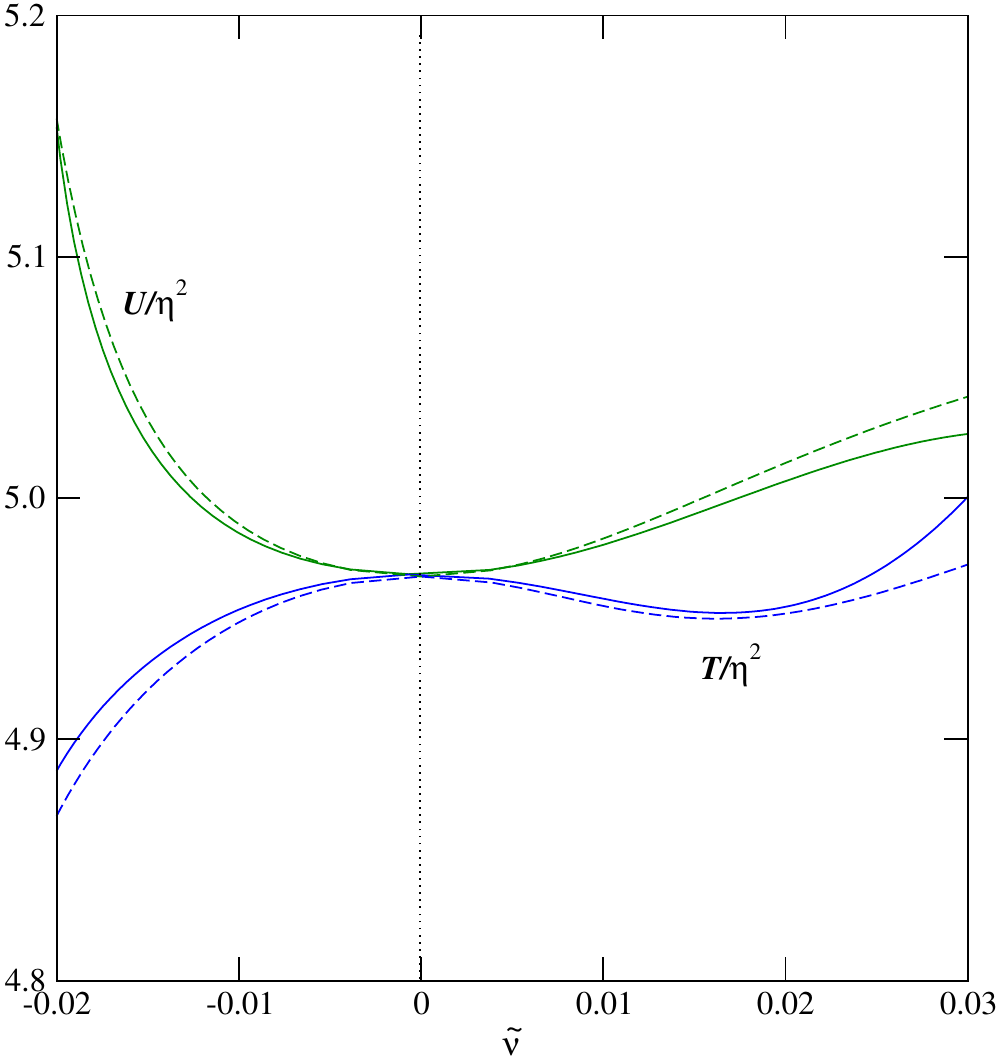}
\includegraphics[width=5.25cm]{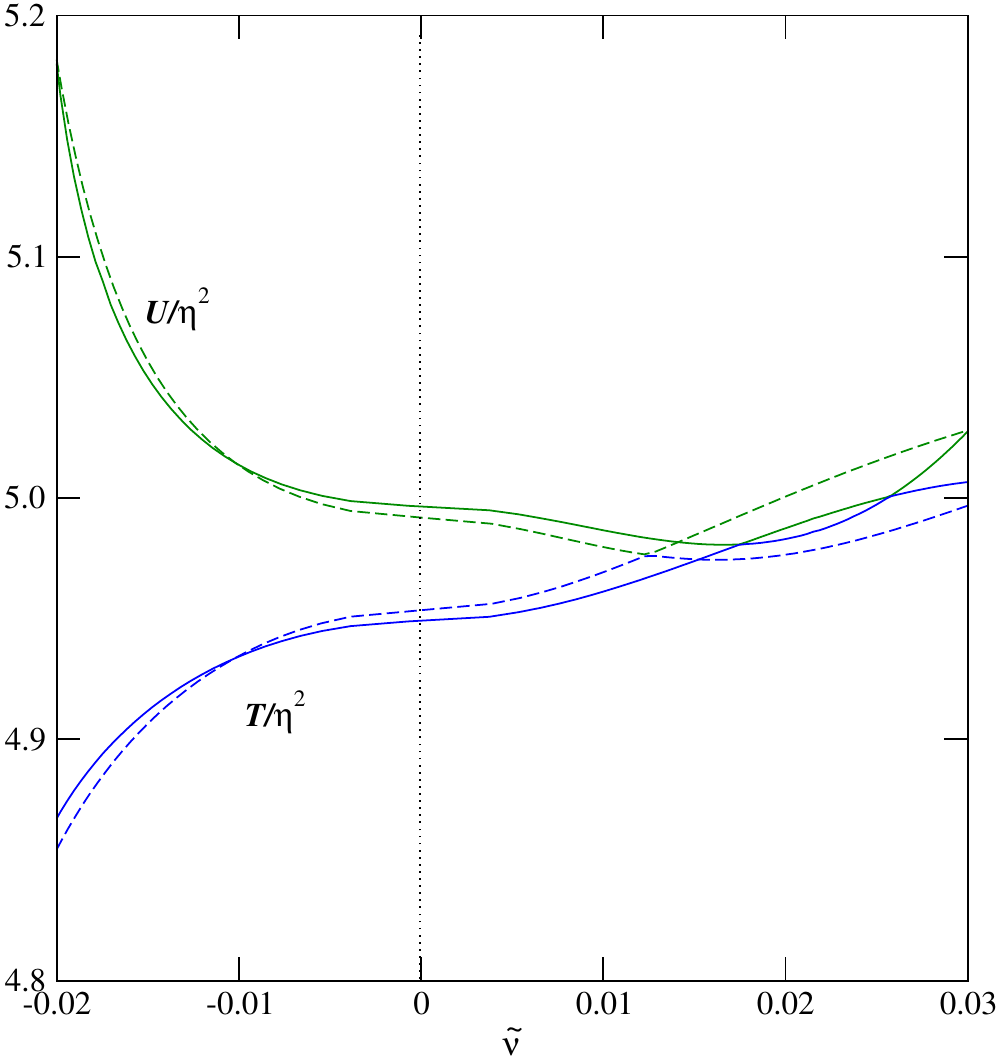}
\includegraphics[width=5.25cm]{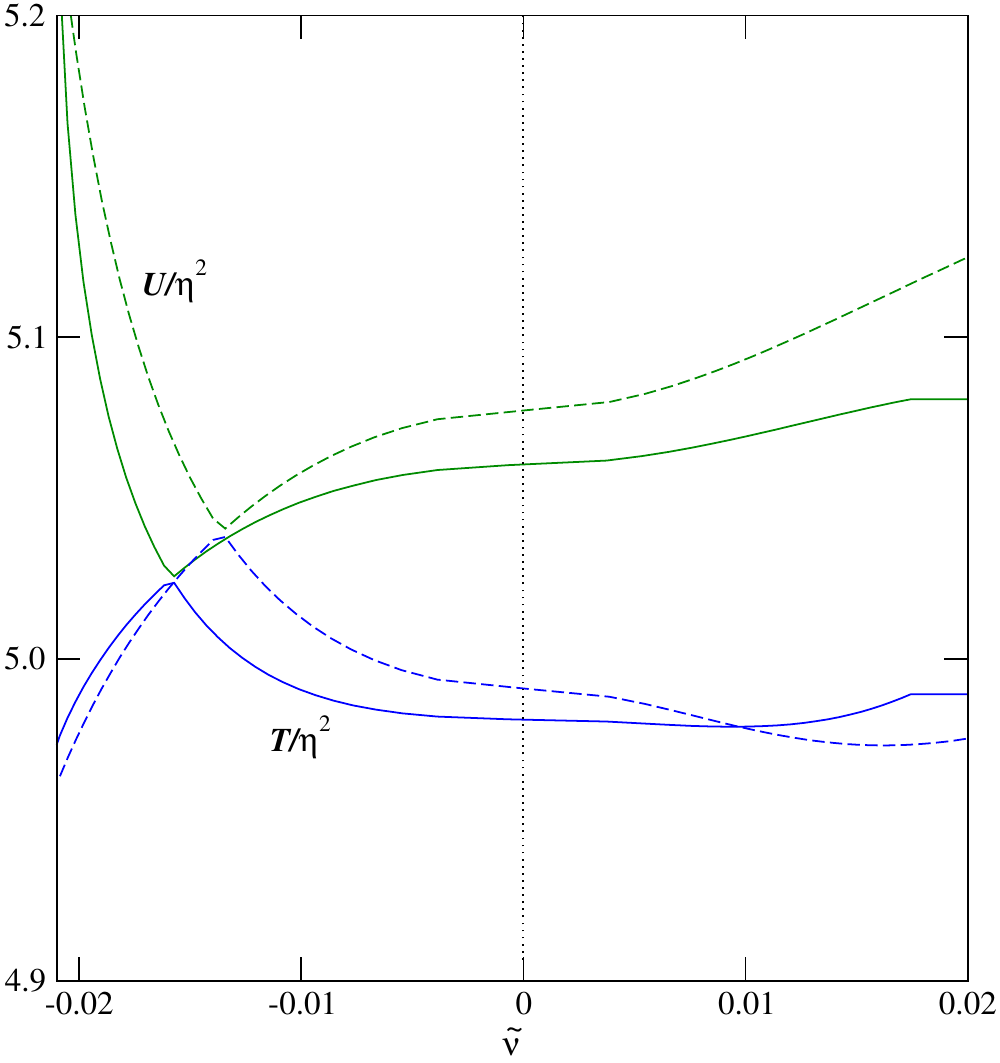}
\includegraphics[width=5.25cm]{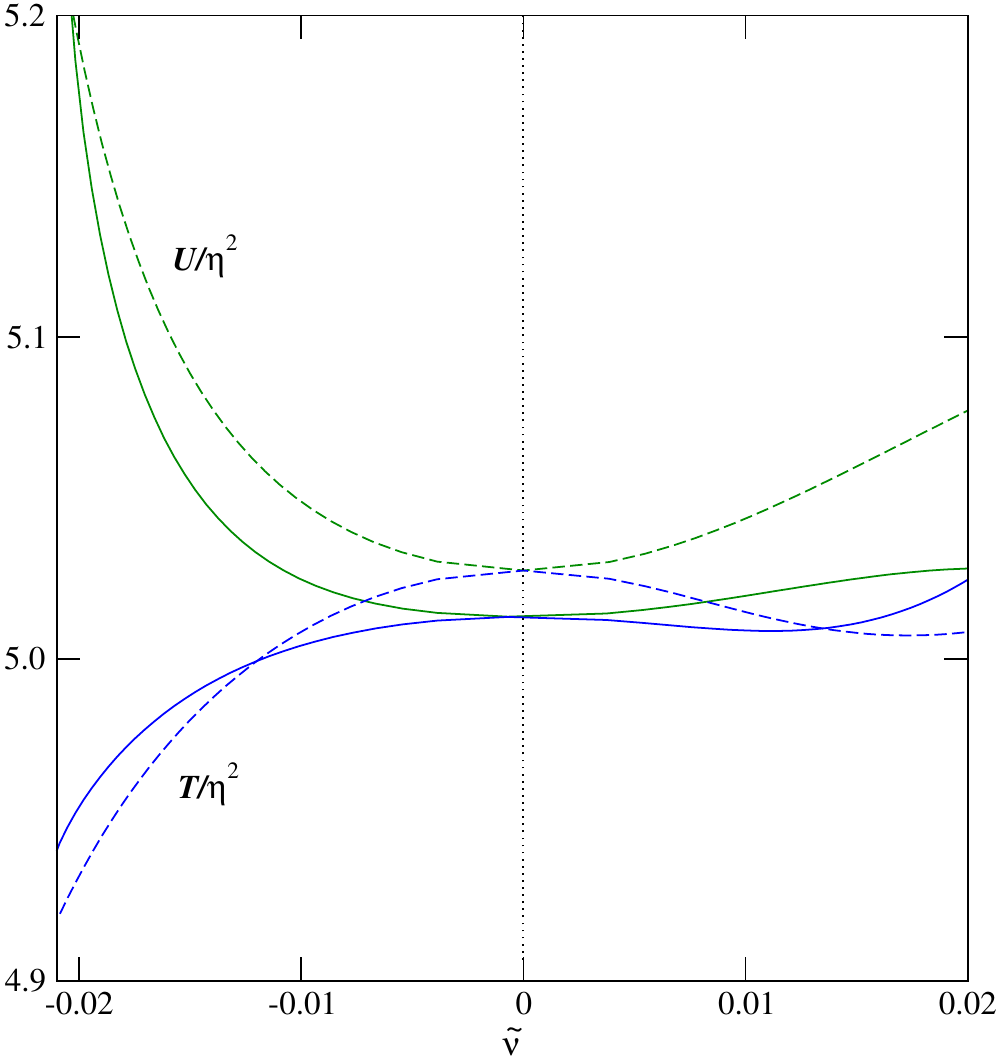}
\includegraphics[width=5.25cm]{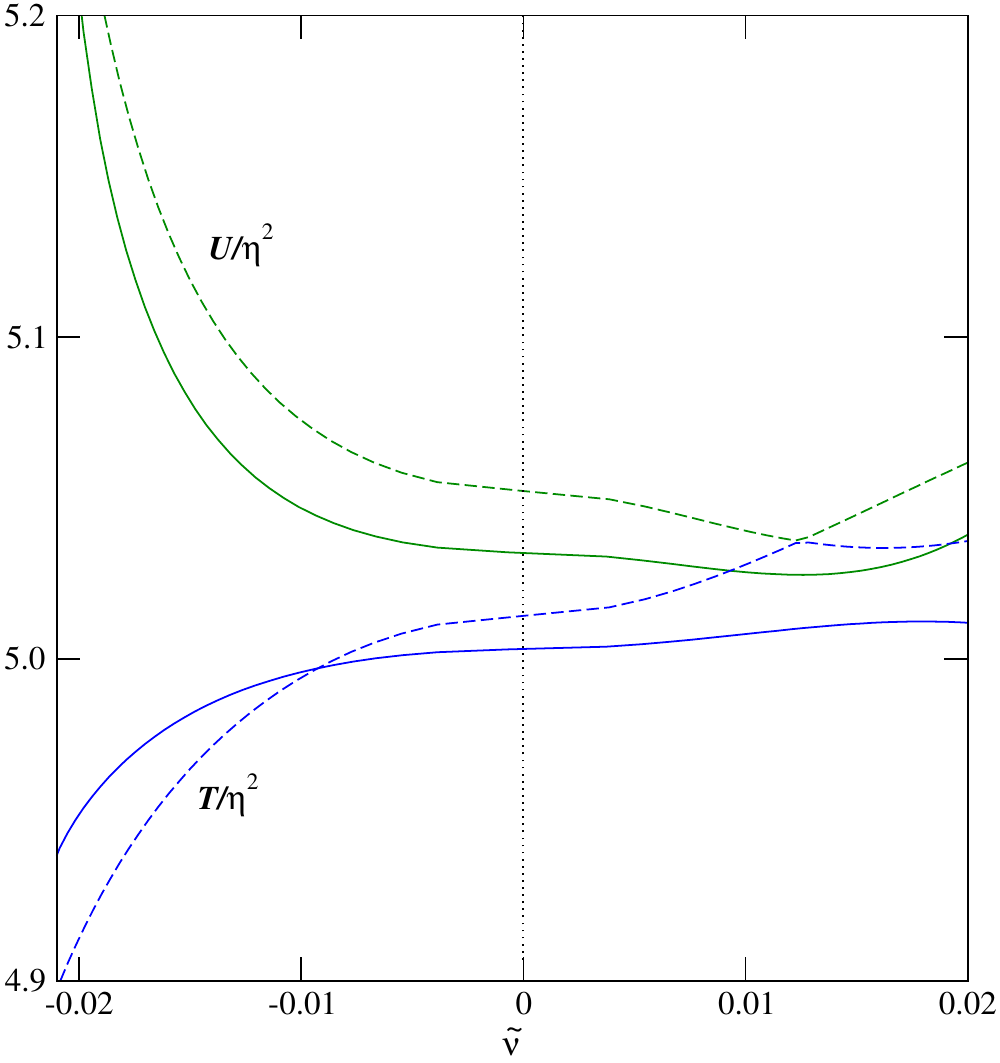}
\caption{Same as Fig.~\ref{fig:sectionsY} but along $w_{\sigma}$ with $w_\phi=-2.25\times 10^{-4}\sim -\gamma_\phi/2$, $0.0$ and $2.25\times 10^{-4}\sim \gamma_\phi/2$.}
\label{fig:sectionsZ}
\end{figure*}

Both $U$ and $T$ are shown in Fig.~\ref{fig:UT2D}, in the $\lb t^{\mu},z^{\mu}\rb$ frame.  For negative and small positive values of $\tilde{w}_{\phi,\sigma}$, there are two condensates in the string.  For larger positive values of either $\tilde{w}_{\phi}$ or $\tilde{w}_{\sigma}$,  one of the two condensates vanishes and the physics reduces to the usual case: the tension diverges to negative values for $\tilde{w}_i\rightarrow -\gamma_i$ while the energy density diverges to positive values (see Figs~\ref{fig:sectionsY} and \ref{fig:sectionsZ}).  When there are two condensates in the string, there exists an additional divergence when both $\tilde{w}_{\phi}$ and $\tilde{w}_{\sigma}$ respectively tend to $-\gamma_{\phi}$ and $-\gamma_\sigma$.  In this limit, the quantity $A$ diverges because it contains an integral in $r$ over a {\it sum} of the squares of $\Phi$ and $\Sigma$ in the potential term and they are themselves divergent~\cite{enon0}.  When $x=0$, $\sqrt{B^2-C^2}$ on the other hand is a slowly growing function in this limit because it is equal to an integral in $r$ over the {\it difference} of the squares of $\Phi$ and $\Sigma$.  Note that in Figs.~\ref{fig:sectionsY} and \ref{fig:sectionsZ}, the full lines represent one dimensional profiles of $U$ and $T$ along $\tilde{w}_{\phi}$ and $\tilde{w}_{\sigma}$ for different values of $\tilde{w}_{\sigma}$ and $\tilde{w}_{\phi}$ respectively.  The dashed lines represent the approximate values of $U$ and $T$ obtained from the analytic model \cite{neutral,models} which we compare to the numerical results in section \ref{simplifiedmodel}.
The timelike and spacelike eigenvectors $u_a$ and $v_a$ are defined respectively through the eigenvalue equations
\begin{equation}
T^{ab} u_b = - U \tilde \eta^{ab} u_b, \ \ \hbox{ with } \ \
\tilde\eta^{ab} u_a u_b = -1, 
\label{equ}
\end{equation}
\noindent and
 \begin{equation}
T^{ab} v_b = - T \tilde \eta^{ab} v_b, \ \ \hbox{ with } \ \
\tilde\eta^{ab} v_a v_b = +1.
\label{eqv}
\end{equation}
The canonical form~(\ref{Tcan}) is recovered provided the four dimensional eigenvectors defined through Eqs.~(\ref{equ}) and (\ref{eqv}) are chosen to be the solutions of Eqs.~(\ref{equ}) and (\ref{eqv}). The eigenvector thus read
\begin{widetext}
\begin{equation}
u^{\mu}=\left[\left( B+\sqrt{B^2-C^2}\right)^2-C^2\right]^{-1/2}
\pmatrix{B+\sqrt{B^2-C^2} \cr C \cr 0 \cr 0},
\label{umu}
\end{equation}
\begin{equation}
v^{\mu}=\left[C^2-\left(B-\sqrt{B^2-C^2}\right)^2\right]^{-1/2}
\pmatrix{B-\sqrt{B^2-C^2} \cr C \cr 0 \cr 0}.
\label{vmu}
\end{equation}
\end{widetext}
Again, given that $B^2-C^2\ge 0$, the quantities under the square roots in $u^{\mu}$ and $v^{\mu}$ are positive definite, a result which will be useful below.  In addition, the combination $B^2-C^2$ depends only on the state parameters and on the $\mcl{K}_i$'s and is thus a Lorentz scalar. Therefore Eq.~(\ref{B2C2K}) holds in whatever frame.  One may therefore work in a frame in which $C\to 0$, in which case Eq.~(\ref{B2C2K}) simply gives the value of $B$.  Note that in order to see that this limit is indeed well-defined in Eqs~(\ref{umu}) and (\ref{vmu}), these two expressions need to be rewritten in a slightly different form from the one presented here.  In the limit $C\to 0$, one then finds that they are nothing but the unit vectors $t^\mu$ and $z^\mu$ of Eq.~(\ref{tmuzmu}).  This frame is the generalization of that frame for which, in the single current case, the phase gradient of the current carrier depends either on time or on space, but not on both, \ie, the boosted frame in which either the frequency or the momentum of the trapped scalar field is removed.
\subsection{Available ranges of variation of the underlying parameters}
\label{arvup}
There exists a finite $w_i$-range within which two currents can appear.  In this range $\mcl{K}_i\ne 0$ ($i=1,2$) and $x$ is well-defined.  The coupling between $\Phi$ and $\Sigma$ acts (nonlinearly) as a positive mass term for both these fields and quite generically (for any given set of masses and coupling constants) will reverse the condensation of either of the two fields outside the appropriate $w_i$-range.  The string then behaves like a one-current-carrying string.\\

The range of variation of $x$ can be constrained using the spacelike or timelike character of the two currents. Setting generic two-dimensional normalized spacelike and timelike vectors $S_\mu$ and $T_\mu$ respectively, with
 \begin{eqnarray}
  S_\mu (\gamma) &\equiv& t_\mu \sinh \gamma + z_\mu \cosh\gamma ,\nonumber \\
  T_\mu (\gamma) &\equiv& t_\mu \cosh \gamma + z_\mu \sinh\gamma ,
\end{eqnarray}
\noindent where $\gamma$ is a constant, the currents $c^i_\mu$ can be chosen proportional to either $S_\mu$ or $T_\mu$, depending on their nature. If the two currents are of a different kind, with, \eg, $c_\mu^{(\Phi)}$ spacelike and $c_\mu^{(\Sigma)}$ timelike, we can take $c_\mu^{(\Phi)}=\sqrt{w_\phi} S_\mu (\gamma_\phi)$ and $c_\mu^{(\Sigma)} = \sqrt{-w_\sigma} T_\mu(\gamma_\sigma)$, leading to $x=-\sqrt{-w_\phi w_\sigma} \sinh(\gamma_\phi-\gamma_\sigma)$, which provides no further restriction on the range.  On the other hand, if both currents are spacelike, we have $c_\mu^{(\Phi)} = \sqrt{w_\phi} S_\mu(\gamma_\phi)$ and $c_\mu^{(\Sigma)} = \sqrt{w_\sigma} S_\mu(\gamma_\sigma)$, and we then find that $x=\sqrt{w_\phi w_\sigma} \cosh (\gamma_\phi-\gamma_\sigma)$, and is therefore larger than $\sqrt{w_\phi w_\sigma}$.  Similarly, if both currents are timelike, we set $c_\mu^{(\Phi)} = \sqrt{-w_\phi} T_\mu(\gamma_\phi)$ and $c_\mu^{(\Sigma)} = \sqrt{-w_\sigma} T_\mu(\gamma_\sigma)$, leading to the result that $x=-\sqrt{w_\phi w_\sigma} \cosh(\gamma_\phi-\gamma_\sigma)$, which is now smaller than $-\sqrt{w_\phi w_\sigma}$. In short, for two currents having the same character, the range of variation of $x$ is restricted to
\begin{equation}
|x|\geq \sqrt{w_{\phi}w_{\sigma}}.
\label{xbound} 
\end{equation}
One of the currents, $c_\mu^{(\Phi)}$ say, can be lightlike. In this case, it must read $c_\mu^{(\Phi)} = m_\phi (\epsilon t_\mu +\epsilon' z_\mu)$, where we have fixed the arbitrary normalization to the mass of the corresponding current-carrier for reasons of dimensions, while $\epsilon^2=\epsilon^{\prime 2} = 1$. If $c_\mu^{(\Sigma)}$ is spacelike, \ie, $c_\mu^{(\Sigma)} =\sqrt{w_\sigma} S_\mu (\gamma_\sigma)$, we obtain $x=\epsilon' m_\phi\sqrt{w_\phi}\ex^{-\epsilon\epsilon'\gamma_\sigma}$, while if it is timelike, with $c_\mu^{(\Sigma)} = \sqrt{-w_\sigma}T_\mu(\gamma_\sigma)$, the third state parameter is then $x=-\epsilon m_\phi \sqrt{-w_\phi} \ex^{-\epsilon\epsilon'\gamma_\sigma}$. Finally, both currents can be lightlike, with $c_\mu^{(\Phi)} = m_\phi(\epsilon_\phi t_\mu + \epsilon'_\phi z_\mu)$ and $c_\mu^{(\Sigma)} =m_\sigma (\epsilon_\sigma t_\mu + \epsilon'_\sigma z_\mu)$, leading to $x=- m_\phi m_\sigma (\epsilon_\phi\epsilon_\sigma -\epsilon'_\phi\epsilon'_\sigma)$.\\

We shall see in detail in Section \ref{transversestability} that the range of $x$ is further restricted by imposing the stability of the string under transverse perturbations.\\

It is important to realize at this stage that the macroscopic dynamics of a string on which two condensates can appear is more complicated than its one condensate counterpart. In particular, the number of degrees of freedom is itself a dynamical variable: as either state parameters $w_\phi$ and $w_\sigma$ evolve along the string or with time (they both are in principle functions of the string internal coordinates~\cite{2D,3D,formal1,formal2,formal3,formal4}), the corresponding fields may switch back and forth between condensating to non-condensating situations, with the consequence that the number of state parameters may jump discontinuously, being equal to either one or three (we are assuming that the underlying parameters are such that for $w_\phi=w_\sigma=0$, both condensates are present). This seems to forbid any kind of macroscopic treatment such as proposed in Ref.~\cite{formal1,formal2,formal3,formal4}. We shall see however that because the string remains essentially a one-dimensional object, its classical stability can be investigated provided a generalization of the usual framework is made. This is was we do in the following sections.
\section{Elastic string dynamics}
\label{secdyn} 
The dynamics of a current-carrying elastic string depends on the string's internal degrees of freedom but not on its geometry.  It has been extensively studied in \cite{formal1,formal2,formal3,formal4}. Because its Lagrangian depends only on the two state parameters $w_i$, a string with two condensates can be studied using the same formalism.  The purpose of this section is to extend the existing formalism for the dynamics of a single-condensate elastic string to an $N$-condensate string.
\subsection{Preliminary geometric definitions}
The string worldsheet is defined as the two-dimensional surface swept by the string during its time evolution.  It has timelike and spacelike directions associated with timelike and spacelike internal coordinates $\xi^0$ and $\xi^1$. The metric on the string worldsheet reads
\begin{equation}
h_{ab}=\displaystyle g_{\mu\nu} x^\mu_{,a} x^\nu_{,b} 
\end{equation}
where the subscripts ``${}_{,a}$'', ``${}_{,b}$'', ... denote derivation with respect to
$\xi^{a}$, $\xi^{b}$, ...; the inverse metric is $h^{ab}$.  The embedding of the two-dimensional metric in four dimensional spacetime is defined through the string worldsheet's first fundamental tensor as
\begin{equation}
  \eta^{\mu\nu}=h^{ab} x^\mu_{,a} x^\nu_{,b}.
\end{equation}
Its mixed form ${\eta^\mu}_\nu$ is identified with the tangential projector on the string worldsheet.  The orthogonal projector is then defined through
\begin{equation}
{\perp^\mu}_\nu={g^\mu}_\nu-{\eta^\mu}_\nu.
\label{perp}
\end{equation}
With these definitions, it is possible to embed all fields on the worldsheet in four dimensional spacetime, provided that the 4D covariant derivative taken along the string worldsheet $\nabla_a$ is replaced by a new longitudinal covariant derivative $\nablas_\mu$ which projects out the  meaningless variations transverse to the string worldsheet,
\begin{equation}
   \nablas_\mu
   ={\eta_\mu}^\nu \nabla_\nu.
\end{equation}
We also introduce the anti-symmetric fundamental tensor $\varepsilon^{\mu\nu}$ through the relation
\begin{equation}
\varepsilon^{\mu\rho} {\varepsilon_\rho}^{\nu}=\eta^{\mu\nu}.
\end{equation}
Note that $\varepsilon$ is defined up to an overall sign.  The curvature tensor,
\begin{equation}
 {K_{\mu\nu}}^\rho = {\eta^\sigma}_\mu \nablas_\nu
  {\eta^\rho}_\sigma,
\end{equation}
is tangent to the worldsheet in its first two covariant indices $\mu$ and $\nu$ and orthogonal to it in its contravariant index $\rho$. Since the projector ${\eta^\mu}_\nu$ defines the space tangent to the string worldsheet, ${K_{\mu\nu}}^\rho$ is symmetric in its first two indices.  This is the Weingarten identity~\cite{formal1,formal2,formal3,formal4}
\begin{equation} 
{K_{[ \mu\nu ]}}^\rho=0.
\label{intcond}
\end{equation}
Finally, using two orthonormal basis vectors $u^\mu$ and $v^\mu$ tangent to the worldsheet and chosen timelike and spacelike respectively, the fundamental tensors take the simple form
\begin{eqnarray}
\eta^{\mu\nu} & = & -u^ \mu u^\nu+v^\mu v^\nu\\
\varepsilon^{\mu\nu} & = & \mbox{} \pm \mbox{} (u^\mu v^\nu -u^\nu
v^\mu).
\label{expr}
\end{eqnarray}
\subsection{The elastic string model}
As stated above, an elastic string is described by an effective Lagrangian which only depends on its {\it internal} degrees of freedom. In the case at hand, there are $N$ internal degrees of freedom given by the $N$ fields $\psi_i$ associated with the $N$ currents living on the string worldsheet. The Lagrangian
\begin{equation} 
\mathcal{L}=\mathcal{L}(\chi_{ij}) 
\label{Lagrangian}
\end{equation}
may thus {\it a priori} depend on any of the scalars 
\begin{equation}
\chi_{ij}=\eta^{\mu\nu} \left(\nablas_\mu \psi_i\right)\left(\nablas_\nu \psi_j\right),
\label{chij}
\end{equation}
\ie, the full set of state parameters that form an $N\times N$ symmetric matrix. The Lagrangian will thus be written in the remainder of the paper as a function of the symmetric matrix $\bm{\chi}=(\chi_{ij})$ and denoted by ${\cal L}(\bm{\chi})$.\\

For instance, in the case of the two currents presented in the previous sections, Eqs.~(\ref{paramscalar}) and (\ref{defx}) show that $\chi_{ii}=w_i$ and $\chi_{\phi\sigma}=\chi_{\sigma\phi}=x$, whereas Eq.~(\ref{lag2}) immediately shows that the Lagrangian of this particular model only depends on the diagonal entries $w_i$ of the matrix $\bm{\chi}$.\\

The equations of motion of the elastic string can be obtained directly by varying the Lagrangian with respect to the worldsheet coordinates $x^\mu (\xi^a)$ and to the internal fields $\psi_i$.  However, it is easier and more useful to write the dynamical equations as conservation equations because the physically meaningful unknowns are the string conserved currents $c^{i\mu}$, not the internal fields $\psi_i$.\\

There are $2N+2$ independent degrees of freedom: six for the first two currents which define the string worldsheet through its tangent space and therefore satisfy the Weingarten identity (\ref{intcond}), and then two for each of the other $N-2$ currents which live in this same tangent space. $2N+2$ equations are therefore needed.\\

The $N$ conserved currents inside the string
 \begin{equation}
c^{i\mu}=\frac{\delta \mathcal{L}}{\delta \left(\nablas_\mu
  \psi_i\right)}=\frac{\delta \mathcal{L}}{\delta \chi_{jk}} \left({\delta^i}_j
\nablas^\mu \psi_k +{\delta^i}_k\nablas^\mu \psi_j\right),
\label{cidef}
\end{equation}
where ${\delta^i}_j$ is the Kronecker delta, can be expressed in terms of the phase gradients through
\begin{equation}
c^{i\mu} ={\mathcal{K}}^{ij} \nablas^\mu \psi_j,
\end{equation}
with
\begin{equation}
 {\mathcal{K}}^{ij}=2\frac{\delta \mathcal{L}}{\delta \chi_{ij}}.
\label{Kij}
\end{equation}
Note that Eq.~(\ref{Kij}) generalizes Eq.~(\ref{defK1}).  These currents are an obvious choice for a first set of $N$ conservation equations. Their conservation equations read
 \begin{equation} 
\nablas_\mu (c^{i\mu})=0.
\label{eqc}
\end{equation}
The stress energy momentum tensor $T^{\mu\nu}$ defined by Eq. (\ref{Tcan}) or equivalently by
 \begin{equation}
   T_{\mu\nu}= {\cal L}
   \eta_{\mu\nu} -2 \frac{\delta {\cal L}}{\delta \eta^{\mu\nu}} = {\cal
     L} \eta_{\mu\nu}-{\mathcal{K}}^{ij} (\nablas_\mu \psi_i)
   (\nablas_\nu \psi_j) ,
\label{Tdef}
\end{equation}
satisfies the conservation equation
 \begin{equation}
\nablas_\mu T^{\mu\nu}=0,
\label{consT}
\end{equation}
and provides 4 additional conservation equations.  The first two are given by the transverse part of this relation, namely
\begin{equation}
 {\perp_\nu}^\rho \nablas_\mu T^{\mu\nu}=0,
\label{transT}
\end{equation}
defined as its projection orthogonal to the worldsheet, is associated with the geometry of the string worldsheet.  The remaining two equations are given by the longitudinal part of Eq.~(\ref{consT}),
\begin{equation} {\eta_\nu}^\rho \nablas_\mu
  T^{\mu\nu}=0,
\label{longT}
\end{equation}
and is associated with the internal degrees of freedom of the string.  Finally, there exists an irrotationality condition on each of the $N$ gradient fields $\nablas_\mu \psi_i$,
\begin{equation}
\mbox{} \varepsilon^{\mu\nu}
\nablas_{\mu}(\nablas_\nu \psi_i)=0.  \mbox{}
\label{irc}
\end{equation}
The vector $\nablas_\mu (\varepsilon^{\mu\nu})$ being purely orthogonal (see Eq.~(\ref{expr})), these irrotationality conditions can be turned into conservation equations through an integration by parts,
\begin{equation}
\nablas_\mu (d_i^\mu)=0,
\label{eqd}
\end{equation}
\noindent  where 
 \begin{equation}
d_i^\mu=\varepsilon^{\mu\nu} \nablas_\nu \psi_i .
\label{dcurrents}
\end{equation}
\noindent The previous considerations result in the existence of $2N$ conserved currents and a conserved tensor for a total of $2N+4$ conservation equations. These are two more than needed. This is because, as will be shown next, there exists a redundancy between the two longitudinal stress energy tensor conservation equations and the current conservation equations. In the well known case of one current only, \ie, for $N=1$, the two current conservation equations are exactly equivalent to the longitudinal part of the stress energy tensor. On the other hand, in the general case, the natural choice is to keep the $2N$ conservation equations only.\\

We now show that the conservation equations (\ref{eqc}) and (\ref{dcurrents}), or equivalently (\ref{irc}), imply the longitudinal conservation of the stress energy tensor. Rewriting the stress energy tensor as
\begin{equation} T^{\mu\nu}={\cal
    L}\eta^{\mu\nu}-c^{j\mu}({\varepsilon^\nu}_\rho d_j^\rho) ={\cal
    L}\eta^{\mu\nu}-({\varepsilon^\mu}_\rho d_i^\rho) c^{i\nu},
\end{equation}
we have
\begin{equation} \eta_{\mu\rho}\nablas_\nu
  T^{\mu\nu}=\nablas_{\rho}{\cal L}-\varepsilon_{\rho\mu} c^{i\nu}
  \nablas_\nu d_i^\mu .
\label{eqlongT}
\end{equation}
Furthermore, the gradient of Eq.~(\ref{Lagrangian}) reads
\begin{equation}
\nablas_{\rho}{\cal L}= \frac{1}{2}{\mathcal{K}}^{ij} \nablas_\rho
\left[\eta^{\mu\nu} \left(\nablas_{\mu} \psi_i
  \right)\left(\nablas_{\nu} \psi_j\right)\right]= c^{i\mu}
\varepsilon_{\mu\nu} \nablas_\rho d_i^\nu,
\end{equation}
so that Eq.~(\ref{eqlongT}) becomes
\begin{equation} \eta_{\mu\rho}\nablas_\nu T^{\mu\nu}=c^{i[ \mu }
  {\eta^{\nu ]}}_\rho {\varepsilon_\mu}^\sigma (\nabla_\nu
  d_{i\sigma}).
\end{equation}
Finally, $c_i^{[ \mu} \eta^{\nu ] \rho}$ is tangential and antisymmetric in its $\mu$ and $\nu$ indices and is thus proportional to the antisymmetric tangential tensor $\varepsilon^{\mu\nu}$,
\begin{equation} c^{i [ \mu} \eta^{\nu ]
    \rho}=({\varepsilon^\rho}_\tau
  c^{i\tau})\varepsilon^{\mu\nu}.
\end{equation}
Eq. (\ref{eqlongT}) therefore reduces to
\begin{equation} \eta_{\mu\rho}\nablas_\nu
  T^{\mu\nu}=(\varepsilon_{\rho\tau} c^{i\tau})(\nablas_\nu d_i^\nu),
\label{redlongT}
\end{equation}
and as expected implies
\begin{equation}
\eta_{\mu\rho}\nablas_\nu T^{\mu\nu}=0.
\end{equation}
\subsection{Equations of state}
It is clear from the previous section that the equations of motion can be reduced to the $2N+2$ conservation equations (\ref{eqc}), (\ref{transT}) and (\ref{eqd}).  As stated before, the currents $d_i^\mu$ are the physically most meaningful quantities and, from their definition (\ref{dcurrents}), yield the matrix of state parameters
\begin{equation} \chi_{ij}=-\eta^{\mu\nu} d_{i\mu}
  d_{j\nu}.
\label{chid}
\end{equation}
They also appear explicitly in $N$ of the conservation equations (\ref{eqd}).\\

On the other hand, the $N+2$ other equations do not depend explicitly on the $d_i^\mu$'s. However, since they determine the matrix of state parameters through (\ref{chid}), which in turn determines the Lagrangian and its derivatives, it is certainly possible to express the stress energy tensor $T^{\mu\nu}$ and the other conserved currents $c^{i\mu}$ as functionals of the unknowns $d_{i\mu}$. These  functionals are therefore additional identities that form a set of equations of state. In general, they will be determined by the underlying field theory model (\ie, a generalization to $N$ currents of what was done in the first four sections of this paper), and will characterize the current-carrying string model under consideration.\\

The $N^2$ equations of state expressing
\begin{equation} 
{\cal K}^{ij}={\cal K}^{ij} (\bm{\chi})
\label{eqstc}
\end{equation}
\noindent are necessary and sufficient to determine the conserved
currents
\begin{equation}
 c^{i\mu}={\cal K}^{ij} \varepsilon^{\mu\nu}  d_{j\nu}, 
\label{c=f(d)}
\end{equation}
\noindent from the other currents $d_{i\mu}$, and the symmetric matrix $\bm{\chi}$ given by (\ref{chid}).\\

As for the stress energy tensor, it is a tangential, and thus two-dimensional, symmetric tensor which is therefore determined by three parameters. The stress energy tensor is usually expressed in the diagonal form (\ref{Tcan}) where $u^\mu$ and $v^\mu$ are a basis of respectively timelike and spacelike orthonormal vectors, whereas the eigenvalues $U$ and $T$ can be respectively interpreted as the string energy density and tension. Thus, the natural choice for the three parameters defining the stress energy tensor are $U$, $T$, and the position of the basis $(u^\mu,v^\nu)$ in the tangent space. The latter is akin to a hyperbolic angle $\psi$ with a given direction such as that of one of the currents $d_{i\mu}$, $d_{1\mu}$ for example. Note that this only defines the vectors $u^\mu$ and $v^\nu$ up to a sign, but this sign does not change the stress energy tensor (\ref{Tcan}). There will thus be three extra equations of state expressing these three parameters as a function of the matrix $\chi$.  In conclusion, there are $N+3$ equations of state which will be chosen here as (\ref{chid}), (\ref{eqstc}) and
\begin{equation}
U=U(\bm{\chi}),\ T=T(\bm{\chi}),\ \psi=\psi(\bm{\chi}),
\label{eqstT}
\end{equation}
\noindent where the hyperbolic angle $\psi$ is such that
\begin{equation}
d_{1\mu}\propto\cosh(\psi)u^\mu +\sinh(\psi) v^\mu, 
\end{equation}
if $d_{1\mu}$ is timelike, and
\begin{equation}
d_{1\mu}\propto\sinh(\psi)u^\mu +\cosh(\psi) v^\mu,
\end{equation}
if it is spacelike.
\subsection{Duality}
In the case of one current, there exists a duality which allows to exchange the roles of the currents $c_1^{\mu}$ and $d_1^\mu$ \cite{formal1,formal2,formal3,formal4}. The existence of $2N$ conserved currents instead of the expected $N$ suggests that there similarly should be a duality between the two sets of $N$ conserved currents $c^{i\mu}$ and $d_i^\mu$.  This is a duality under the global exchange of the two sets of currents, not just of two given currents $c^{i\mu}$ and $d_i^\mu$. More precisely, we are looking for a dual Lagrangian $\tilde{\cal L}$, of the same general form as ${\cal L}$ (but function of different scalar fields $\tilde{\psi}$), yielding the same equations of motion, but such that the two sets of $N$ conserved  currents are exchanged. Using tildes to distinguish all quantities derived from the dual Lagrangian, it is sufficient to require that
\begin{eqnarray}
\tilde{c}_{i\mu} & = & d_{i\mu} 
\label{tilde1} \\
\tilde{d}^{i\mu} & = & c^{i\mu} 
\label{tilde2} \\
\tilde{T}^{\mu\nu} & = & T^{\mu\nu}
 \label{tilde3}.
\end{eqnarray}
Using the notations of Eqs.~(\ref{Lagrangian}), (\ref{chij}) and (\ref{Kij}) as well as the definitions (\ref{cidef}) and (\ref{dcurrents}) with tildes for the dual model, Eq.~(\ref{tilde2}) becomes
\begin{equation}
 {\varepsilon_\mu}^\nu \nablas_\nu \tilde{\psi}^i =
  {\cal K}^{ij} \nablas_\mu \psi_j
\label{intertildex}
\end{equation}
\noindent which yields in turn that
\begin{equation}
\tilde{\chi}^{ij}=-{\cal K}^{ik}\chi_{kl}{\cal
    K}^{lj}=-(\bm{{\cal K}\chi {\cal
    K}})^{ij}
\label{tildex}
\end{equation}
\noindent where the right hand side is to be understood as matrix multiplications.  Then, substituting Eq.~(\ref{intertildex}) into (\ref{tilde1}) yields
\begin{equation}
\nablas_\mu \psi_i={\varepsilon_\mu}^\nu \tilde{\cal K}_{ij}
\nablas_\nu \tilde\psi^j=\tilde{\cal K}_{ij}{\cal K}^{jk}\nablas_\mu
\psi_k,
\end{equation}
\noindent which in turn implies
\begin{equation} 
\bm{\tilde{\mathcal{K}}} =\bm{{\mathcal{K}}}^{-1},
\label{tildek}
\end{equation}
\noindent where the right hand side is again to be understood as a matrix inverse.\\

Finally, the last equation (\ref{tilde3}) can be expanded using Eq.~(\ref{Tdef}) and the previous expressions for $\tilde{\cal K}$ and $\nablas \tilde\psi _i$ as
\begin{equation} (\tilde{\cal L}-{\cal L})\eta_{\mu\nu} = -{\cal
    K}^{ij} (\nablas_\rho \psi_i) (\nablas_\sigma \psi_j)
  ({\eta_\mu}^\rho {\eta_\nu}^\sigma -{\varepsilon_\mu}^\rho
  {\varepsilon_\nu}^\sigma) .
\label{inter}
\end{equation}
\noindent  The identity 
\begin{equation}
 {\eta_\mu}^\rho
  {\eta_\nu}^\sigma-{\varepsilon_\mu}^\rho {\varepsilon_\nu}
  ^\sigma=\eta_{\mu\nu} \eta^{\rho\sigma}-\varepsilon_{\mu\nu}
  \varepsilon ^{\rho\sigma},
\end{equation}
\noindent which can be derived for instance by using a pair of orthonormal vectors and Eqs.~(\ref{expr}), enables to simplify Eq.~(\ref{inter}) to
\begin{equation}
\tilde\mathcal{L} = \mathcal{L}-\mathcal{K}^{ij}
  \chi_{ij}={\cal L}- \tr \left( \bm{\mathcal{K}\chi}\right), 
\label{tildel} 
\end{equation}
\noindent where in the latter expression, the matrix trace $\tr$ has been taken.\\

Thus a model with Lagrangian $\tilde\mathcal{L} \left( \tilde\chi_{ij} \right)$ given by Eqs.~(\ref{tildex}), (\ref{tildek}) and (\ref{tildel}) has, by construction, the same $2N$ conserved currents and stress energy tensor, and thus the same equations of motions as the initial model described by the Lagrangian ${\cal L}$. However, these three equations are not completely independent since one must satisfy consistency equations derived from Eq.~(\ref{Kij}), namely
\begin{equation} \tilde\mathcal{K}_{ij} = 2\frac{\dd \tilde\mathcal{L}}
{\dd\tilde\chi^{ij}} = \left( \bm{\mathcal{K}}^{-1}\right)_{ij},
\end{equation}
\noindent This equation can straightforwardly be derived from Eqs.~(\ref{tildel}) and (\ref{tildex}).\\

Thus, the model described by the Lagrangian $\tilde\mathcal{L} \left( \bm{\tilde\chi} \right)$ derived from the initial model by Eqs~(\ref{tildex}) and (\ref{tildel}) produces the exact same elastic string dynamics.  This is precisely the duality found previously \cite{formal1,formal2,formal3,formal4} for an elastic string with a single current.
\section{The elastic domain}
The elastic $N-$current-carrying string model introduced in preceding sections is obtained from the field theory solution of a straight and static string solution and further assuming that, at the macroscopic level, the string is locally straight and static in a rotated and boosted frame in which $u^\mu$ and $v^\mu$ defined in Eq.~(\ref{Tcan}) are aligned with the $t$ and $z$ axes. This is usually an excellent approximation since the string curvature is of the order of the Hubble scale, while its thickness (to which the latter curvature must be compared to evaluate the ``straightness'' of the string) is of the order of the Compton length of the lightest particle involved (typically a current carrier).\\

However, this approximation is only meaningful if the static straight solution is dynamically stable, because otherwise, the string curvature, however negligible to begin with, will drive the string away from the straight and static solution. The conditions under which the string remains straight and static will define the domain of elasticity of the string.  Evaluating string stability at the field theory level is a complicated task. Instead, we shall determine whether the static straight string solution is dynamically stable in the elastic string model.  If it is, the description is self-consistent.  If it is not, a more detailed calculation must be performed at the microscopic level.\\

In the straight and static solution, the currents $d_i^\mu$ are constant along the string worldsheet.  This implies that the other currents $c^{i\mu}$ and the stress energy tensor $T^{\mu\nu}$ are also constant. This ensures that the equations of motion (\ref{eqc}), (\ref{transT}) and (\ref{eqd}) are satisfied.\\

Here, we shall take the static timelike Killing vector $t^\mu$ and the static spacelike Killing vector $z^\mu$ introduced in (\ref{tmuzmu}) to describe the time coordinate and the direction along which the string lies.  The stress-energy tensor (\ref{Tcan}) then reads
\begin{equation}
T^{\mu\nu}= U t^\mu t^\nu-T z^\mu z^\nu.
\label{Tmunu0}
\end{equation}
\noindent It constrains the frame $(t^\mu,z^\mu)$ through $\psi$ in Eq.~(\ref{eqstT}).\\

To test the dynamical stability of the straight and static `background' solution thus defined, we Fourier expand the first order perturbations of the equations of motion, and derive the characteristic modes.  If all the characteristic modes are real, the unperturbed solution is stable, otherwise it is unstable.
\subsection{Stability of the transverse modes}
As explained earlier, there are only two transverse degrees of freedom and the corresponding equations are given by Eq.~(\ref{transT}), which can be rewritten as
\begin{equation}
{\perp_\mu}^{\rho}\left(u^\nu \nabla_\nu u^\rho
    -\frac{T}{U} v^\nu \nabla_\nu v^\rho\right) =0,
\label{trans1}
\end{equation}
\noindent where ${\perp_\mu}^{\rho}$ is the projector orthogonal to the string worldsheet defined in Eq.~(\ref{perp}).  The string worldsheet is defined implicitly from its tangent space (generated by the eigenvectors $u^\mu$ and $v^\mu$). Eq.~(\ref{intcond}) therefore needs to be satisfied by the eigenvectors.  It yields
\begin{equation} 
{\perp_\mu}^{\rho}(u^\nu \nabla_\nu v^\rho -v^\nu
  \nabla_\nu u^\rho) =0.
\label{trans2}
\end{equation}
\noindent Transverse perturbations of $d_{i\mu}$ do not change their mutual scalar products $-\chi_{ij}$ to first order and thus, the energy density $U$ and tension $T$ of the string are not affected by transverse perturbations of the worldsheet.  Conversely, the current conservation equations are unaffected to first order by transverse perturbations. Thus, the transverse modes are decoupled and it is sufficient to consider transverse perturbations of the eigenvectors $u^\mu$ and $v^\mu$, together with Eqs (\ref{trans1},\ref{trans2}).\\

This system of equations is exactly the same as in the case of a
string carrying a single current~\cite{formal1,formal2,formal3,formal4}.  Given the usual definition \cite{carterPLB89} $\delta\lb Q\rb\equiv e^{i\chi}\epsilon\lb Q\rb$ with $\nabla_{\mu}\chi=k_{\mu}$, $k=k_{\mu}v^{\mu}$, $\omega=-k_{\mu}u^{\mu}$, where $Q$ denotes $u^{\mu}$ or $v^{\mu}$ and $\epsilon\lb Q\rb\ll 1$, one finds that transverse perturbations propagate at velocity
\begin{equation}
\cT^2=\frac{\omega^2}{k^2}=\frac{T}{U}, 
\end{equation}
as can also be checked directly from Eqs.~(\ref{trans1}) and (\ref{trans2}).  This result was to be anticipated, since the conservation equations are identical to those of the single string case and their projection orthogonal to the worldsheet do not depend on perturbations of the conserved currents but only on $\delta\lb u^{\mu}\rb$ and $\delta \lb v^{\mu}\rb$. Thus, the dynamical stability of the transverse perturbation of the straight string requires simply that the tension be positive,
\begin{equation}
 T\geq 0.
\label{transcons}
\end{equation}
\subsection{Stability of the longitudinal modes}
\label{longstab}
We now consider the longitudinal modes, \ie, those propagating in an internal way along the string worldsheet.  We expect the results to differ from those of the single string case since, contrary to the transverse conservation equations, the longitudinal conservation equations depend on perturbations of the conserved currents.  The presence of more than one current implies current-current interactions.  In addition, contrary to the single current case in which $T^{\mu\nu}$ is automatically diagonal, we loose the freedom to align the currents with our prefered frame $\lb u^{\mu},v^{\mu}\rb$. As a result, the well-known result $\cL^2=-\dd T/\dd U>0$ is not expected to be recovered in the general case of $N$ currents.\\

In order to describe these modes, we once again assume that the string remains straight and the problem is effectively two-dimensional in the $\lb t^\mu,z^\mu\rb$ plane.  In this case, the unknowns are the string currents $d_i^\mu$.  The corresponding equations can be chosen to be the conservation equations (\ref{eqc}) and (\ref{eqd}), together with the equations of state (\ref{eqstc}).  As in the single current case, the worldsheet is not perturbed by longitudinal modes. The tensors $\eta^{\mu\nu}$ and $\varepsilon^{\mu\nu}$ are therefore not affected by longitudinal perturbations.\\

We expand the perturbations of the $N$ currents $d_{i\mu}$ in Fourier modes as $\delta d_{i\mu}\ex^{i\omega_\mu x^\mu}$. The equations of motion for the longitudinal perturbations are then derived from Eq.~(\ref{eqc}) as
\begin{equation}
 {\cal K}^{ij}(\omega_\mu \varepsilon^{\mu\nu} \delta
  d_{j\nu})-2\frac{\delta{\cal K}^{ij}}{\delta
    \chi_{kl}}\eta^{\rho\sigma}\delta d_{k\rho} d_{l\sigma}
  (\omega_\mu \varepsilon^{\mu\nu} d_{j\nu})=0,
\label{eqlong}
\end{equation}
\noindent and from Eq. (\ref{eqd}) as
\begin{equation}
 \eta^{\mu\nu} \omega_\mu \delta d_{i\nu}=0.
\label{eqlong1}
\end{equation}
\noindent At this order in perturbations, a mode associated with one of the currents thus couples to all other unperturbed currents but not to their perturbated part.  Thus, the string will be stable against longitudinal perturbations as long as the dispersion relation for a given mode is a real.  Unfortunately, the system of $2N$ equations given by the two expressions above is quite complicated and there is {\it a priori} no way to solve it in the general case.  In particular, as announced at the begining of this section, the simple form $\cL^2=-\dd T/\dd U>0$ is not recovered.  The main simplification which can be achieved is to halve the number of equations and unknowns by noting that, from Eq.~(\ref{eqlong1}), all the perturbations $\delta d_{i\mu}$ must be orthogonal to $\omega_\mu$, and thus colinear to each other. However, this is not enough to solve the system algebraically or even to reduce the problem to a standard eigenvalue problem.  Thus, to get explicit constraints of the elastic domain associated with the longitudinal equations, we shall restrict ourselves to the case of two currents.
\section{The elastic domain for the two currents model}
\label{sectwocurr}
We shall now concentrate explicitly on the elastic string model that pertains to the microscopic model of the first sections.  In this case, the Lagrangian (\ref{Lagrangian}) that stems from Eq.~(\ref{lag2}) depends only on the state parameters $w_i=\chi_{ii}$ ($i=1,2$, with the identification $\phi\rightarrow 1$ and $\sigma\rightarrow 2$), but not on the off-diagonal component $x$.  Note that in the general case a dependence of $\mcl{L}$ on $\chi_{12}=x$ is \apriori allowed. In the case at hand, while Eq.~(\ref{Tcan}) or Eq.~(\ref{Tdef}) do not depend on $x$, Eq.~(\ref{eqstT}) {\em does} depend on it, and one must be careful to take this dependence into account.  The other relevant parameters of the problem, $c^{i\mu}$ and ${\cal K}_i$, can similarly be identified with the ones defined in Eqs.~(\ref{cimu}) and (\ref{KphiKsigma}) respectively with the change of indices $\phi\rightarrow 1$ and $\sigma\rightarrow 2$.
\subsection{Stability of the transverse modes}
\label{transversestability}
The propagation of transverse perturbations along the string takes place, independently of the string's internal structure, with a velocity~\cite{formal1,formal2,formal3,formal4} $\cT^2 =T/U$. Given that $U>0$, the stability of a string subjected to transverse perturbations requires $T>0$.   This is equivalent to a constraint on $x$, which can be obtained in the microscopic theory using Eqs~(\ref{T}), (\ref{B2C2a}) and (\ref{B2C2b}), and reads
\begin{equation}\label{xlimK}
 x\leq x_{\mrm{lim}}=\frac{1}{\sqrt{\mcl{K}_{\phi}\mcl{K}_{\sigma}}}\lsb A^2 
-\left(\frac{1}{2}w_\phi \mathcal{K_\phi}-\displaystyle\frac{1}{2}w_\sigma
\mathcal{K}_\sigma\right)^2\rsb^{1/2}.
\label{consttrans}
\end{equation}
This inequality is not automatically possible, for it demands $\displaystyle A^2 > \frac{1}{4}\left( w_\phi^2 {\cal K}_\phi^2 + w_\sigma^2 {\cal
    K}_\sigma^2\right)$, a condition which has no counterpart in the
single current case.
\subsection{Stability of the longitudinal modes}
\label{stablong}
The four longitudinal equations can be chosen as either the conservation of the four currents $c^{\mu}_{1,2}$ and $d^{\mu}_{1,2}$, or as the conservation of the two currents $c^{\mu}_{1,2}$ and of the longitudinal part of the stress energy tensor. Indeed, from Eq.~(\ref{redlongT}) we know that the conservation of the four currents implies the conservation of the longitudinal part of the stress energy tensor, but also that the converse is true, since
\begin{equation}
c_{2,1}^\mu \nablas_\nu {T_\mu}^\nu  = c_{2,1}^\mu
\varepsilon_{\mu \sigma} c_i ^\sigma\nablas_{\nu} d_i^\nu
= (-1)^{1,2} c_{2}^\mu \varepsilon_{\mu\sigma} c_{1}^\sigma
\nablas_{\nu} d_{1,2}^\nu.
\end{equation} 
Of course, because of the duality relations (\ref{tilde1}-\ref{tilde3}), we can symmetrically choose to keep the conservation of the two currents $d_{1,2 \mu}$ and of the longitudinal part of the stress energy tensor instead.\\

As in section \ref{longstab}, we work with the four conservation equations (\ref{eqc}) and (\ref{eqd}) together with the identity (\ref{c=f(d)}) and the equations of state (\ref{eqstc}).  The latter reduce to two equations only since the Lagrangian (\ref{lag2}) only depends on the two state parameters $w_{1,2}$. The corresponding dynamical equations perturbed to first order are then given by Eqs.~(\ref{eqlong},\ref{eqlong1}).
\subsubsection{General case}
\label{generalcase}
We first will consider the cases when both fields are condensed in the string and $x\not= 0$ or $w_1w_2\not=0$. The other cases, corresponding to having only one field condensed in the string and one field set to zero or having all four currents colinear and lightlike, will be studied separately.\\

In the case at hand, the conserved currents have both timelike and spacelike components in the basis $\lb u^{\mu},v^{\mu} \rb$ and it is therefore convenient to introduce the following two lightlike vectors
\begin{equation} e_{i\mu}^+=\frac{1}{2} \left[ d_{i\mu}-(-1)^i
    {\varepsilon_\mu}^\nu d_{i\nu} \right], 
\end{equation}
where the overall sign of ${\varepsilon_\mu}^\nu$ is chosen such that $e_{i\mu}^+=d_{i\mu}$ if $x\not= 0$ and $w_1w_2=0$. Another two lightlike vectors are chosen such that
\begin{equation}
\eta^{\mu\nu} e_{i\mu}^- e_{i\nu}^+=-\frac{1}{2}, \ \eta^{\mu\nu}
e_{i\mu}^- e_{i\nu}^-=0.
\end{equation}
When $w_1w_2\not= 0$, these vectors are given explicitly by
\begin{equation} e_{i\mu}^-=\frac{1}{2w_i} \left[ d_{i\mu}+(-1)^i
    {\varepsilon_\mu}^\nu d_{i\nu}\right].  
\end{equation}
These four lightlike vectors are obviously related since there are only two null directions in the longitudinal plane. In fact, one can show that
\begin{equation}
w_1 d_{2\mu} = x d_{1\mu}-\mathrm{sign}(x) \sqrt{x^2-w_1w_2}({\varepsilon_\mu}^\nu d_{1\nu}),
\label{rel1}
\end{equation}
and
\begin{equation}
w_2 d_{1\mu} = x d_{2\mu}+\mathrm{sign}(x)
\sqrt{x^2-w_1w_2}({\varepsilon_\mu}^\nu d_{2\nu}),
\label{rel2} 
\end{equation}
where the sign of $\varepsilon^{\mu\nu}$ was chosen appropriately when $w_1w_2\not= 0$, and the sign function was used.  Note that the term under the square root is always positive since, from Eq. (\ref{xbound}), when $w_1w_2\ge 0$, one must have $x^2\ge w_1w_2\ge 0$. When $x=0$ (and thus $w_1w_2\leq 0$), the sign of $x$ plays no role in these identities and can be fixed arbitrarily at $+1$.  From Eqs~(\ref{rel1}) and (\ref{rel2}) one can then derive the simple relation
\begin{equation}
e_{1\mu}^\pm = \lambda^{\pm 1} e_{2\mu}^\mp 
\label{e+-}
\end{equation}
with
\begin{equation}
\lambda = x+\mathrm{sign}(x) \sqrt{x^2-w_1w_2},
\label{lambdadef}
\end{equation}
Note again that $\lambda\not= 0$ in the case we are considering.  With these definitions, the currents can be rewritten as 
\begin{equation}
d_{i\mu}=e_{i\mu}^+ +w_i e_{i\mu}^-.
\end{equation}
Perturbations can be expanded in Fourier modes $\omega^\mu$ and are written as
\begin{equation}
\delta d_{i\mu}=(\delta d_i^+ e_{i\mu}^+ +\delta
  d_i^- e_{i\mu}^-)\ex^{i\omega_\mu x^\mu},
\end{equation}
with no summation on the repeated index $i$.  The variation of the state parameters is given by Eq.~(\ref{chid}) as
\begin{equation} 
\delta w_i=(\delta d_i^-+w_i \delta
  d_i^+)\ex^{i\omega_\mu x^\mu}.
\label{deltaw} 
\end{equation}
The other two currents $c^{i\mu}$ can be deduced from the identity (\ref{c=f(d)}) and
\begin{equation} 
  \frac{\delta c^{i\mu}}{{\cal
      K}^{ii}}=\varepsilon^{\mu\nu} \delta d_{i\nu}+\frac{\partial \ln
    ({\cal K}^{ii})}{\partial w_j} \delta w_j \varepsilon^{\mu\nu}
  d_{i\nu},
\end{equation}
with no summation on the repeated index $i$, while $\delta w_j$ given by Eq. (\ref{deltaw}). Note that since we assumed that two currents were condensed in the string, ${\cal K}^{ii}\not= 0$.\\

The equation of motion for the longitudinal modes are given by Eqs.~(\ref{eqlong}( and (\ref{eqlong1}), and yield a homogeneous linear system of equations,
\begin{widetext}
\begin{eqnarray}
&&X_1^+ \delta d_1^+ +X_1^- \delta d_1^-=0,\nonumber\\
&&X_2^+ \delta d_2^+ +X_2^- \delta d_2^-=0,\nonumber\\
&&\lsb X_1^++L_{1,1}(X_1^+-w_1 X_1^-)w_1\rsb \delta d_1^++
\lsb-X_1^-+L_{1,1}(X_1^+-w_1 X_1^-)\rsb\delta d_1^- +
L_{1,2}(X_1^+-w_1 X_1^-) (w_2\delta d_2^++\delta d_2^-)=0,\nonumber\\
&&L_{2,1}(X_2^+-w_2 X_2^-)(w_1 \delta d_1^+ +\delta d_1^-)+\lsb X_2^++
L_{2,2} (X_2^+-w_2 X_2^-)w_2\rsb\delta d_2^++
\lsb -X_2^-+L_{2,2}(X_2^+-w_2 X_2^-)
\rsb\delta d_2^-=0,\nonumber\\
&&
\label{long1}
\end{eqnarray}
\end{widetext}
where
\begin{equation}
X_i^\pm=\eta^{\mu\nu}\omega_\mu {e_i^\pm}_\nu,\,\,\,\,\, L_{i,j}=
\frac{\partial \ln ({\cal K}^{ii})}{\partial w_j}.
\label{long4}
\end{equation}
The dispersion equation is obtained by equating the determinant to zero and is a homogeneous fourth degree polynomial in the two components of $\omega_{\mu}$ which does not appear to have any obvious solutions. It can be written down in a compact form as
\begin{equation}
\sum_{i=0}^4 c_i (X_2^-)^{4-i}(X_1^-)^i=0.
\label{polychar}
\end{equation}
Noting that $X_{1,2}^+=\lambda X_{2,1}^-$ from Eq. (\ref{e+-}), the coefficients $c_i$ are found to be
\begin{widetext}
\begin{eqnarray}
c_0 & = & D\lambda^2 w_2^2 \\
c_1 & = & -2\lambda (Dw_1 w_2^2+Dw_2\lambda^2+\lambda^2 L_{1,1}+w_2^2 L_{2,2})\\
c_2 & = & D(w_1^2w_2^2+4\lambda^2w_1w_2+\lambda^4)+4\lambda^2(1+L_{1,1}w_1+L_{2,2}w_2)\\
c_3 & = & -2\lambda \left(Dw_1^2 w_2+Dw_1\lambda^2+w_1^2 L_{1,1}+\lambda^2 L_{2,2}\right)\\
c_4 & = & D\lambda^2 w_1^2, 
\end{eqnarray}
\end{widetext}
\noindent where $D$ is the determinant of the matrix $L_{i,j}$ and $\lambda$ was defined in Eq. (\ref{lambdadef}).  We define
\begin{equation} 
  \xi=\frac{X_1^{-}}{X_2^{-}}=\frac{-\zeta
    e_{1t}^-+e_{1z}^-}{-\zeta e_{2t}^-+e_{2z}^-} 
\end{equation}
\noindent where $\zeta=\omega_t/\omega_z$ is the dispersion relation.  The string will remain stable under perturbations in the range of parameter space in which $\zeta$ is a real.  Since the $e_{i\mu}^{\pm}$ are real, it suffices that $\xi$ be real.  In terms of $\xi$, Eq.~(\ref{polychar}) reads
\begin{equation}
c_0+c_1\xi+c_2\xi^2+c_3\xi^3+c_4\xi^4=0.
\label{fourthorderpoly}
\end{equation}
Algebraic solutions to this equation are easily obtained using the well-known Ferrari procedure.  Only $L_{1,1}$ and $L_{2,2}$, which depend {\it implicitly} on the $w_i$'s, must be determined numerically from the integrated quantities of Section~\ref{macro}.  In order to determine the regions of parameter space in which the string is indeed stable, one must therefore input the values of $L_{1,1}$ and $L_{2,2}$ obtained numerically in the analytic forms of the roots of (\ref{fourthorderpoly}) obtained with the Ferrari procedure.  We do so in the range $w_{\mrm{min}}\leq w_i\leq w_{\mrm{max}}$ and $-x_{\mrm{lim}}\leq x \leq x_{\mrm{lim}}$ where the upper and lower limits on $w_i$ are given by Eqs~(\ref{w_upper}) and (\ref{w_lower}) and the limits on $x$ are given by the condition for stability of the transverse modes.  Figs~\ref{fig:stability1}, \ref{fig:stability2}, \ref{fig:stability3} and \ref{fig:stability4} provide a comparison of the regions of stability obtained using the constraint $x_{\mrm{lim}}$ and the solutions of Eq.~(\ref{fourthorderpoly}) $\in \Reals$ for the cases of zero, weak and moderate coupling $\tilde{g}$.  The region of stability in the $x$-direction is given by the conditions for stability of the transverse modes while that in the $\tilde{w}_i$ ($i=1,2$) directions is provided by requiring that the roots of Eq.~(\ref{fourthorderpoly}) be real.  As shown in the figures, stability can be enhanced in the $x$ direction as one goes to stronger coupling, but it is always significantly reduced in the $\tilde{w}_1$ and $\tilde{w}_2$ directions.  This is especially true when one of the currents is spacelike: in Figs~\ref{fig:stability2} and \ref{fig:stability3}, the string is entirely unstable for a coupling greater or equal to $\tilde{g}_3$.  Note that the region of stability of the analytic model of Section \ref{simplifiedmodel} is also shown in these figures.  It is discussed at the end of the next section.  Finally, we point out once again that the stability conditions obtained by solving Eq.~(\ref{fourthorderpoly}) is not valid when one or both currents are lightlike, \ie, for $\tilde{w}_1\tilde{w}_2=0$. The stability in these cases is discussed in the next subsections.
\subsubsection{Special case of a static condensate}
This case is characterized by $x=0$ and either $w_1=0$ or $w_2=0$. It cannot be included in the general case because $\lambda=0$ and thus, from Eq.~(\ref{e+-}), $e_{i\mu}^+=0$.  Let us call generically $i$ the index for which $w_i=0$, and $j$ the other index. This case corresponds to $\omega_i=k_i=0$ in Eqs~(\ref{ansatzPhi}) and (\ref{ansatzSigma}), and thus to a static condensate. As a consequence, it is clear that $d_{i\mu}=0$. Using Eq.~(\ref{eqlong})  one finds that the perturbation equations for the two currents decouple. The two characteristic modes $\omega_\mu$ for the static condensate are simply given by the two light-like directions in the longitudinal plane. Thus, these perturbations are always stable.  The stability condition for the other current reduces to
\begin{equation} 
1+2w_j L_{j,j} \geq 0. 
\label{cL2} 
\end{equation}
It is interesting to note that the characteristic polynomial (\ref{polychar}) found in the previous general case has two obvious real solutions corresponding to the two lightlike directions $X_1^-=0$ and $X_1^+=0$, and two other solutions which are real exactly when the condition (\ref{cL2}) is satisfied. Thus, the general stability condition from the first case extends also to this limiting case.
\subsubsection{Special case of only one current condensate}
In this case, the effective Lagrangian (\ref{lag2}) depends only on one of the state parameters.  There are only two longitudinal degrees of freedom and the equations derived in Subsection 1 of Section~\ref{stablong} do not apply.  Calling generically $i$ the index of the remaining condensate, only $w_i$, ${\cal K}^{ii}$ and $L_{i,i}$ are nonzero. The perturbation equations derived from Eqs.~(\ref{eqlong}) are exactly the same as the ones derived for the decoupled nonstatic condensate in the previous case. Thus the stability condition is also given by the inequality (\ref{cL2}).\\

Note that this case has been studied in detail in previous works~\cite{formal1,formal2,formal3,formal4} and it is in fact well known that the speed of propagation of longitudinal perturbations is given by
\begin{equation}
\cL^2=-\frac{\dd T}{\dd U}=(1+2w_iL_{i,i})^{\mbox{\scriptsize sign}(w_i)},
\label{cL2d}
\end{equation}
The stability condition is simply $\dd T/\dd U<0$ or equivalently that $1+2w_i L_{i,i}\geq 0$.
\subsubsection{Special case of two static condensates}
This occurs when $x=w_1=w_2=0$. In this case, there is no current in the string, from (\ref{B2C2K}), $B^2-C^2=0$, thus from Eqs~(\ref{U}) and (\ref{T}) $U=T$, and the stress energy tensor is Lorentz invariant.\\

This limit corresponds effectively to a currentless Nambu-Goto string which is stable and although it is an interesting model in it own right, it is {\it not} an elastic string model.
\section{Stability of the simplified model}
\label{simplifiedmodel}
\begin{figure*}
\includegraphics[width=6.5cm]{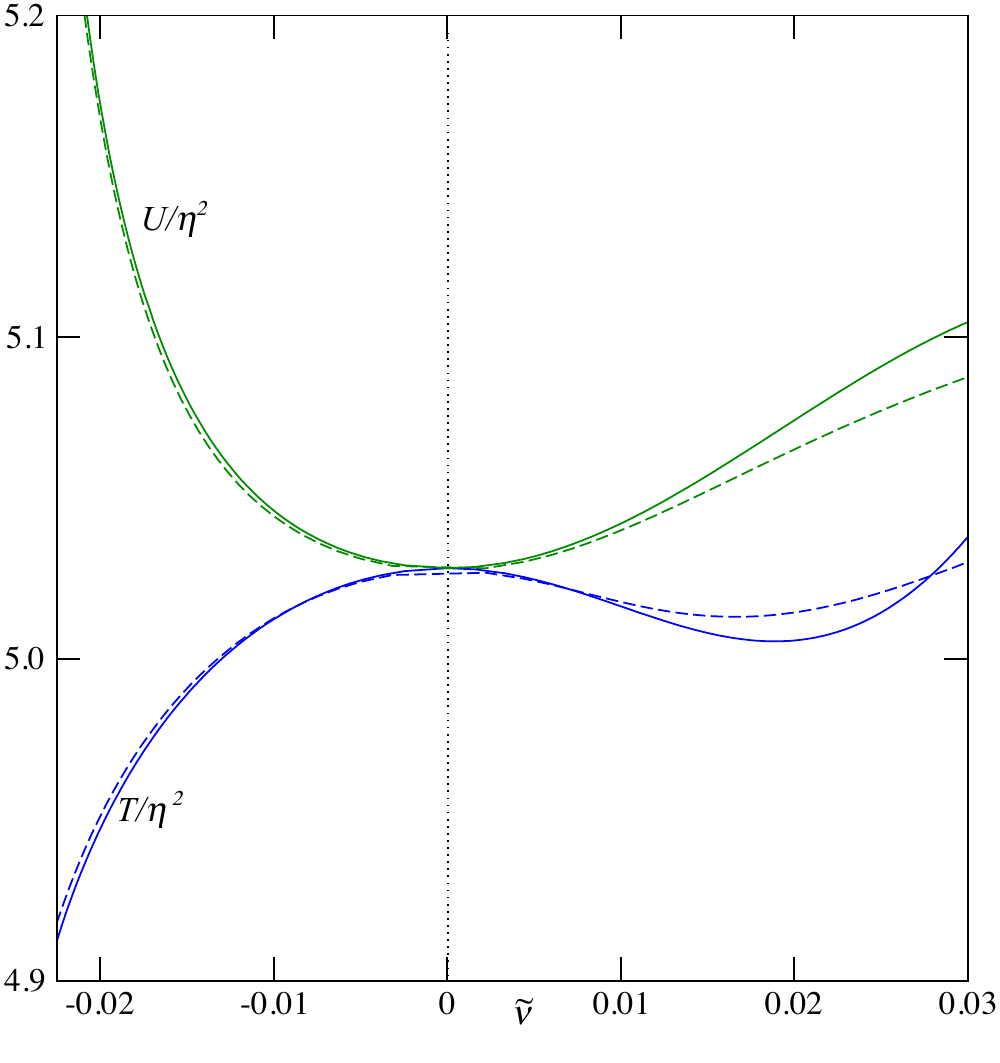}
\includegraphics[width=6.5cm]{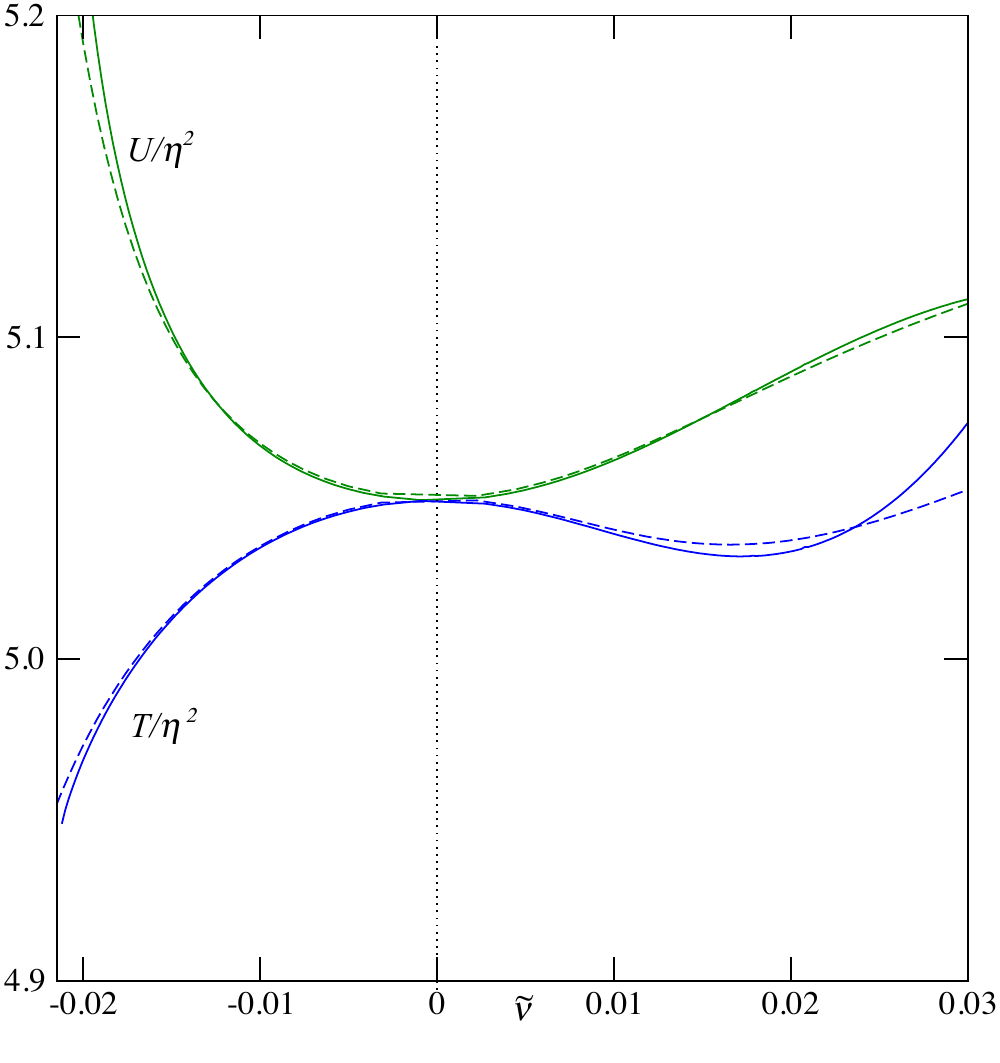}
\includegraphics[width=7.5cm]{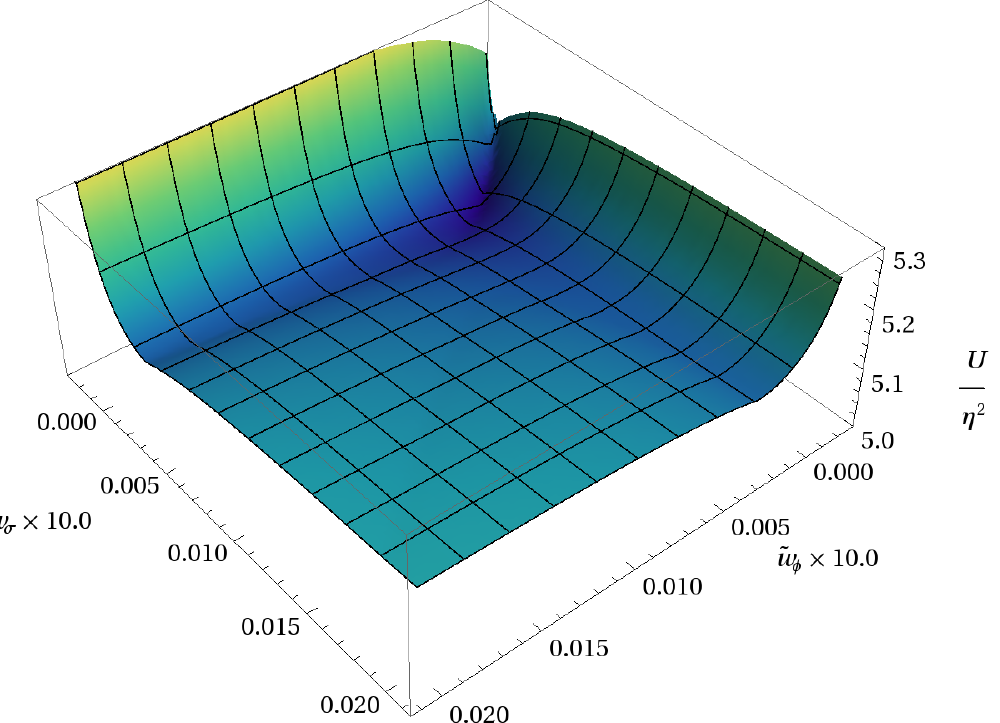}
\hspace{0.5cm}
\includegraphics[width=7.5cm]{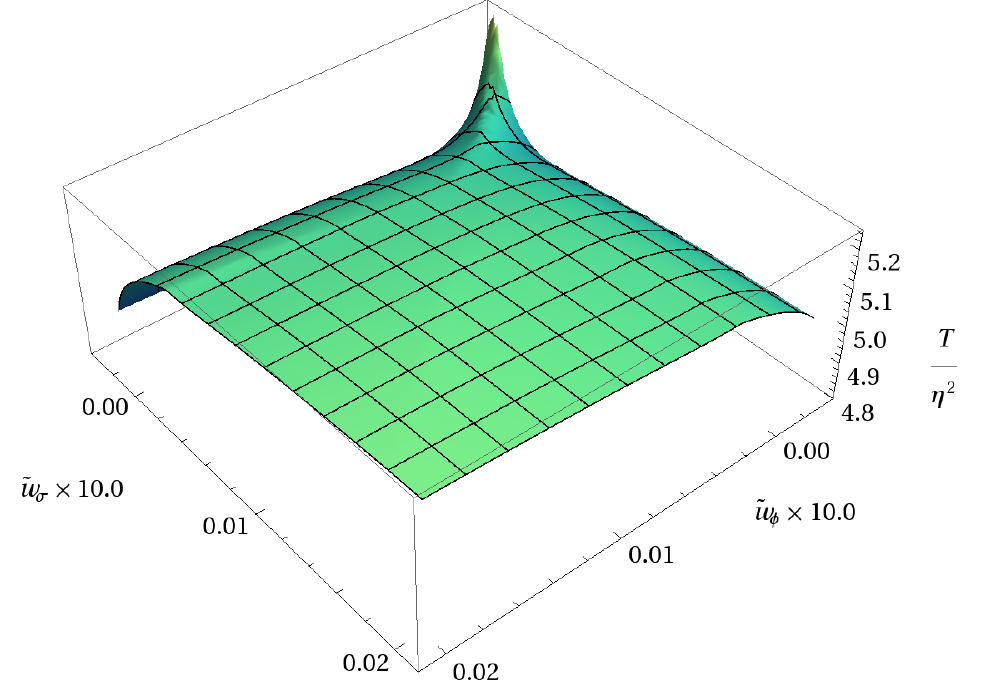}
\caption{(Top) Analytic models {\it vs}.~numerically determined $U$ and $T$ for a string with one current. The microscopic parameters used to compute $U$ and $T$ numerically from the complete interacting case (solid lines) are those used for $\Phi$ (left) and $\Sigma$  (right) throughout the rest of this work.  The mass parameters $m$ and $m_i$ (see Eqs.~(\ref{magneticmodel}) and (\ref{electricmodel})) used to compute $U$ and $T$ in the analytic model (dashed lines) were adjusted to match the normalization of $U$ and $T$ obtained numerically. (Bottom) Analytic computation of $U$ (left) and $T$ (right) for a string with two currents using the mass parameters $m$ as in the single current case shown in the upper part of the figure.}
\label{fig:modelfit}
\end{figure*}

In this section, we test the possibility that while both fields are condensed in the string, they can be assumed (in some approximation) to be independent of each other.  In this case the Lagrangian reads
\begin{equation}
\mathcal{L}=\mathcal{L}_1(\bm{\chi}_1)+\mathcal{L}_2(\bm{\chi}_2)+m^2,
\label{redLag}
\end{equation}
\noindent where $\mathcal{L}_i(\bm{\chi}_i)$ is an effective Lagrangian which can be identified with that of a string with a single condensate.  We include an additional $m^2$ term in order to compensate the bare Goto-Nambu energy contribution which, as we shall see, comes in twice, \ie, once for each of the Lagrangians $\mathcal{L}_1$ and $\mathcal{L}_2$.  As shown in Refs~\cite{neutral,models}, the energy $U_i$ and tension $T_i$ as a function of the state parameter $w_i$ of a string with a single current are accurately reproduced by a Lagrangian of the following form
  \begin{equation}
\mathcal{L}_i(\bm{\chi}_i) =  -m^2-\frac{w_i}{2}
\left( 1+\frac{w_i}{m_i^2}\right)^{-1} \qquad \mrm{if} \qquad w_i>0,\\
\label{magneticmodel}
\end{equation}
\ie, in the magnetic case, and 
\begin{equation}
\mathcal{L}_i(\bm{\chi}_i) = -m^2-\frac{m_i^2}{2} 
\ln\left( 1+\frac{w_i}{m_i^2}\right)\qquad \mrm{if} \qquad w_i<0,
\label{electricmodel}
\end{equation} 
\noindent \ie, in the electric case.  In Eqs.~(\ref{magneticmodel}) and (\ref{electricmodel}), $m$ and $m_i$ are adjustable mass parameters of the order of the Higgs mass $m_h=\sqrt{\lambda}\eta$ and of the order of the condensate mass parameters $m_{\phi}$ and $m_{\sigma}$ respectively.  The energy per unit length $U_i$ and tension $T_i$ are given by 
\cite{formal1,formal2,formal3,formal4}
\begin{equation}
T_i=-\mcl{L}_i+w_iK^i\qquad \mrm{and}\qquad U_i=-\mcl{L}
\qquad \mrm{if}\qquad w_i>0
\end{equation}
and
\begin{equation}
U_i=-\mcl{L}_i+w_iK^i\qquad \mrm{and}\qquad T_i=-\mcl{L}\qquad
\mrm{if}\qquad w_i<0.  
\end{equation}
 A comparison between the numerical integration for a string with a single condensate and the analytic model is provided in Fig.~\ref{fig:modelfit} and shows remarkable agreement in most of the stable range given by Eq.~(\ref{cL2}) and the condition that $T>0$.
In practice, for the simplified model of Eq.~(\ref{redLag}), \ie, in the situation where more than one field condensates onto the string,  the mass parameters are shifted from the mass parameters one would get for a string with a single condensate. We will assume in this section that for the two fields to condensate, a minimum condition is that the associated currents be dynamically stable separately, \ie, $c_{i _\mathrm{T,L}}^2\geq 0$ for each \cite{formal1,formal2,formal3,formal4}. Now, assuming that both fields condensate in the string, one can consider the dynamical stability of this simplified string model against perturbations in the associated currents.  Because of the decoupled form of the Lagrangian (\ref{redLag}), it is clear that the conserved currents $c_i^\mu$ and $d_i^\mu$ depend only on the $\mathcal{L}_i(\bm{\chi}_i)$. This means that the longitudinal perturbation equations decouple and  reduce to twice (once for each current) the usual longitudinal equations for a string with only one current.  Thus, the stability condition is just $c_{i_\mathrm{L}}^2\geq 0$, a condition which we took to be necessary to have two current condensates.  This result can be verified from Eqs.~(\ref{long1}-\ref{long4}) by noting that for the Lagrangian of Eq.~(\ref{redLag}), $L_{1,2}=L_{2,1}=0$, and $c_{i_\mathrm{L}}^2$ is given by Eq.~(\ref{cL2d}).\\

The stability of the string against transverse perturbations is simply given by (\ref{transcons}).  The stress energy tensor associated with Eq.~(\ref{redLag}) is the sum of the stress energy tensors of each component
  \begin{equation}
T^{\mu\nu}=T_1^{\mu\nu}+T_2^{\mu\nu}+m^2\eta^{\mu\nu}, \
\label{Tred}
\end{equation}
\noindent where $T_i^{\mu\nu}$ are the stress energy tensors for a string with a single condensate,
 \begin{equation} 
T_i^{\mu\nu}=U_i u_i^\mu u_i^\nu -T_i v_i^\mu  v_i^\nu, 
\end{equation}
\noindent in which $U_i$ and $T_i$ are the energy density and tension in the string.\\

The energy density and tension of the simplified model for the two condensate case are shown in the lower part of Fig.~\ref{fig:modelfit} and are compared to the numerical results obtained for the interacting model in Figs.~\ref{fig:sectionsY} and \ref{fig:sectionsZ} for the three values of the coupling $\tilde{g}$ considered in this work.  In all three cases, the effective Higgs mass parameter $m$ of the model (dashed lines) was adjusted so as to best fit the numerical results obtained for the full interacting model.  On the other hand, better agreement with the numerical results of the interacting model was found by keeping the effective scalar field mass parameters $m_i$ equal to the best fit value obtained in the single condensate case.  As expected, agreement of the model with the numerical results is very satisfactory for $\tilde{g}_1$ and $\tilde{g}_2$ but is unsatisfactory for $\tilde{g}_3$.  This serves to confirm that the model can only be used at weak coupling. We further note that (somewhat co\"incidentally) the model appears to better agree with the weakly coupled case (\ie, for $\tilde{g}_2$) than with the uncoupled case (\ie, for $\tilde{g}_1$).  Finally, we stress that the simplified model and the fully interacting model but with $\tilde{g}_1=0$ should not be taken to be two descriptions of precisely the same physics since in the latter, there is an interplay between the two scalar fields through their couplings to the Higgs and associated $U(1)^{\mrm{local}}$ gauge field.  As a result, one should always expects a discrepency between the two.\\

We now turn to a more detailed discussion of the stability of transverse and longitudinal modes in the simplified model.  The vectors $u_i^\mu$ and $v_i^\mu$ are respectively timelike and spacelike unit vectors associated with current $i$.  In order to diagonalize $T^{\mu\nu}$, we write the stress energy tensor (\ref{Tred}) in the basis $(u_1^\mu, v_1^\mu)$. The eigenvalues $U_i$ and $T_i$ are related by the equations of state in the usual way~\cite{models} of a one condensate string, and the other eigenvectors $(u_2^\mu,v_2^\mu)$ can be determined from $(u_1^\mu,v_1^\mu)$ and $x$ with Eqs (\ref{rel1}) and (\ref{rel2}).  As stated before, the stability of the transverse modes is given by $T>0$ and this constraint on $T$ can be used to obtain a constraint on the range of $x$.  This is what we do in the remainder of this work.  The longitudinal modes decouple and their stability, as in the single current case, is given by $1+2\omega_iL_{i,i}\geq 0$.  For the models of Eq.~(\ref{magneticmodel}) and Eq.~(\ref{electricmodel}) this reduces to $-\gamma\leq\tilde{w}\leq \gamma/3$ (or $-m_i^2\leq w_i<m_i^2/3$) in both possible cases, \ie, $w_1w_2\ge 0$ and $w_1w_2\le 0$.
\subsubsection{Case I: $w_1w_2\ge 0$}
\label{2tlc}
In this case, $|x|\geq\sqrt{w_1w_2}$, and $d_i^\mu=\sqrt{-w_i}u_i^\mu$ for timelike currents and $d_i^{\mu}=\sqrt{w_i} v_i$ for spacelike currents. Using Eq.~(\ref{rel1}), one has
\begin{eqnarray}
u_2^\mu & = & \frac{1}{\sqrt{w_1w_2}}\lb xu_1^\mu+
\mrm{sign}\lb x\rb \sqrt{x^2-w_1w_2}v_1^\mu \rb, \cr
&&\label{u2}\\
v_2^\mu & = & \frac{1}{ \sqrt{w_1w_2}}\lb\mrm{sign}\lb
x\rb\sqrt{x^2-w_1w_2} u_1^\mu+x v_1^\mu \rb.\cr
&&\label{v2} 
\end{eqnarray}
\noindent Using Eq.~(\ref{Tred}), the stress energy tensor in the basis $(u_1^\mu,v_1^\mu)$ is given by the sum $T^{\mu\nu}+m^2 \eta^{\mu\nu}$ where
\begin{widetext}
\begin{equation}
T^{\mu\nu}=\left(
 \begin{array}{cc}\displaystyle U_1+\frac{x^2U_2-\lb x^2-w_1w_2\rb T_2}{w_1w_2} 
& \displaystyle\lb U_2-T_2\rb\frac{|x|\sqrt{x^2-w_1w_2}}{w_1w_2}\\
    \displaystyle\lb U_2-T_2\rb \frac{|x|\sqrt{x^2-w_1w_2}}{w_1w_2}
    &\displaystyle -\left[ T_1+\frac{x^2T_2-\lb x^2-w_1w_2\rb
      U_2}{w_1w_2}\right]
\end{array}
\right)
\label{Tdev}.
\end{equation}
In order to address the first stability condition, $T\geq 0$), we compute $U$ and $T$. They are simply given by the eigenvalues of the stress energy tensor.  We therefore solve
\begin{equation}
\det\lsb T^{\mu\nu}+\lb m^2-\lambda\rb \eta^{\mu\nu}\rsb 
= X^2+(U_1+U_2+T_1+T_2)X+ \lsb\lb U_1+U_2\rb \lb T_1+T_2\rb-
\frac{y^2}{w_1w_2}\lb U_1-T_1\rb \lb U_2-T_2\rb\rsb=0,
\label{eigeq1}
\end{equation} 
\end{widetext}
where $X=\lambda-m^2$ and $y^2=x^2-w_1w_2\geq 0$.  In order to satisfy $T\ge 0$, one finds that
\begin{equation}
y^2\leq w_1w_2\frac{(U_1+U_2-m^2)(T_1+T_2-m^2)}{\lb U_1-T_1 \rb \lb U_2-T_2 \rb}.
\label{limy2A}
\end{equation}
\subsubsection{Case II: $w_1w_2\le 0$} 
In this case, and without loss of generality, we take $d_1^\mu=\sqrt{-w_1}u_1^\mu$ and $d_2^\mu=\sqrt{w_2}v_2^\mu$.  We then have
\begin{eqnarray}
u_2^\mu & = & \frac{1}{\sqrt{-w_1w_2}}\lb \mrm{sign}\lb x\rb
\sqrt{x^2-w_1w_2} u_1^\mu+x v_1^\mu \rb, \cr
&&\\
v_2^\mu & = & \frac{1}{\sqrt{-w_1w_2}}\lb xu_1^\mu+
\mrm{sign}\lb x\rb\sqrt{x^2-w_1w_2}v_1^\mu \rb.\cr
&&
\end{eqnarray}
Following the procedure outlined in the preceding section, we find
\begin{equation}
y^2\leq |w_1w_2|\frac{\lb U_1+T_2-m^2\rb \lb T_1+U_2-m^2\rb}
{\lb U_1-T_1\rb \lb U_2-T_2\rb}.
\label{limy2B}
\end{equation}
Alternatively, from Eq.~(\ref{limy2A}) and Eq.~(\ref{limy2B}), stable configurations can be expressed as a limit on $x$.  For $w_1w_2\geq 0$,
\begin{equation} x^2\leq
  x_\mathrm{lim}=\frac{w_1w_2(U_1-m^2+T_2)(U_2-m^2+T_1)}{\lb U_1-T_1
    \rb \lb U_2-T_2\rb},
\label{cond2b1}
\end{equation}
and for $w_1w_2\leq 0$,
\begin{equation}
x^2\leq x_\mathrm{lim}=\frac{|w_1w_2|\lb T_1+T_2-m^2\rb\lb U_1+U_2-
m^2\rb}{\lb U_1-T_1 \rb \lb U_2-T_2\rb}.
\label{cond2b2}
\end{equation}
\begin{figure*}
\includegraphics[width=12cm]{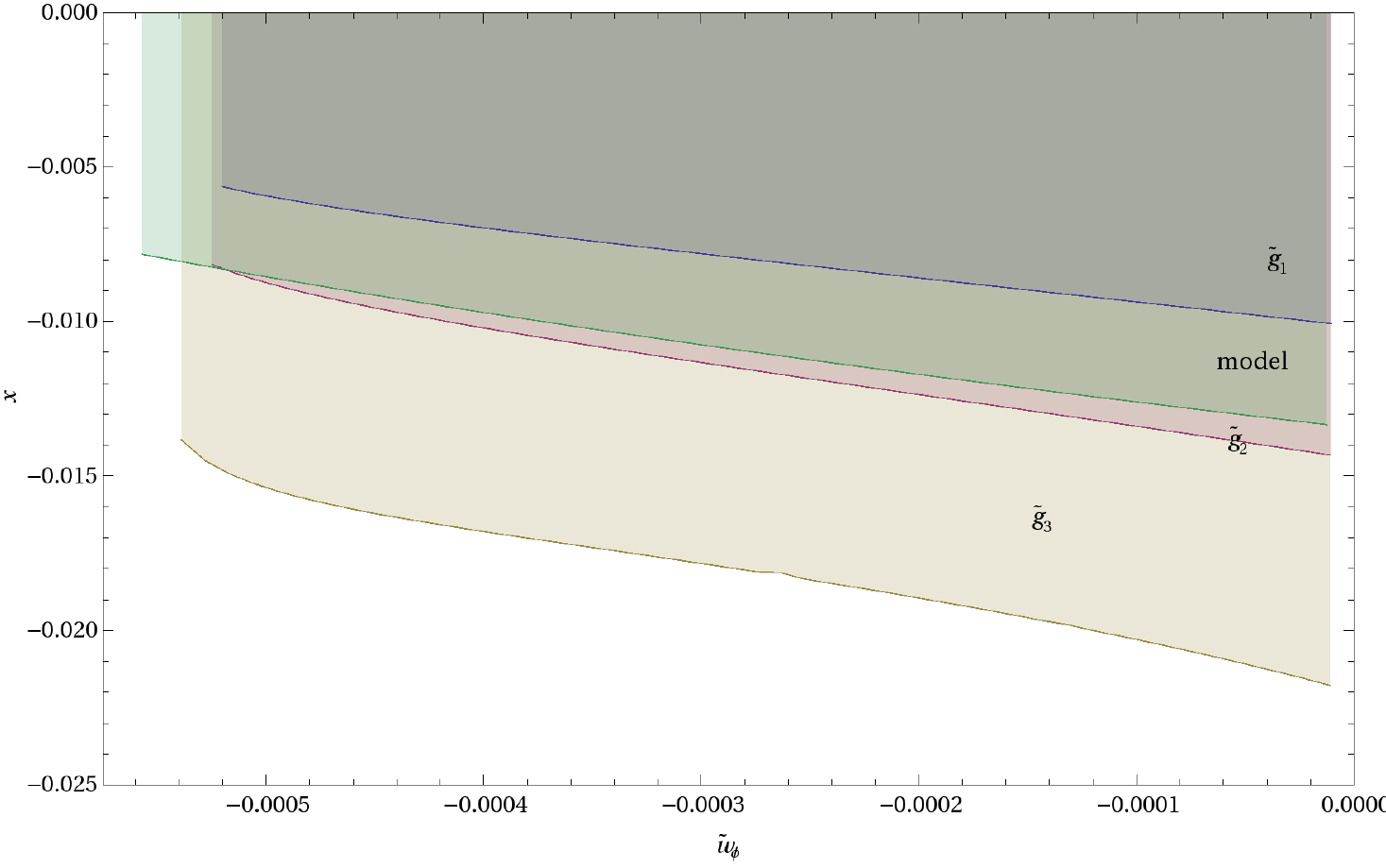}
\caption{Region of stability for $\tilde{w}_{\sigma}\sim -\gamma_{\sigma}/2$ and $\sigma_{\phi}<0$ (\ie, for timelike currents) for all three values of $\tilde{g}$ and compared with the region of stability obtained with the analytic model.}
\label{fig:stability1}
\end{figure*}
\begin{figure*}
\includegraphics[width=12cm]{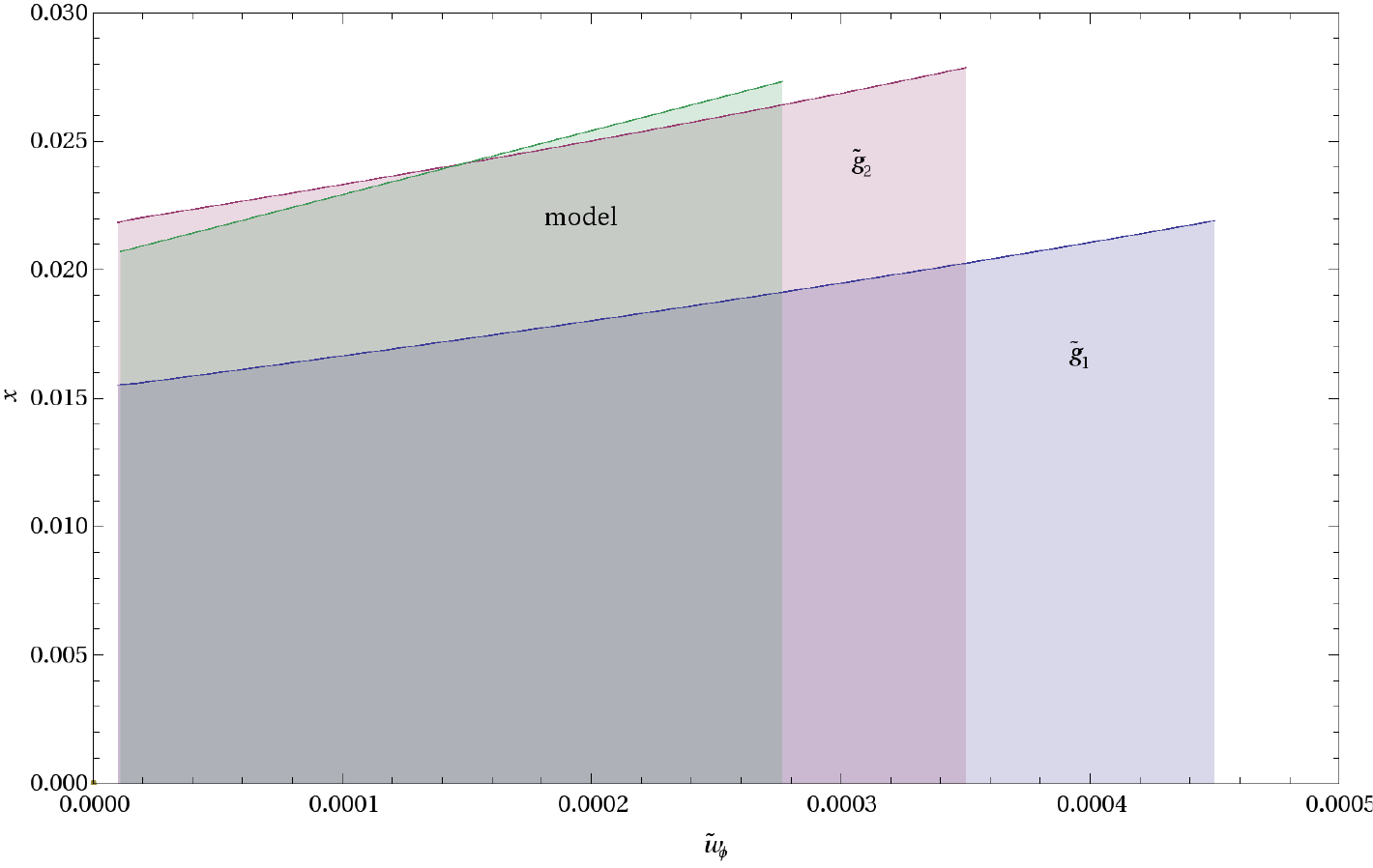}
\caption{Same as Fig.~\ref{fig:stability1} but for $\tilde{w}_{\sigma}\sim \gamma_{\sigma}/2$ and $\sigma_{\phi}>0$ (\ie, for spacelike currents).  Note that in the case $\tilde{g}_3$, the string is unstable in this range of parameters.}
\label{fig:stability2}
\end{figure*}
\begin{figure*}
\includegraphics[width=12cm]{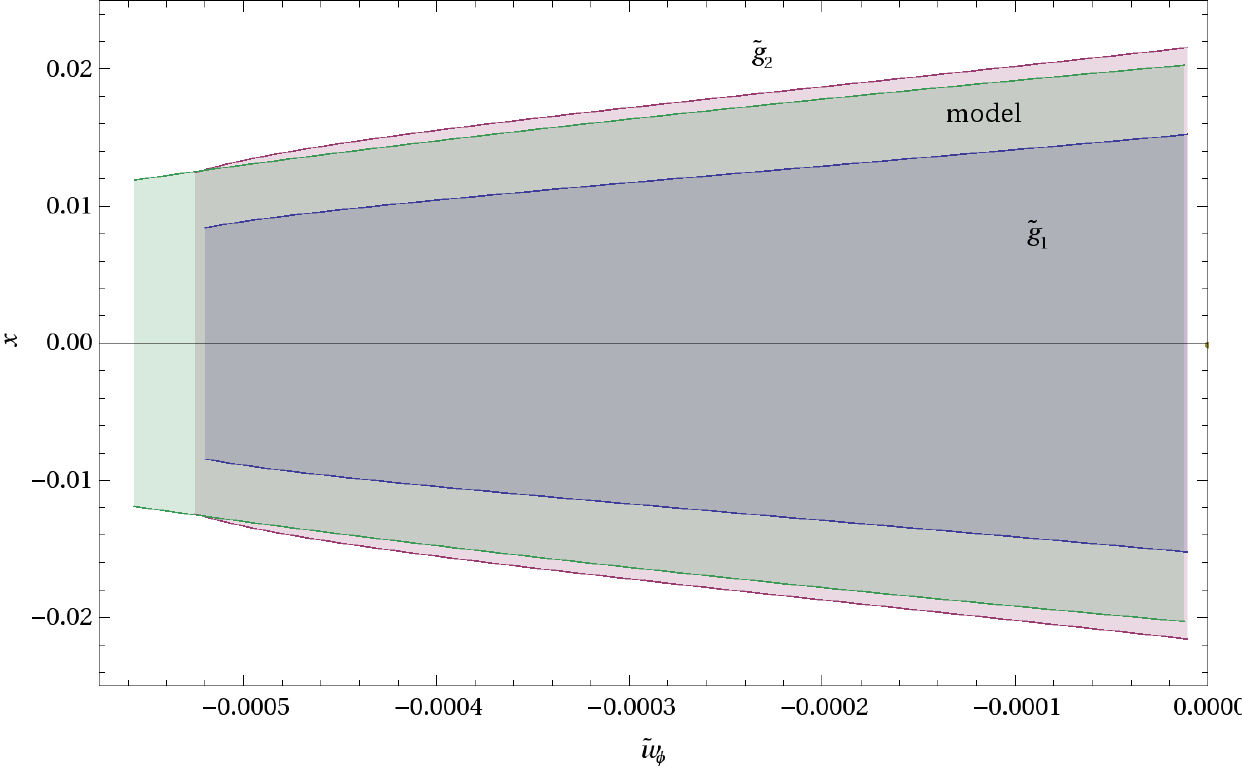}
\caption{Same as Fig.~\ref{fig:stability1} but for $\tilde{w}_{\sigma}\sim \gamma_{\sigma}/2$ and $\sigma_{\phi}<0$ (\ie, for one timelike and one spacelike current). As in the previous figure, the case of $\tilde{g}_3$ is unstable in this parameter range.}
\label{fig:stability3}
\end{figure*}
\begin{figure*}
\includegraphics[width=12cm]{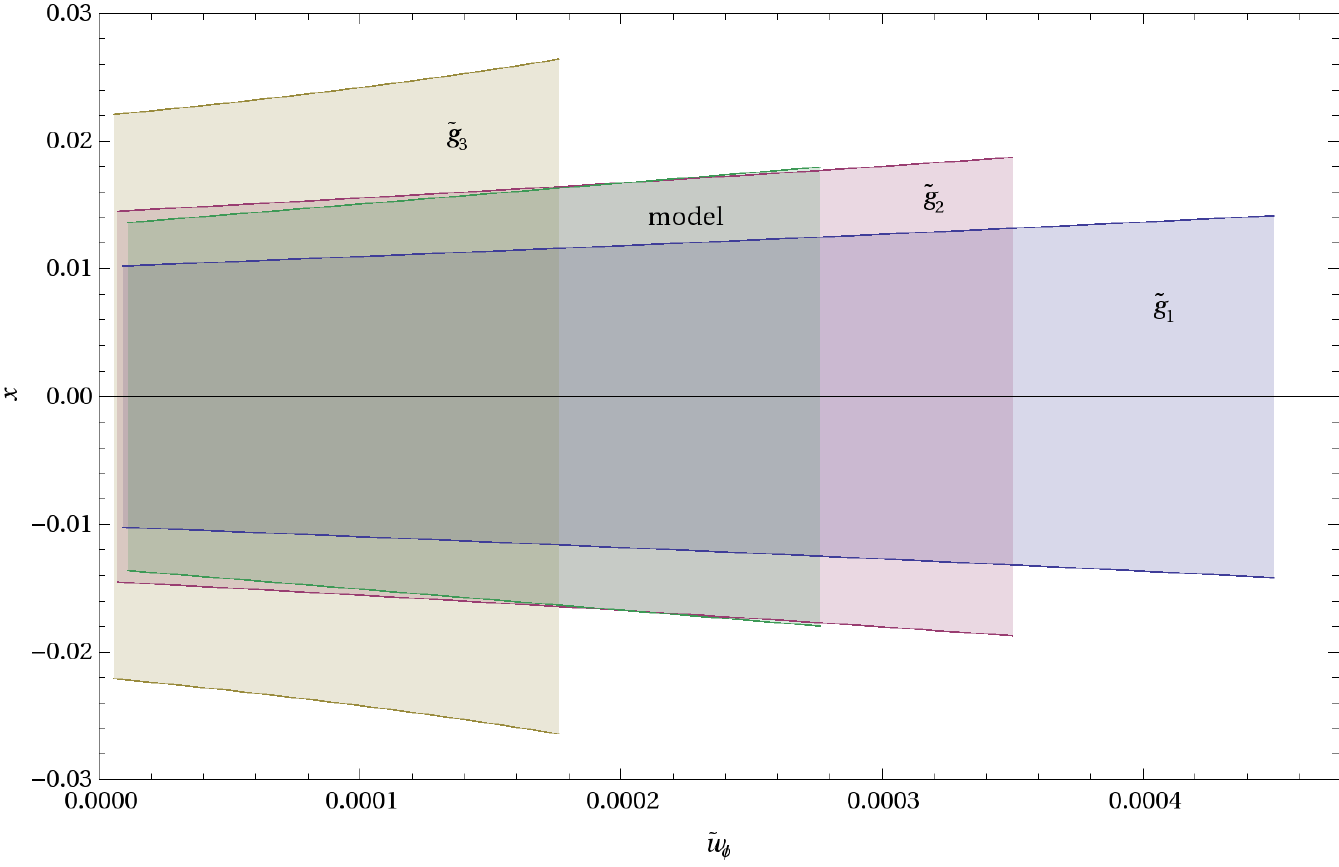}
\caption{Same as Fig.~\ref{fig:stability1} but for $\tilde{w}_{\sigma}\sim -\gamma_{\sigma}/2$ and $\sigma_{\phi}>0$ (\ie, roles reversed with respect to the previous figure).}
\label{fig:stability4}
\end{figure*}
Figs.~\ref{fig:stability1}, \ref{fig:stability2}, \ref{fig:stability3} and \ref{fig:stability4} provide a comparison of the regions of stability of the fully interacting model obtained using the constraint $x_{\mrm{lim}}$ given by Eq.~(\ref{consttrans}) and the solutions of Eq.~(\ref{fourthorderpoly}) $\in \Reals$ with the region of stability obtained for the simplified model using the constraints given by Eqs~(\ref{cond2b1}) and (\ref{cond2b2}) for $\tilde{g}_1$, $\tilde{g}_2$ and $\tilde{g}_3$.  In the simplified model, both $T_i$ and $U_i$ are proportional to $m^2$.  As a result, $x_{\mrm{lim}}$ is independent of $m$.  It is interesting to note that, at least in principle, the coupling $g$ indirectly plays a role in the physics of the decoupled model if the best fit $m_i$'s differ in going from $\tilde{g}_1$ to $\tilde{g}_3$.  If, however, the $m_i$'s of the analytic model are kept the same for all values of $\tilde{g}$ (which is the case here), the stability of the simplified model will be strictly the same independently of the value of $g$.  In this case, the simplified model will reproduce the stability of the interacting full theory only for weak coupling.  This is indeed the conclusion that can be drawn from the figures, where the regions of stability computed using simplified model approach those of the fully intereacting model.
\section{Conclusions}
In the Witten bosonic neutral superconducting string model, the conserved current is given by the phase gradient of a single $U(1)^{\mrm{global}}$ scalar field condensate.  At the field theory level and also in the integrated macroscopic worldsheet formalism~\cite{formal1,formal2,formal3,formal4}, such a string can be described by a single state parameter given by the square of the phase gradient of the neutral field once the microscopic mass parameters and couplings are fixed.  In this work, we extended both the existing microscopic and macroscopic descriptions of the neutral string to the case of a string carrying several of these currents.\\

In the first part of this paper, we described how to generalize the single-current microscopic structure of a string to the two-current case (\ie, to the case of a string in the presence of two $U(1)^{\mrm{global}}$ fields) and calculated the integrated quantities needed to describe the macroscopic evolution of a string worldsheet endowed with two currents, \ie, the energy per unit length, the tension, and the total currents.  We pointed out that at the field theory level, only the usual state parameters given by the square of the phase gradients are needed to fully describe the dynamics.  However, and in contrast with the other known case of a string carrying two currents~\cite{2c}, in addition to the two Lorentz-invariant state parameters required in the microscopic description, the scalar product of the phase gradients of the different fields is needed in order to describe the energy per unit length, tension and integrated currents relevant in the worldsheet description.  This difference can be explained by noting that the field equations cannot depend on this third state parameter, denoted $x$, because $x$ represents the (Lorentz) angle between the worldsheet currents, and this could only have a dynamical effect on the microstructure if there existed an interaction between the scalar field phases. Such an interaction is excluded by the assumed symmetry. On the other hand, the macroscopic properties of the string needed to describe the worldsheet dynamics are non-local quantities and therefore can and do depend on the parameter $x$.

In this first part of the paper, we also solved numerically the field equations using a relaxation method and obtained the parameter range in which both fields condensate onto the string.  As could be expected, when the $U(1)^{\mrm{global}}$ fields are strongly coupled at the microscopic level, the region of parameter space in which both fields condense onto the string is greatly reduced.  In addition, we computed the energy per unit length, tension and the integrated current as a function of the first two state parameters.\\

It was found that, at the microscopic level, the results obtained in the single current case extend to this more complicated case.  In particular, the phase frequency threshold~\cite{neutral} that appears for a timelike current for one condensate exists here as well.  As a result, in the region of the parameter space in which both currents play a significant part in the string physics, divergences are observed in the energy per unit length, tension and integrated currents.  In particular, the divergences observed as one (but not the other) state parameter reaches its phase frequency threshold are similar to the  ones observed in the single condensate string.  However, a divergence of the tension $T$ to {\it positive} values is observed when the phase frequency threshold of {\it both} currents is approached.  This feature does not exist in the single condensate string: as the phase frequency threshold in the single condensate string is approached, the tension diverges {\it negatively}.  This feature enlarges the range of validity of the no-spring conjecture~\cite{NoSpring,ProcNoSpring}; it would affect the electromagnetically supported loop configuration~\cite{emLoops} and their long-range gravitationnal properties~\cite{GravStringPuyPP}. Finally, exact relations describing the dependence of the various quantities of interest on the third state parameter were obtained.  This was made possible only because, as already mentioned, the parameter $x$ is absent at the microscopic level.\\

In the second part of this paper, we presented a general extension of Carter's worldsheet formalism to describe the worldsheet dynamics of a string endowed with $N$ condensates.  We derived the conservation equations associated with the $2N$ string currents and obtained a duality relation identical to the one that exists in the one condensate string.  We then worked out the stability conditions of an $N$-condensate string.  At first order in perturbations, the criterion for stability of transverse modes of a string carrying $N$ currents is identical to that of a string carrying a single current: the propagation speed of perturbations must be positive.  It reduces to the same requirement as in the single condensate string, namely that the string tension be positive.  This conclusion can be reached by inspection, simply by substituting in the extrinsic equations of motion of the string the ansatz for a perturbation.  At first order, perturbations in the longitudinal modes do not couple.  The string will therefore be stable if the propagation speed of these modes is positive.  However, for a general $N$-condensate string, this condition does not reduce to $\mrm{d}T/\mrm{d}U<0$ as in the single condensate string.  Instead, one obtains a set of coupled equations for the perturbations of the $N$ currents (a coupling of the {\it perturbation} of a current with index $i$ to {\it all} other unperturbed currents with index $j\ne i$). Furthermore, in the worldsheet formalism, there exists an additional set of state parameters, namely the scalar product of all pairs of distinct phase gradients that, in the $N=2$ condensate string, are naturally identified with the parameter $x$ of the microscopic field model.  These new parameters naturally complement the ones obtained when going from one to two currents. 

Given that there exists {\it a priori} no procedure to solve the set of longitudinal equations in the case of $N$ currents, we carried out the numerical analysis in an application to a two-condensate string.  In this case, the condition for stability of the transverse modes, namely $T>0$, can be turned into a restriction on the range of possible variations condition on the third state parameter.  The perturbation equations in the longitudinal direction reduces to a set of four coupled equations in which case the condition for stability turns into a constraint on the solutions of a $4^\mathrm{th}$ order polynomial equation.  This condition is that its roots, which are functions of the dispersion relation of a given mode, be real.  Algebraic solutions of this equation were obtained using the well-known Ferrari procedure.  Combining the constraint on $x$ with the constraints on the roots of the  $4^\mathrm{th}$ order polynomial equation completes the stability analysis of the two-condensate string.  The regions of stability were shown in Figs~\ref{fig:stability1}, \ref{fig:stability2}, \ref{fig:stability3}, and \ref{fig:stability4}.\\

In the third part of this work, we investigated whether the study of decoupled fields whose integrated Lagrangian can be reproduced using simple approximate analytic formulae is suitable to reproduce the physics in the two-coupled fields case.  In this approximation, the Lagrangian is simply given by the sum of two single-current string Lagrangians.  As expected, we found that even in this approximation, the energy per unit length and tension depend on both the single-current state parameters and $x$.  Nevertheless, even when the nonlinear coupling constant is set to zero in the fully interacting Lagrangian, the simplified model should not be taken as a description of precisely the same physics because in the fully interacting Lagrangian, there is an interplay between the two scalar fields through their couplings to the Higgs and associated $U(1)^{\mrm{local}}$ gauge field.  As a result, one should always expect a discrepency between the two.  In short, the full dynamical evolution of a two-current string, and thus presumably of an $N-$current string, is not expected to be fully reproduced by such a model.  However, the ability to give an approximate description of the physics in the interacting case with a fully analytic model is attractive and as shown in Figs~\ref{fig:sectionsY}, and \ref{fig:sectionsZ}, the energy per unit length and tension are very well approximated by the analytic description of uncoupled fields as long as the coupling $g$ is small.  Larger deviations are however observed for stronger couplings.\\

We completed the study of the approximate model with an analysis of its stability against transverse and longitudinal perturbations in the currents.  It was found that stability of the string against longitudinal perturbations yields the constrainst $\mrm{d}T/\mrm{d}U<0$.  This is of course identical to the result obtained for a string with a single condensate but different from the result obtained the second part of the paper in which  coupled fields were considered.  On the other hand, the stability of the string against transverse perturbations yields a constraint on $x$, as in the fully coupled case.  We finally compared the regions of stability predicted by the analuytic model of uncoupled fields to the ones obtained in the interacting theory in Figs~\ref{fig:stability1}, \ref{fig:stability2}, \ref{fig:stability3}, and \ref{fig:stability4} and found good agreement at weak coupling only.  We therefore conclude that it is possible to address the stability of string endowed with coupled fields in a fully analytic way with satisfactory accuracy at weak coupling only.

\acknowledgments 
We thank B.~Carter, G.~Comer, A.~C.~Cordero and R.~Prix for enlightening
discussions.

\bibliography{references}

\end{document}